\documentclass{article}
\usepackage[T1]{fontenc}
\usepackage[latin1]{inputenc}
\usepackage{graphics}
\usepackage{graphicx}
\usepackage{longtable}
\usepackage{color}
\usepackage{pslatex}
\usepackage{hyperref}
\usepackage{sverb}
\usepackage{subfigure}

\usepackage{multirow}
\usepackage{adjustbox}
\usepackage{indentfirst}
\usepackage{mathrsfs}

\usepackage{rotating}

\usepackage{graphicx}
\usepackage{subfigure}
\usepackage{longtable}
\usepackage{multirow}
\usepackage{textcomp}
\usepackage{authblk,colortbl}
\usepackage{units}
\usepackage{amsmath,amssymb,amsfonts} 

\usepackage{authblk}

\usepackage{rotating}
\definecolor{colortwo}{rgb}{1,0,0.0}

\definecolor{colorfour}{rgb}{0.6,0.3,0}

\begin{document}

\begin{titlepage}
 
\begin{flushright} 
{ \bf IFJPAN-IV-2020-10 
} 
\end{flushright}
 
\vskip 30 mm
\begin{center}
  {\bf\huge Adequacy of Effective Born for  electroweak effects
       }\\
  \vskip 4mm
    {\bf\huge and   {\tt TauSpinner  } algorithms for LEP, Tevatron, HL-LHC and FCC simulated samples. }
\end{center}
\vskip 13 mm

\begin{center}
   {\bf  E. Richter-Was$^{a}$ and Z. Was$^{b}$}\\
   \vskip 3 mm
   {\em $^a$ Institute of Physics, Jagellonian University, ul. Lojasiewicza 11, 30-348 Krak\'ow, Poland} \\
   {\em $^b$  IFJ-PAN, 31-342, ul. Radzikowskiego 152, Krak\'ow, Poland}
\end{center}
\vspace{1.1 cm}
\begin{center}
{\bf   ABSTRACT  }
\end{center}
Matching and comparing  the measurements of past and future experiments calls for consistency checks of
 calculations used for their interpretation. On the other hand, 
new calculation schemes of the field theory   
can be beneficial for precision, even if they  may obscure comparisons with earlier results.
Over the years concepts of {\it Improved Born}, {\it Effective Born}, as well as of
effective couplings, in particular of $\sin^2\theta_W^{eff} $ mixing angle for electroweak
interactions, have evolved.
\\
In our  discussion we use  four {\tt DIZET} electroweak  library versions;
that of today and of the last 30 years. They were used for phenomenology
of practically all HEP accelerator
experiments. Versions differ by incremental updates of certain types of
corrections which became available with time.
We rely on the codes published and archived with the 
{\tt KKMC} Monte Carlo program for $e^+e^- \to f \bar f n(\gamma)$.
All these  versions became recently  available
for the  {\tt TauSpinner}
algorithm of simulated event reweighting as well. Such reweighting can be
performed after events are generated and stored in data files.
To this end  {\tt DIZET} is first invoked, and its results
are used. Documentation of  {\tt TauSpinner} upgrade,  to version 2.1.0, and of
its arrangement for semi-automated electroweak effects benchmark plots are provided.
Some details of the tool, which can be used to
complete simulation sample with electroweak effects, are given.

Focus  is placed on the numerical results,
on the different approximations introduced
in Improved Born to obtain Effective Born, suited better to match
with QCD corrections.
The $\tau$ lepton polarization $P_{\tau}$, forward backward asymmetry $A_{FB}$
and parton level total cross section $\sigma^{tot}$ are used to monitor the
size of electroweak effects and  effective $\sin^2\theta_W^{eff}$ picture
limitations for precision physics.
Collected results  include:
(i) feasibility of {\it Effective Born} approximation and 
$\sin^2\theta_W^{eff}$, (ii)  differences between
versions of electroweak libraries  
 and (iii) parametric ambiguities due to e.g.
 $m_t$ or  $\Delta \alpha_h^{(5)}(s)$. These results  can be considered
 as  examples only, but allow one to evaluate the adequacy of {\it Effective Born} with respect to
 {\it Improved Born}. Definitions are addressed too.

\vskip 1 cm


\vfill
{\small
\begin{flushleft}{   IFJPAN-IV-2020-10
December 2020
}
\end{flushleft}
}
 
\vspace*{1mm}
\footnoterule
\noindent
{\footnotesize 
This project was supported in part from funds of Polish National Science
Centre under decisions DEC-2017/27/B/ST2/01391.\\
Partly supported by 
the CERN FCC Design Study Program. 
Majority of the numerical calculations were performed at the PLGrid Infrastructure of 
the Academic Computer Centre CYFRONET AGH in Krakow, Poland.
}
\end{titlepage}
\section{Introduction}
One of  precision high energy physics great achievements in  electroweak (EW)
sector
measurements is establishing that quantum field theory can be indeed
used to calculate predictions that match  measurement  predictions \cite{ALEPH:2010aa,ALEPH:2005ab}.
To handle results and  interpretation, the concept of idealized observables
was very useful \cite{Davier:1992nw}.
Over many years the effective EW mixing angle $\sin^2\theta_W^{eff}$
(of the process $e^+e^- \to Z/\gamma^* \to f \bar f$, where $f$ denote leptons
or quarks)
was a prime candidate to evaluate sensitivity of observations of the
EW sectors and in fact established itself as the idealized
observable~\cite{ALEPH:2005ab}. This universal quantity
is not of the principal nature and to remain useful, 
confirmation that the  necessary approximation still
holds is needed. Already in the past,
limits for the validity of the assumptions made were investigated.
Let us point to an early reference \cite{Gambino:1993dd}, where dependence
on flavour and the flavour-dependent $\sin^2\theta_W^{eff}$ were elaborated.

In the calculation scheme used in LEP times, EW corrections calculated
at one-loop level were improved with selected, dominant higher order terms
and embedded in the Improved Born Approximation \cite{Bardin:1999yd}. This
formulation was a cornerstone for LEP measurements \cite{ALEPH:2010aa}. 
The Effective Born means redefining coupling
constants to the values incorporating dominant contributions from the higher
order (loop) corrections (at fixed energy, e.g. $\sqrt{s}$ or $M_Z$). Such
a redefinition usually does not break properties
required for factorizing out calculation of  strong interaction effects.
In contrary, Improved Born where couplings are accompanied
with energy- and angle-dependent, complex form-factors if used directly, can
easily damage strong interaction gauge invariance
  and force the necessity of simultaneous calculation
of the EW and QCD effects, without conveniences of factorization.
The effective coupling  definitions rely on properties of EW loop
corrections, which could invalidate the concept
of  Effective Born and  $\sin^2\theta_W^{eff}$ definition
as well.
The concept has evolved  over time
\cite{Barroso:1987ae,Bardin:1999gt,Hollik:2005va}. In fact, approach
variants, the controversies and differences in conventions can be now
easily identified and avoided.
One should keep in mind that such conventions were used in data
analyzes of the past.
With the improving measurements precision, one has to readdress
the validity of assumptions and approximations necessary for 
definitions and usefulness of Effective 
and  Improved  Born.

To address the above points 
one needs to investigate first if Improved
or Effective Born can be separated with sufficient precision from complete
calculations including strong interactions. 
References \cite{Richter-Was:2016mal,Richter-Was:2016avq} were devoted to studying
how the  strong interaction separates in  LHC processes of $W$ and $Z$ boson
production and decay. The study was necessary for validation of
{\tt TauSpinner} event reweighting
algorithm~\cite{Czyczula:2012ny,Przedzinski:2018ett} in its implementation
of EW effects \cite{Richter-Was:2018lld}.
Similar evaluations were completed in the past
for $e^+e^-$ collisions
in context of Monte Carlo generators \cite{koralz4:1994,kkcpc:1999,Jadach:2000ir} and for
semi-analytical calculations in \cite{Bardin:1999yd}.
One of the important numerical assumptions was that in all these applications,
numerical differences between Improved and Effective Born are, around the $Z$ pole,
not too large. 
The one-loop genuine weak corrections were usually calculated for all these projects
with the help of  {\tt DIZET} library \cite{Bardin:1989tq}. Important was the inclusion of
some higher order strong interaction or QED effects. One has always  to check if
 necessary higher order contributions to weak loop effects can be (and are)
introduced into Effective or Improved Born and thus partly re summed as well.
For  {\tt DIZET} archivization and evolution see Ref.~\cite{Arbuzov:2020} and references therein.

Obviously, whenever precision is expected to 
improve, assumptions behind calculations need to be revisited.  The {\tt TauSpinner} algorithms can be
helpful in that respect and used to evaluate if for a given observable
some classes of the corrections are necessary or can be ignored.  
Independently if old calculations are sufficient or if the new one may
be necessary, it is useful to establish which of the effects need to be taken
into account.

We focus on
discussion/evaluation of:
\hskip 2mm  (i) the suitability of Effective
versus Improved Born approximation (also the usefulness of $\sin^2\theta_W^{eff}$),
\hskip 2 mm
(ii) differences between results of {\tt DIZET} library versions in use
over the last 30 years,
\hskip 2mm
(iii) ambiguities due to parametric uncertainties or  due to (sometimes)
missing contributions.
We explain minor, but useful for studies of 
EW effects, {\tt TauSpinner} extensions
with respect to Ref.~\cite{Richter-Was:2018lld}, too.

In the scope of the paper, we present numerical results, either
from semi-analytical calculations
(predominantly for $e^+e^- \to l^+l^-$ processes)
or from reweighting of the Monte Carlo event samples
(predominantly for $ pp \to Z/\gamma^* \to l^+l^-$ processes at 8~TeV center
of mass collisions).
In the second case we  use Powheg Monte Carlo $Z+ \; jets$ events as described
in Refs.~\cite{Richter-Was:2016mal,Richter-Was:2018lld}.

Section \ref{sec:effBorn} recalls basic definitions necessary for the
introduction of the Improved Born practical comments on $\sin^2\theta_W^{eff}$
definitions are provided. Simplifications enabling introduction
of Effective Born are explained as well. 
The purpose of Section~\ref{sec:conditions} is  to recall first numerical
results on non-EW effects which are instrumental for  the concept of
Effective Born and effective couplings.
Both Improved Born and Effective Born require interpolation of the
$2\to 2$ kinematics and evaluation of its  scattering angle. This requires
careful optimalization in the presence of QED/QCD initial and final state
emissions; see references above, which discuss
the issue.
In this context, redefinitions of couplings are convenient as
such couplings  can be usually used in strong interactions amplitudes without
complication  
of  gauge dependence restoration. Alternatively, re-weighting to Improved Born
can be advocated following e.g. the {\tt TauSpinner} solution of 
Ref.~\cite{Richter-Was:2018lld}. The particular choice will need to be decided by the user
precision requirements. That is why we do not give any guidelines here.
Definitions of simplified test observables used all over the paper are provided.

Section \ref{sec:FromIBtoEF} is devoted to the comparison of  Improved Born and  Effective
Born. In particular, results useful to evaluate
precision
of Effective  Born and $\sin^2\theta_W^{eff}$ approximations with respect
to Improved Born are provided.
Most of the numerical details, 
 are delegated to Appendices.  Ref. \cite{Arbuzov:2020}
summarizes numerically most important upgrades of EW library
{\tt DIZET}.  Some numerical comparisons of its variants are collected
in Section \ref{sec:onLIbraries}.
Numerical results for parametric uncertainties are given in 
 Subsection \ref{sec:numerical}. Summary, Section \ref{sec:summary}, closes the paper.

 Technical and physics content details of {\tt DIZET}
library versions, including used by us in these cases  initialization parameters are collected
in Appendix~\ref{app:6XXEW}. Alternative EW projects and calculation
schemes are not discussed and we address interested readers to the 
documentation of {\tt KKMC} \cite{kkcpc:1999} or even older
{\tt KORALZ} documentation \cite{koralz4:1994}.
Appendix  \ref{app:B} supplements the paper with
technical details on the  {\tt TauSpinner}
re-weighting algorithm.
Details of 
variants of Born definitions and activating them flags available in {\tt TauSpinner} are given in
Appendix~\ref{app:variantseff}. 
Appendix \ref{app:Zwidth} addresses the important but auxiliary point of the $Z$
propagator with running or fixed width
and provides corresponding 
numerical
results.  Appendix \ref{app:kkmc} enumerates versions of {\tt DIZET} library,
which are available from Ref.~\cite{Arbuzov:2020}.

\section{Improved Born and electroweak form-factors.}
\label{sec:effBorn}
In the Improved Born Approximation, the complete $O(\alpha)$ EW corrections,
supplemented by selected higher order terms, are handled with form-factor corrections, dependent on (s,t), multiplying couplings
and propagators of the usual Born expressions.
Let us continue with definition
of Improved Born used in  {\tt TauSpinner}. It is
detailed in Ref.~\cite{Richter-Was:2018lld} but we will recall
it with Eq.~(\ref{Eq:BornEW}) for the process $e^+e^- \to f \bar f$.
The formula can be used also in the case when initial and final state are
interchanged. The $z$ component of the fermion isospin $T_3^{e,f}$,
EW mixing angle
$s_W^2=\sin^2\theta_W$, $c_W^2=1-s_W^2$ and electric charge $q_{e,f}$ are used,
as usual,
for the coupling constant calculations. The Mandelstam variables
$s=(p_{e^+}+p_{e^-})^2$  and  $t=(p_{f}-p_{e^-})^2$ are used for the kinematical
dependence. The Fermi coupling $G_\mu$, QED coupling constant $\alpha$, the $Z$ boson mass $M_Z$ and
width $\Gamma_Z$ complete basic notations. Definitions of EW
form-factors   
$ { K_{e}(s,t)},{ K_{f}(s,t)},{ K_{ef}(s,t)}$, for $\ell=e,\mu,\tau$, 
photon vacuum polarization ${ \Pi_{\gamma \gamma}(s)}$ and  ${ \rho_{\ell f}(s,t) }$, are as used in \cite{Richter-Was:2018lld}.
It is important that they are only weakly dependent on $t$, and the $s$
dependence is not sizable as well\footnote{At the LO EW
  $K_{e}(s,t)=K_{f}(s,t)=\rho_{\ell f}(s,t)=1, \; \Pi_{\gamma \gamma}(s)=0$ and
  $\frac{G_\mu M_Z^2 \Delta^2}{\sqrt{2}8\pi\alpha}=1$. We use
  $s^2_W=1-M_W^2/M_Z^2$ for the on-mass-shell
  definition, while $\sin^2\theta_W^{eff}$
  is used for 
  the effective value corresponding to the ratio of couplings at the $Z$-pole
  $v^{eff}_f/a^{eff}_f= 1-4q_f \sin^2\theta_W^{eff\ f}$.}.

\begin{eqnarray}
  \label{Eq:BornEW}
  ME_{Born+EW}   = &  {{\cal N}\frac{\alpha}{s}}\bigl\{& [\bar u \gamma^{\mu} v   g_{\mu \nu}  \bar v \gamma^{\nu} u] \cdot ( q_e \cdot q_f)  \cdot  { \Gamma_{V_{\Pi}}} \cdot {\chi_{\gamma}(s)} \nonumber \\
       && + [\bar u \gamma^{\mu} v g_{\mu \nu} \bar \nu \gamma^{\nu} u  \cdot  ( v_e \cdot v_f  \cdot vv_{ef}) +  \bar u \gamma^{\mu} v g_{\mu \nu} \bar \nu \gamma^{\nu} \gamma^5 u  \cdot  (v_e \cdot a_f) \\
       && +  \bar u \gamma^{\mu} \gamma^5 v g_{\mu \nu} \bar \nu \gamma^{\nu}  u  \cdot  (a_e \cdot v_f) + \bar u \gamma^{\mu} \gamma^5  v  g_{\mu \nu}\bar \nu \gamma^{\nu} \gamma^5  u  \cdot  (a_e \cdot a_f) ]
        \cdot { Z_{V_{\Pi}}} \cdot {\chi_Z (s)}\; \bigr\}, \nonumber
\end{eqnarray}

\begin{eqnarray}
  v_e && = (2 \cdot T_3^e - 4 \cdot q_e \cdot s^2_W \cdot { K_e(s,t)})/\Delta, \nonumber \\
  v_f && = (2 \cdot T_3^f - 4 \cdot q_f \cdot s^2_W \cdot { K_f(s,t)})/\Delta, \nonumber  \\
  a_e && = (2 \cdot T_3^e )/\Delta, \hskip 12 mm s^2_W=(1-c_W^2)= 1-M_W^2/M_Z^2, \nonumber \\
  a_f && = (2 \cdot T_3^f )/\Delta, \hskip 12 mm \Delta = 4 s_W c_W, \nonumber\\
  \chi_Z(s) &&=   \frac{G_{\mu} \cdot M_{z}^2  \cdot \Delta^2 }{\sqrt{2} \cdot 8 \pi \cdot \alpha}\cdot \frac{s}{s - M_Z^2 + i \cdot \Gamma_Z \cdot s/M_Z},  \nonumber \\
  \Gamma_{V_{\Pi}} && = \frac{1}{ 2 - (1 +  { \Pi_{\gamma \gamma}(s)})}, \hskip 8 mm
   Z_{V_{\Pi}} = { \rho_{\ell f}(s,t)},\ \hskip 8 mm  \chi_{\gamma}(s) = 1,\nonumber
\end{eqnarray}
\begin{eqnarray}
  vv_{ef} =&& \frac{1}{v_e \cdot v_f} [    
    ( 2 \cdot T_3^e) (2 \cdot T_3^f) - 4 \cdot q_e \cdot s^2_W \cdot { K_f(s,t)}  - 4 \cdot q_f \cdot s^2_W \cdot { K_e(s,t)} \nonumber \\
    && + (4 \cdot q_e \cdot s^2_W) (4 \cdot q_f \cdot s^2_W) { K_{ef}(s,t)}] \frac{1}{\Delta^2}. \nonumber 
\end{eqnarray}
\noindent
In the formula $u,\ v$ stand for spinors - fermions wave functions, and $\cal N$ is a normalization factor
which is convention dependent (e.g. for wave functions normalization).

The formula
(\ref{Eq:BornEW}) is not the only possibility for implementation of EW
corrections.
Genuine EW corrections can be, under some conditions, combined
with the ones of QED or strong interactions. 
For example in {\tt KKMC} implementation, the Improved or Effective Born approximation  is not
 used.  EW form-factors are installed into spin amplitudes directly%
 \footnote{
That could be disastrous, as gauge cancellations would be broken.
Spinor techniques of Kleiss-Stirling
\cite{kleiss-stirling:1985} are exploited for {\tt KKMC}
Monte Carlo of $e^+e^-\to l^+l^- n\gamma$ processes, where second order
QED matrix element and coherent exclusive exponentiation is used  \cite{Jadach:2000ir,kkcpc:1999}.
Nonphysical huge contributions
proportional even to
$\sim 1/m_e^2$ could appear. However this is not the case as contributions
to Yennie Frautchi Suura spin amplitude level $\beta_0$, $\beta_1$, ...
terms are calculated explicitly and gauge cancellations are explicitly
performed, before EW form-factors installation.

One needs to keep in  mind this difficulty when other applications are
developed. Even in {\tt KKMC} case this needed to be watched after, in the context
of e.g. IFI interference contribution, where cancellation of real and
virtual corrections play a role and may be obscured by energy angular
dependence of form-factors. Formally of higher order, such
mismatches could substantially impact conclusions if attention was not paid.
 }.
 Care of the gauge cancellation was essential for that.
 The ``running'' of $\Gamma_Z$ for the $\chi_Z(s)$ propagator is used in Eq.~(\ref{Eq:BornEW}) as was commonly the case for LEP physics, but less so for LHC oriented MC's.   
Another possibility is to use Effective Born (not  exact at  one loop level, it will
 be discussed in Section \ref{sec:effective}).
 The main idea
is to simplify formula (\ref{Eq:BornEW}). In particular, include in 
$\sin^2\theta_W^{eff}$ the bulk effect of $K_e(s,t)$ and $K_f(s,t)$ form-factors
 present in front of $s^2_W$, and vacuum polarization
corrections  $\Pi_{\gamma \gamma}$ into a redefinition of $\alpha$.  In fact a real part or
a module, and all calculated at the $Z$ pole.  It is also possible to think of 
$\sin^2\theta_W^{eff}$ as the best result of the fit to the data.
At certain precision level, such distinction may start to play a role
and one should be aware of alternatives to avoid misunderstandings.

Definitions of form-factors follows the one of the {\tt DIZET} library
\cite{Bardin:1989tq,Arbuzov:2020}
and are as used for spin amplitudes in the {\tt KKMC} Monte Carlo as well.
For convenience they are used in the  Improved Born, formula (\ref{Eq:BornEW}),
of {\tt TauSpinner} too. Details of the EW scheme, initialization parameters
are provided in Appendix \ref{app:6XXEW}. In Fig. \ref{Fig:Re_Rho_K_b_wide}
EW form-factors calculated with  {\tt DIZET} library are
drawn.

\begin{figure}[htp!]
\begin{tabular}{ccc}
  \includegraphics[width=0.44\columnwidth]{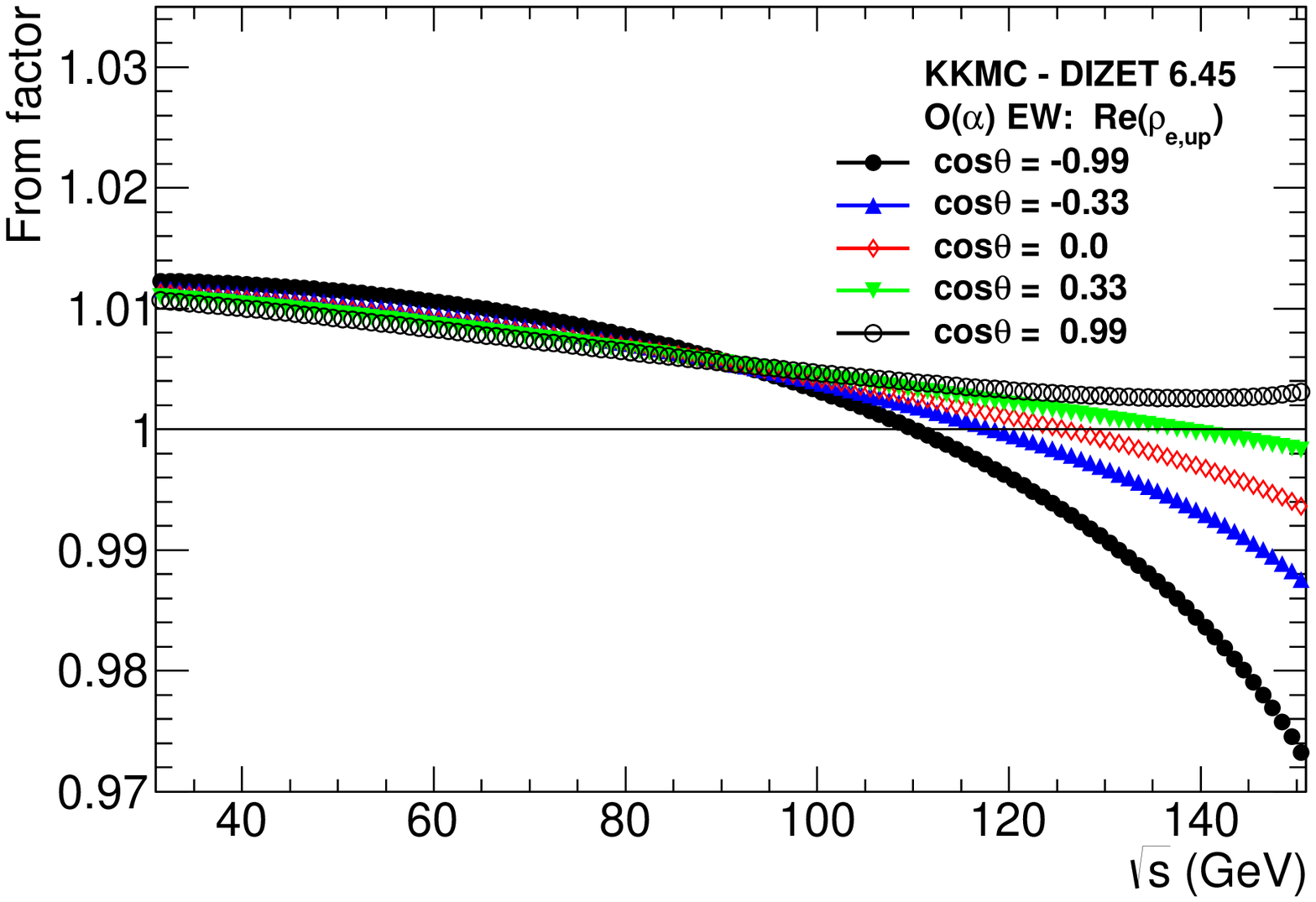} &
  \includegraphics[width=0.44\columnwidth]{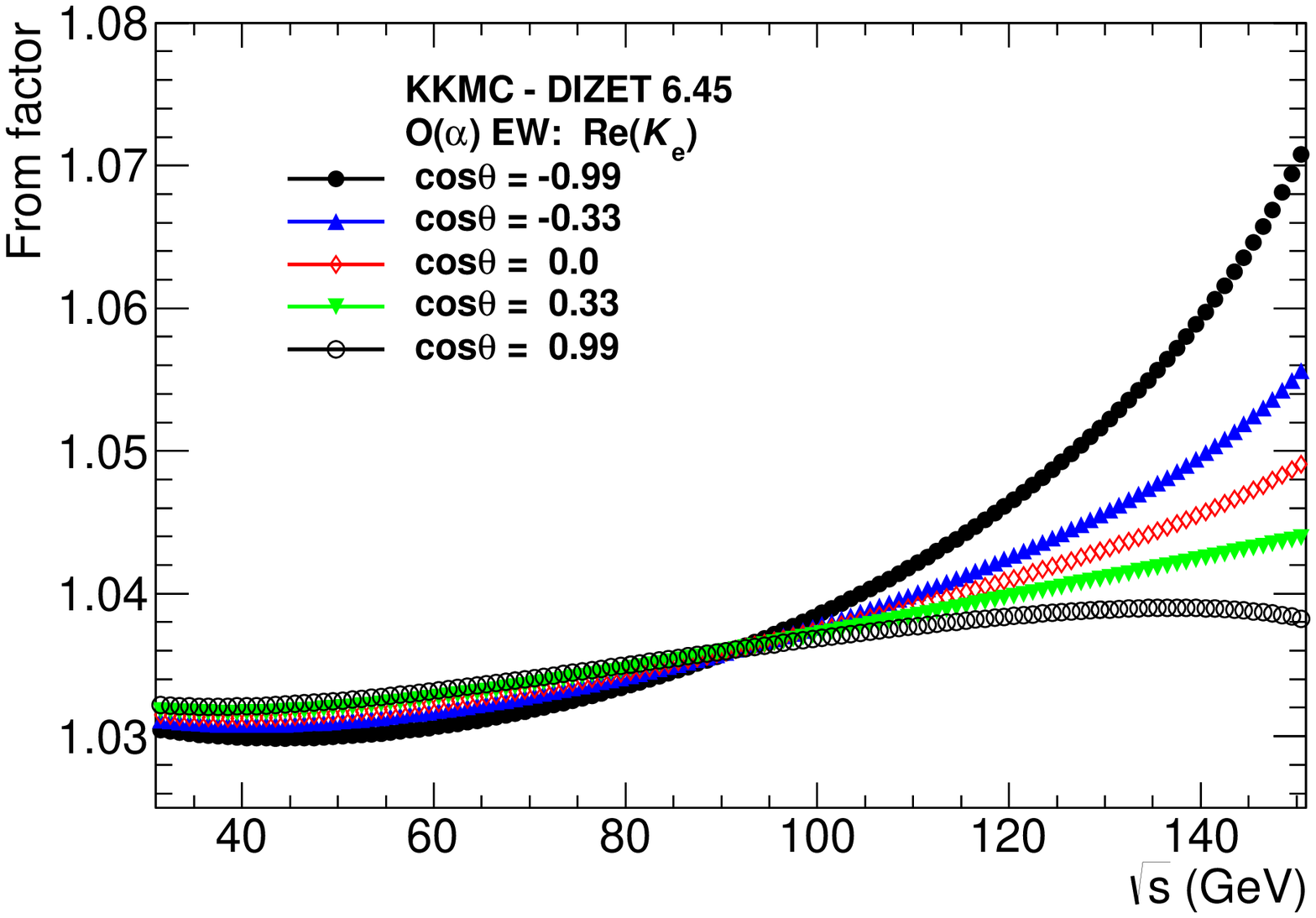}  \\
  \includegraphics[width=0.44\columnwidth]{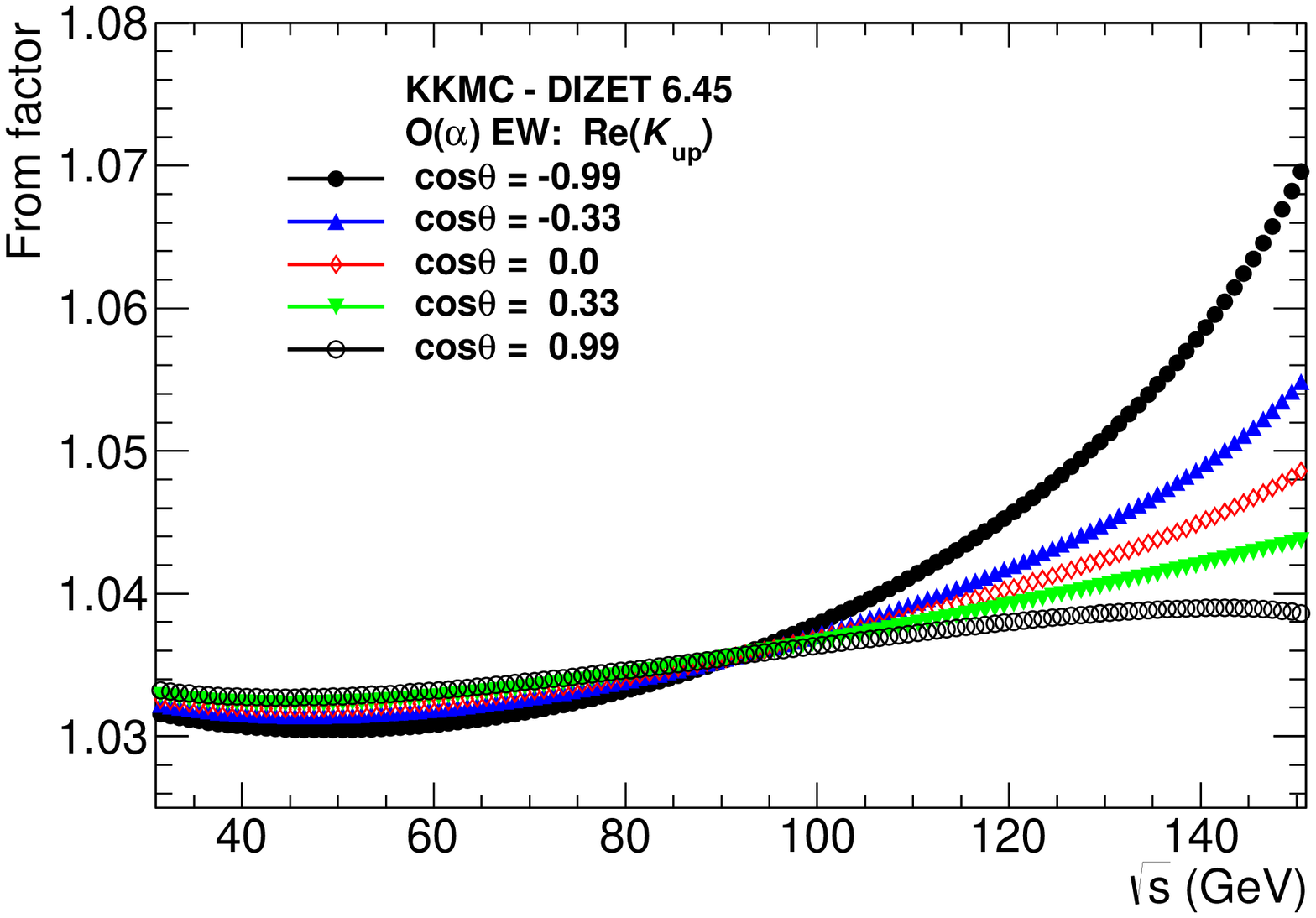} &
  \includegraphics[width=0.44\columnwidth]{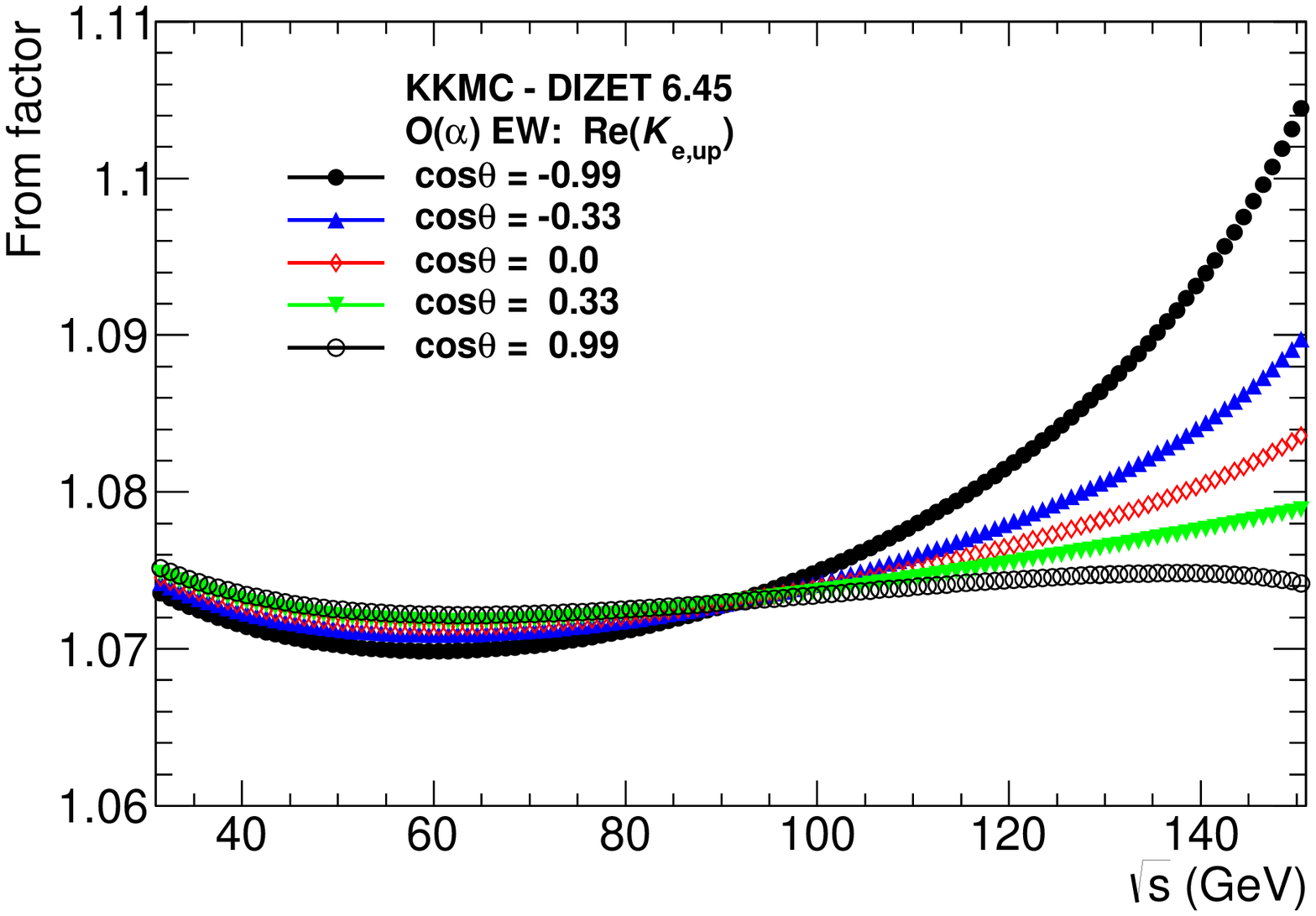} \\
  \includegraphics[width=0.44\columnwidth]{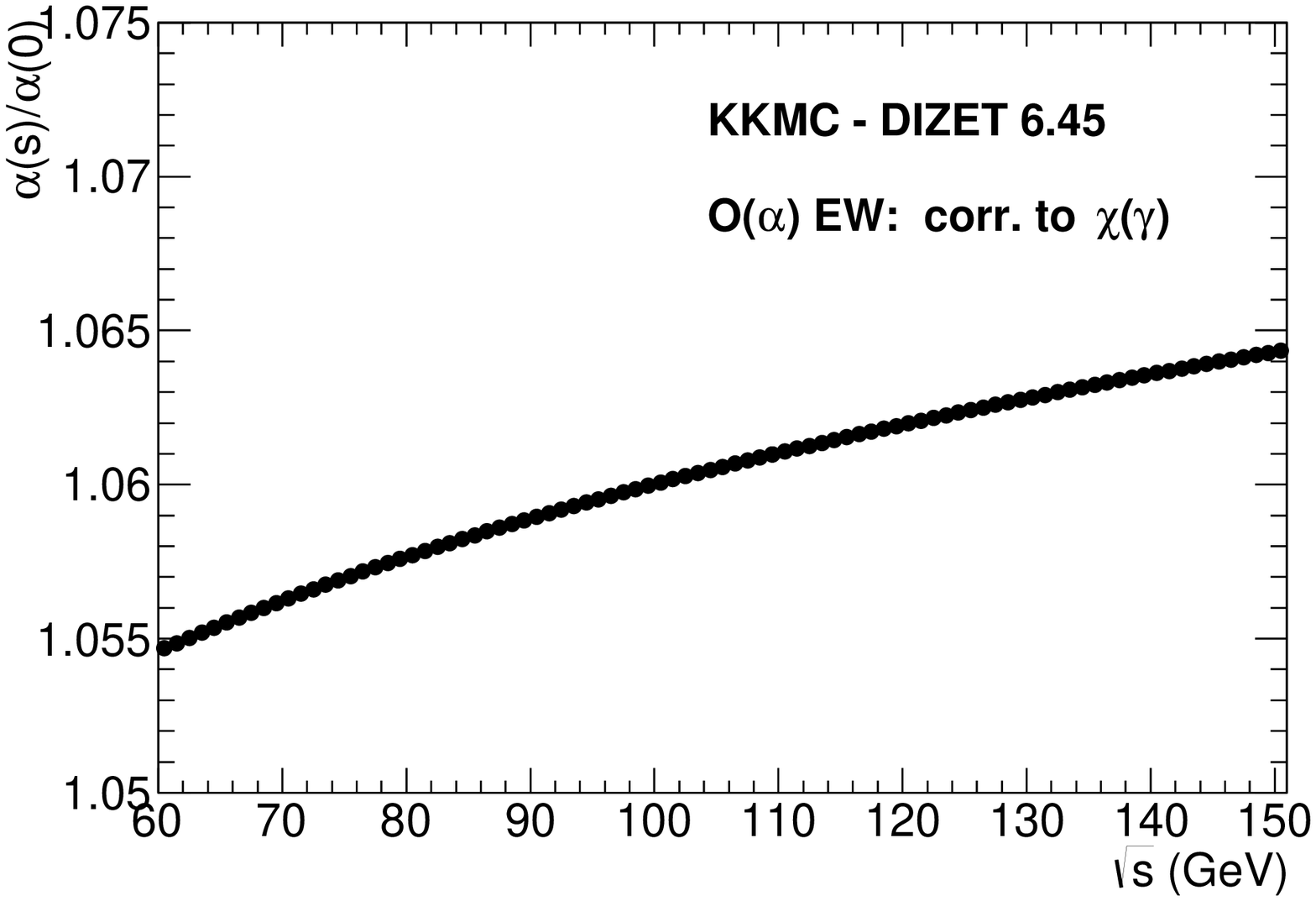} &
\end{tabular}
  \caption{\small 
    Plots from the new {\tt examples/Dizet-example} directory of {\tt TauSpinner}. The real parts of the
    $\rho_{e, up}$, ${\mathscr K}_{e}$, ${\mathscr K}_{up}$ and  ${\mathscr K}_{e, up}$
    EW form-factors of $ee \to Z \to u\bar u$ process,   
    as a function of $\sqrt{s}$ and for the few values of $\cos \theta$
    Note, that ${\mathscr K_{e,up}}$ depends on the flavour of outgoing quarks.
    On the last plot (bottom line) the ratio $\alpha(s)/\alpha(0)= \Gamma_{V_\Pi}$  is also shown.
\label{Fig:Re_Rho_K_b_wide} }
\end{figure}

A disadvantage of the Improved  Born with respect to the Effective one
is that it is not immediate to merge its
formulation into calculations for  strong interactions.
Formula~(\ref{Eq:BornEW})  can be used as
a starting point to explain
relations between Improved Born and Effective Born, where $s$- and $t$-dependent form-factors are avoided.
For the latter, form-factors are often set to unity, but may be
replaced with the real constants without compromising strong interaction
calculations. One should keep in mind that 
numerical values for e.g. $s^2_W$ and $\alpha$ need then to be
chosen differently, also 
 to accumulate dominant numerical contributions of the loop corrections.
Technical details for the arrangements used in the programs  are collected in Table \ref{Tab:gswnames} of
the Appendix \ref{app:variantseff}.

\section{  Factorization requirements for Born amplitudes}
\label{sec:conditions}
In the general case, but especially for  lepton pair production at the LHC,
the definition and use of the quark $2\to 2$ Born-level scattering process  as
a building block for phenomenology, may seem  rather crude and difficult to control. Let us
recall arguments why it is not necessarily the case. 

Properties of Born-level spin amplitudes lead
to  features necessary for its factorization from complete formulae. Quality of such separation is
of a decisive importance. Already a long time ago \cite{Berends:1983mi}, even in
the presence of
hard bremsstrahlung photons, the part of  amplitudes which corresponds to the Born-level
distribution were  identified and separated out.
This was studied in the context of hadronic kinematic configuration 
of $pp$ LHC physics as well \cite{Richter-Was:2016mal,Richter-Was:2016avq}
and for configurations with up to two high $p_T$ jets.
These factorization properties were expected, thanks to e.g. results of \cite{Mirkes:1994dp}.
They require that the  Born cross section is described by spherical
harmonics of the second order.
Indeed, the Born cross section for $f \bar f \to Z/\gamma^* \to \ell^+ \ell^-$
 reads (azimuthal angle dependence can be avoided with appropriate choice of the reference frame):
 
{\small 
 \begin{equation}
   \label{Eq:Bornqqbar}
   \frac{d \sigma^{q \bar q}_{Born}}{d \cos\theta} (s, \cos\theta, p) =
   (1 + \cos^2\theta)\ F_0(s) + 2 \cos\theta\ F_{1}(s) - p [(1+\cos^2\theta )\ F_2(s) + 2 \cos\theta\ F_3(s)],
   \end{equation}
}
where  $p$  denotes polarization  of the outgoing leptons,
 $ \theta$ an angle between incoming quark and outgoing lepton in the rest frame of outgoing leptons.
For a general orientation of the reference frame all second order
spherical harmonics in $\theta$ and $\phi$ angles appear.
Second order spherical harmonics are sufficient also when  transverse spin effects are taken into account.

 The $F_i$ read:
{\small
 \begin{eqnarray}
   F_0(s) && = \frac{\pi \alpha^2}{2 s } [ q^2_f q^2_{\ell}\cdot  \chi^2_{\gamma}(s) + 2 \cdot \chi_{\gamma}(s) Re\chi_Z(s)\ q_fq_{\ell}v_fv_{\ell}
     + | \chi_Z^2(s)|^2 (v_f^2 + a_f^2) (v_{\ell}^2 + a_{\ell}^{2}) ], \nonumber \\
   F_1(s) && = \frac{\pi \alpha^2}{2 s } [ 2 \chi_{\gamma}(s) Re\chi(s)\ q_fq_{\ell}v_fv_{\ell} +  | \chi^2(s)|^2\ 2 v_f a_f 2 v_{\ell} a_{\ell} ], \\
   F_2(s) && = \frac{\pi \alpha^2}{2 s } [ 2 \chi_{\gamma}(s) Re\chi(s)\ q_fq_{\ell}v_fv_{\ell} +  | \chi^2(s)|^2\  (v_f^2 + a_f^2) 2 v_{\ell} a_{\ell} ], \nonumber \\
   F_3(s) && = \frac{\pi \alpha^2}{2 s} [ 2 \chi_{\gamma}(s) Re\chi(s)\ q_fq_{\ell}v_fv_{\ell} +  | \chi^2(s)|^2\  (v_f^2 + a_f^2) 2 v_{\ell} a_{\ell} ], \nonumber
 \end{eqnarray} 
}

\begin{figure}
  \begin{center}                               
{ \hskip -2cm
   \includegraphics[width=8.0cm,angle=0]{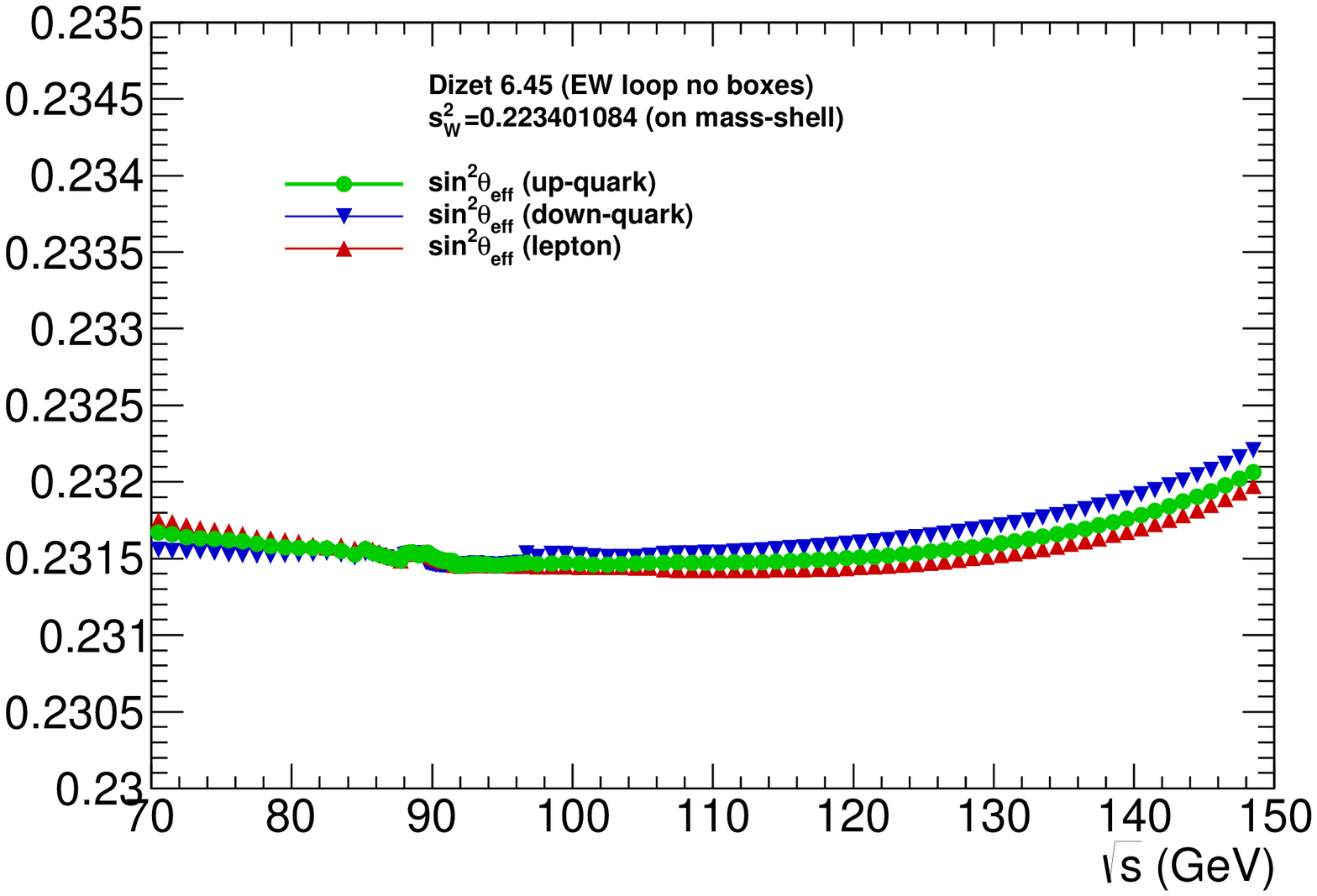}
   \includegraphics[width=8.0cm,angle=0]{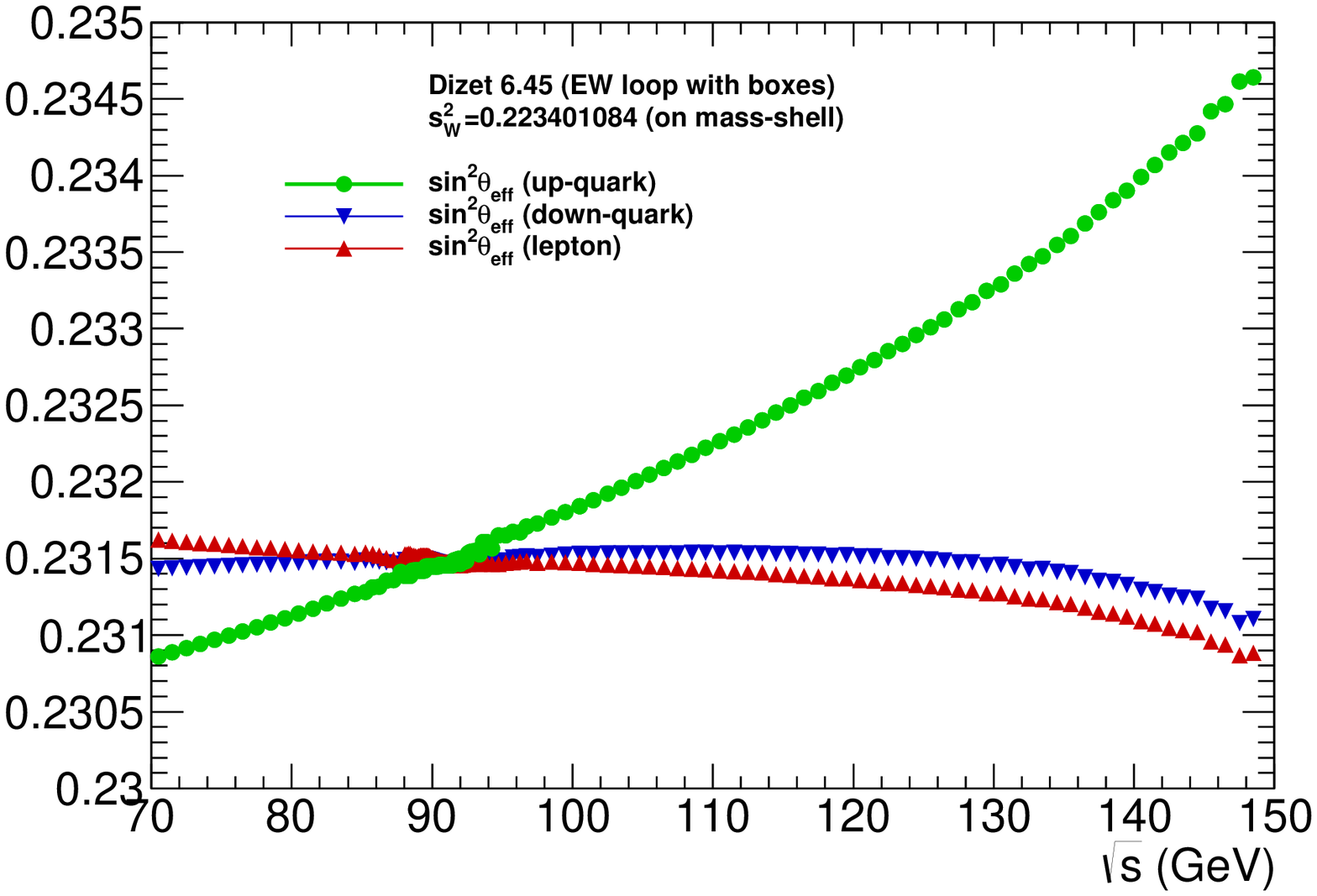}
}
\end{center}
  \caption{\small
    Averaged for $pp\to Zj, \; Z \to l\bar l$ events 
    effective weak mixing angles
    $\sin^2\theta_W^{f\; eff}(s)=Re(K^f(s,t))\ s^2_W$
    as a function of $\sqrt{s}$  and $t$-dependence integrated over,
    without (left-hand plot) and with (right-hand plot) box corrections.
    The ${\mathscr K}^f(s,t)$ form-factor calculated using {\tt DIZET 6.45}
    library and on-mass-shell  $s^2_W=0.22340108$ were used (see Table \ref{Tab:Dizet_6.XX_coupl}).
    The complete formula reads $\sin^2\theta_W^{eff}(s,t)= K^f(s,t))\ s^2_W + I_f^2(s,t)$. Real
    part of $K^f(s,t)$ is used, imaginary part and 
    $I^2_f(s,t)$ are  only about $10^{-4}$.
\label{Fig:sw2effpp} }
\end{figure}

Unfortunately with $s,t$-dependent EW  form-factors of Eq.~(\ref{Eq:BornEW}),
the assumption on spherical harmonics decomposition of the second order only does not hold.
Inevitably the  approximation needs to be re-checked if it matches the required precision.
Checks if approximations do not  deteriorate precision sizably,
can be obtained from semi-analytical calculation or from fits of re-weighted
events distributions. For many
applications it is sufficient to realize that an  assumption
of angle independent form-factors works well at the vicinity of the $Z$ peak.
 This can be observed for  form-factors presented in Fig.  
 \ref{Fig:Re_Rho_K_b_wide}:  lines corresponding to distinct scattering
 angles cross all  at about $\sqrt{s}=M_Z$.  This observation may be not always sufficient.
 If off peak contributions to the observable of interest  are sizable,
 then checks
 with figures indicating the range of $\sqrt{s}$ contributing to observables   
 need to be evaluated.
 To estimate,  $e^+e^-$, or parton level, total cross
 sections and
 asymmetries, like presented  later in
 Fig.~\ref{figTest1}, may be used.
 The $\sin^2\theta_W^{eff}$   represents a typical observable and/or 
   coupling constant.  That is why
   Fig. \ref{Fig:sw2effpp} of  $\sqrt{s}$ and flavour-dependent
   $\sin^2\theta_W^{eff}$  provides a hint on the size of the effect as well.
 For $t$-dependence, EW boxes 
  contribute with correction of isospin  dependent sign. On the other hand,
  the  $\sin^2\theta_W^{eff}$ variations remain below $20 \cdot 10^{-5}$ in the range of $M_Z\pm 5$~GeV.
  The  $t$-dependence
originating from $WW$ and $ZZ$ boxes becomes sizable once $s$ approaches $4M_W^2$,  the $W$-pair production threshold.
 
  Note that even if for a given observable the nearby of the $Z$ peak
dominates   for LEP (or even FCC), this may not be
the case for LHC or linear colliders.  There, off $Z$ peak contributions are 
larger due to PDF or beamstrahlung spread.
This needs to be kept in mind,  in context of the
$\sin^2\theta_W^{eff}$ interpretation as universal idealized observable.
 We recall that off the $Z$ peak   form-factors dependence
 on flavour and scattering angle increase, see Fig.~\ref{Fig:Re_Rho_K_b_wide}.
It is therefore of interest to validate the range 
  where  form-factors angle and energy dependence can be
 safely ignored and the Effective Born Approximation used.

\subsection{Improved Born Approximation and Effective Born}
\label{sec:effective}
In principle it is not possible to absorb fully the  effects of
EW form-factors of Eq.~(\ref{Eq:BornEW}) (which are all complex and
angle/energy dependent) into rescaling of 
constants. In particular, introduction of $\sin^2\theta_W^{eff}$ is prone to
complications and ambiguities (see footnote$^1$ earlier in the text).
Let us now  recall
details of Effective Born amplitude definition, which differs
from formula~(\ref{Eq:BornEW}): 
the  form-factors are replaced by effective  coupling constants
(see also Table \ref{Tab:gswnames} of Appendix \ref{app:variantseff}).

\begin{eqnarray}
  \label{Eq:Borneff}
  ME_{Born-eff}   =  & {\cal N}\frac{\alpha}{s} \bigl\{&[\bar u \gamma^{\mu} v   g_{\mu \nu}  \bar v \gamma^{\nu} u] \cdot ( q_e \cdot q_f)  \cdot  { \Gamma_{V_{\Pi}}} \cdot {\chi_{\gamma}(s)} \nonumber \\
       && + [\bar u \gamma^{\mu} v g_{\mu \nu} \bar \nu \gamma^{\nu} u  \cdot  ( v_e \cdot v_f  \cdot vv_{ef}) +  \bar u \gamma^{\mu} v g_{\mu \nu} \bar \nu \gamma^{\nu} \gamma^5 u  \cdot  (v_e \cdot a_f) \\
       && +  \bar u \gamma^{\mu} \gamma^5 v g_{\mu \nu} \bar \nu \gamma^{\nu}  u  \cdot  (a_e \cdot v_f) + \bar u \gamma^{\mu} \gamma^5  v  g_{\mu \nu}\bar \nu \gamma^{\nu} \gamma^5  u  \cdot  (a_e \cdot a_f) ]
        \cdot { Z_{V_{\Pi}}} \cdot {\chi_Z (s)}\;  \bigr\}\nonumber
\end{eqnarray}

\begin{eqnarray}
  v_e && = (2 \cdot T_3^e - 4 \cdot q_e \cdot s^2_W )/\Delta \nonumber \\
  v_f && = (2 \cdot T_3^f - 4 \cdot q_f \cdot s^2_W )/\Delta \nonumber  \\
  a_e && = (2 \cdot T_3^e )/\Delta, \hskip 12 mm \Delta=4s_w c_w \nonumber \\
  a_f && = (2 \cdot T_3^f )/\Delta, \hskip 12 mm \chi_Z(s) =   \frac{G_{\mu} \cdot M_{z}^2  \cdot \Delta^2 }{\sqrt{2} \cdot 8 \pi \cdot \alpha}\cdot \frac{s}{s - M_Z^2 + i \cdot \Gamma_Z \cdot s/M_Z},  \nonumber \\
  \Gamma_{V_{\Pi}} && = 1, \hskip 8 mm
  Z_{V_{\Pi}} =\mathrm{Re} \rho_{\ell f}(M_Z^2),\ \hskip 8 mm  \chi_{\gamma}(s) = 1,  \nonumber\\
  vv_{ef}&& =1, \hskip 8 mm s^2_W=(1-c_W^2)=\sin^2\theta_W^{eff}(M_Z^2), \;
  \nonumber \\
  \alpha&&=\alpha(M_Z^2)= \frac{\alpha(0)}{2-(1+ \mathrm{Re} \Pi_{\gamma\gamma}(M_Z^2))}.\nonumber
\end{eqnarray}

Note that  $\sin^2\theta_W^{eff}(M_Z^2)$, $\rho_{\ell f}(M_Z^2)$ and
$\alpha(M_Z^2)$   are now used. They absorb dominant parts of EW corrections;
EW form-factors and vacuum polarization corrections.
This useful approximation may take into account bulk
of the EW effects, and couplings of fixed values are used.
There is some level of ambiguity in the numerical values. The best match to Improved Born should correspond
to the values predicted by these calculations.
In particular,  $\sin^2\theta_W^{eff}(M_Z)=$ Re$K(M_Z^2,-M_Z^2/2) s^2_W$,
$s^2_W=1-M_W^2/M_Z^2$, where $M_W$ is a calculated quantity including
EW corrections. The $s=M_Z^2$, $t=-M_Z^2/2$, corresponds to the Born-level with
scattering angle $\theta=0$.
Alternatively, one could use best measured values \cite{ALEPH:2010aa,ALEPH:2005ab} and not rely on the EW calculations.
This may be of importance for calculations which are
focused, e.g., on strong interactions.
Complications due to mixed EW and strong interaction loops in Feynman diagrams i.e.
gauge dependence non-cancellation
can be avoided. The price is  precision limitation with respect
to Improved Born.  
Such an approach was used
for the previous {\tt Tauola/TauSpinner} ME implementation,
with  $\sin^2\theta_W^{eff}(M_Z)$,
$\alpha(M_Z^2)$ as measured at LEP \cite{ALEPH:2005ab} and
$\rho_{\ell f}=1$  for simplification.

To  monitor EW effects we use $e^+e^-$ ($q\bar q$) to pair of leptons,
parton level cross section  $\sigma^{tot}$,
forward-backward asymmetry $A_{FB}$ and $\tau$ lepton polarization $P_\tau$,  as a function of $\sqrt{s}$.

\vskip 1 mm
\noindent
{\bf The $Z/\gamma^*$-boson $e^+e^-$ ( or $q\bar q$) cross section $\sigma^{tot}$}
in the EW LO,
depends only on coupling constants and two parameters $(M_Z, \Gamma_z)$.
The effect on $\sigma^{tot}$ from EW loop corrections are due to corrections to the propagators:
vacuum polarization corrections (running $\alpha$) and $\rho$ form-factor.
This causes a change in relative contributions of the $Z$ and $\gamma$, and change
of the  $Z$-boson vector to axial coupling ratio ($\sin^2\theta_W^{eff}$). They
affect not only $s-$dependence but normalization of the cross section too.

\vskip  1 mm \noindent
{\bf The forward-backward asymmetry $A_{FB}= \frac{\sigma(\cos \theta > 0) - \sigma(\cos \theta < 0)}{\sigma(\cos \theta > 0) + \sigma(\cos \theta < 0)}$ }\\
 is defined in a standard way.
For  $e^+e^-$ collision, an angle  $\theta$ between incoming particle and outgoing lepton is taken. For $pp$ collision the {\it Collins-Soper}
frame~\cite{Collins:1977iv} is used for  angle  $\theta$ definition.
The asymmetry varies strongly with $\sqrt{s}$ around the $Z$ peak because 
it is proportional to product of small   vector  $v_i$ and  $v_f$ couplings of
incoming parton  and outgoing lepton. The product is specially small for $e^+e^-$ initial state. That is why,
off the peak, $s$-channel $Z$-photon exchange interference quickly become sizable.

\vskip 1 mm \noindent
{\bf The $\tau$ polarization $P_\tau =  \frac{\sigma(R) - \sigma(L)}{\sigma(R) + \sigma(L)}$ }
where $\sigma(R/L)$ denote cross section for production of right/left hand
polarized $\tau$, 
       is of interest in itself as it offers independent
data-point for precision EW sector measurements. It is of convenience,
because, in the first approximation, it is linearly proportional to
$\sin^2\theta_W^{eff\; \tau}$, thus it is useful for discussion of systematic
ambiguities. The systematic errors for this measurement differ
from  that of $\sigma^{tot}$ or $A_{FB}$. Predominantly because $P_\tau$ is not
measured directly, but through distribution of $\tau$ decay products only.
These points were recently recalled  in \cite{Banerjee:2019mle}.
On the other hand, relation between  $P_\tau$, $Z$ couplings and $\sin^2\theta_W^{eff}$ is not affected and generally is of the same nature like for $A_{FB}$.
That is why this data point is particularly suitable for discussion
with $e^+e^-$ semi-analytical results.

\vskip 1 mm \noindent
{\bf Test observables and $\sin^2\theta_W^{eff}$. }
It is worth to point out that of the   $e^+e^-$ scattering results, the ones 
for the  $P_\tau$ are particularly convenient in discussion of
$\sin^2\theta_W^{eff}$.
This is because $P_\tau$ is the linearly proportional to small vector $Z$-lepton
coupling, thus to $\sin^2\theta_W^{eff}$ itself. Also, the $P_\tau$
varies with energy in the vicinity of the $Z$ peak relatively slowly.
To a good approximation, as one can easily deduce form formula
(\ref{Eq:Borneff}) any
variation $\delta$ of $P_\tau$ measured at the $Z$ peak
translates into  $\sim\frac{1}{8} \delta$  
shift\footnote{At the $Z$ peak $P_\tau \simeq \frac{2v_e}{a_e}$.} of $\sin^2\theta_W^{eff}$ quite independently of the flavour of incoming
state. This holds not only for $e^+e^-$ but for incoming quarks too. For $A_{FB}$
and $\sigma^{tot}$ similar relations can be obtained, then
$\rho$ and $\alpha$ dependence would need to be taken into the picture.
The  initial state flavour  and much stronger energy dependence would
lead to multitude of cases. That is why
we will use $P_\tau$ as an example to discuss suitability of the
$\sin^2\theta_W^{eff}$ picture and its limitations.

\begin{figure}[htp!]
\begin{tabular}{ccc}
  \includegraphics[width=0.47\columnwidth]{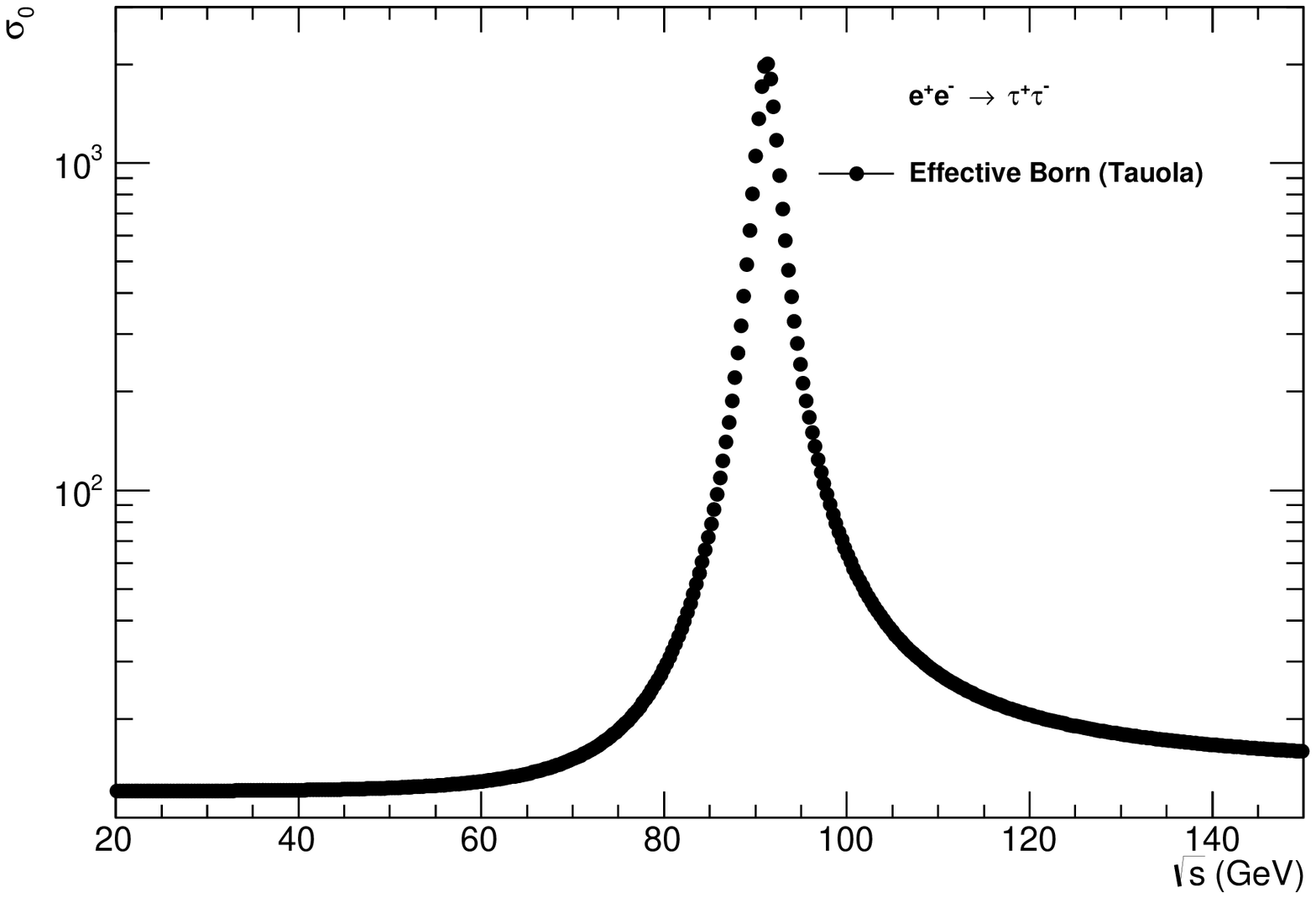} & 
  \includegraphics[width=0.47\columnwidth]{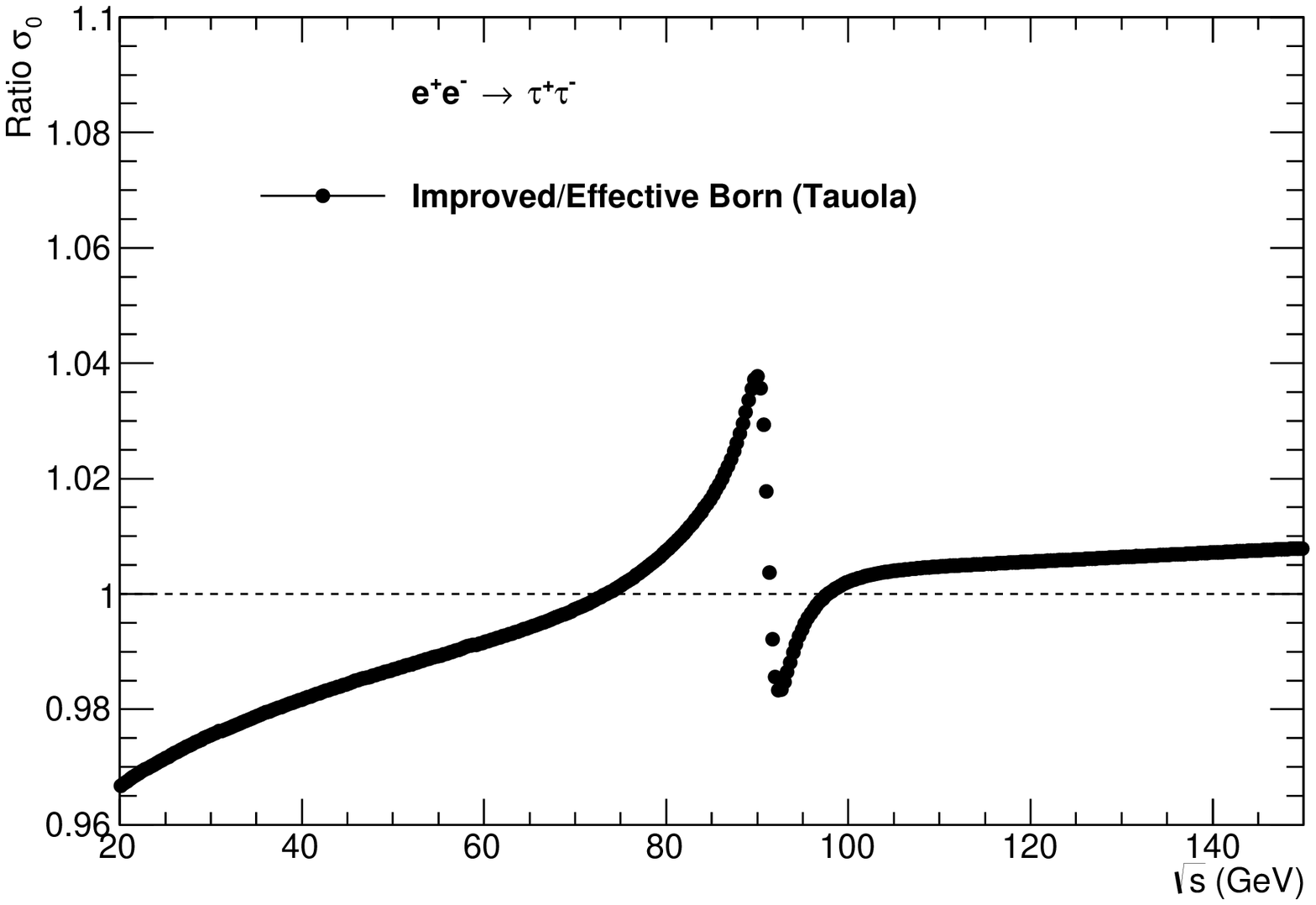} \\
  \includegraphics[width=0.47\columnwidth]{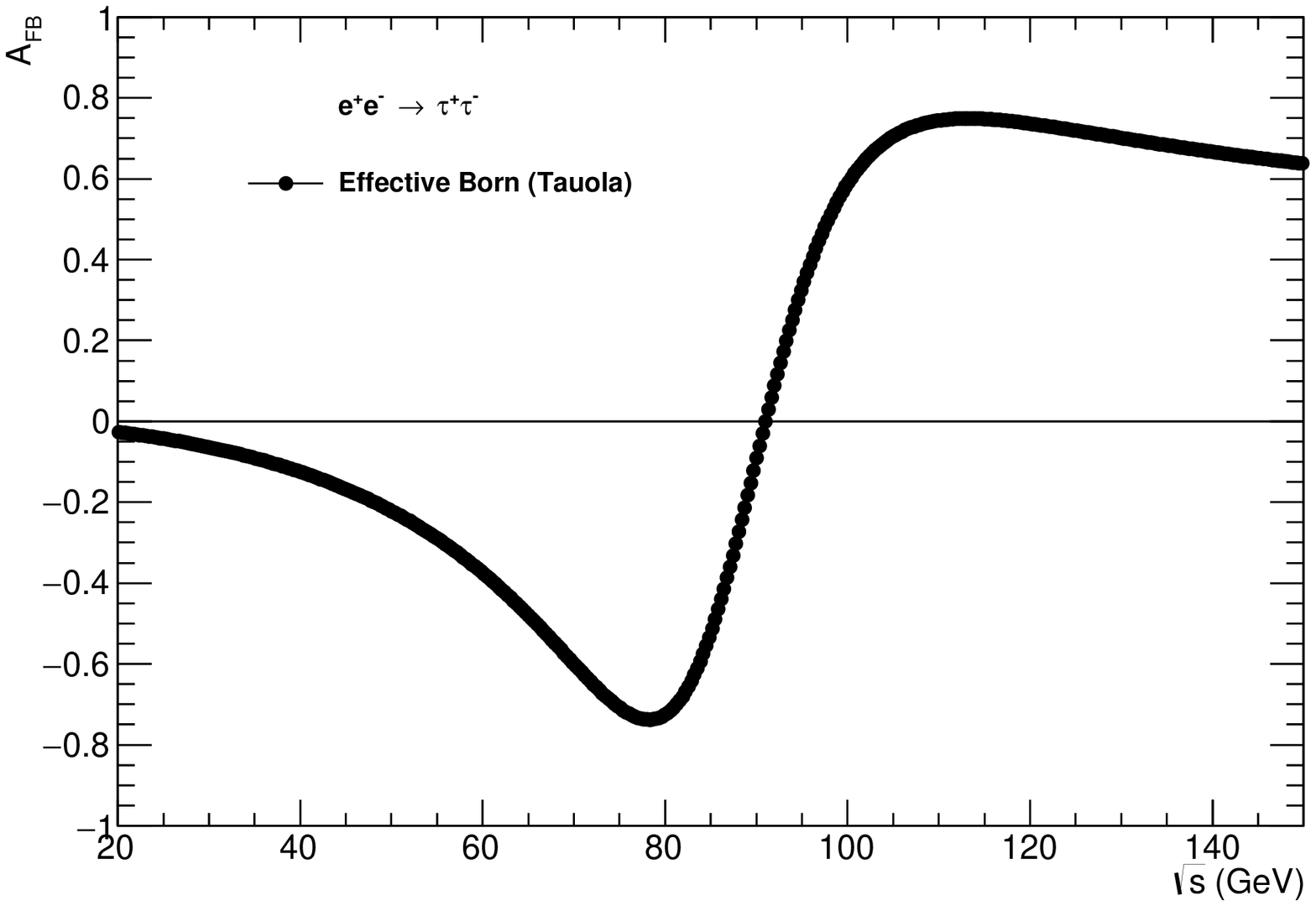} & 
  \includegraphics[width=0.47\columnwidth]{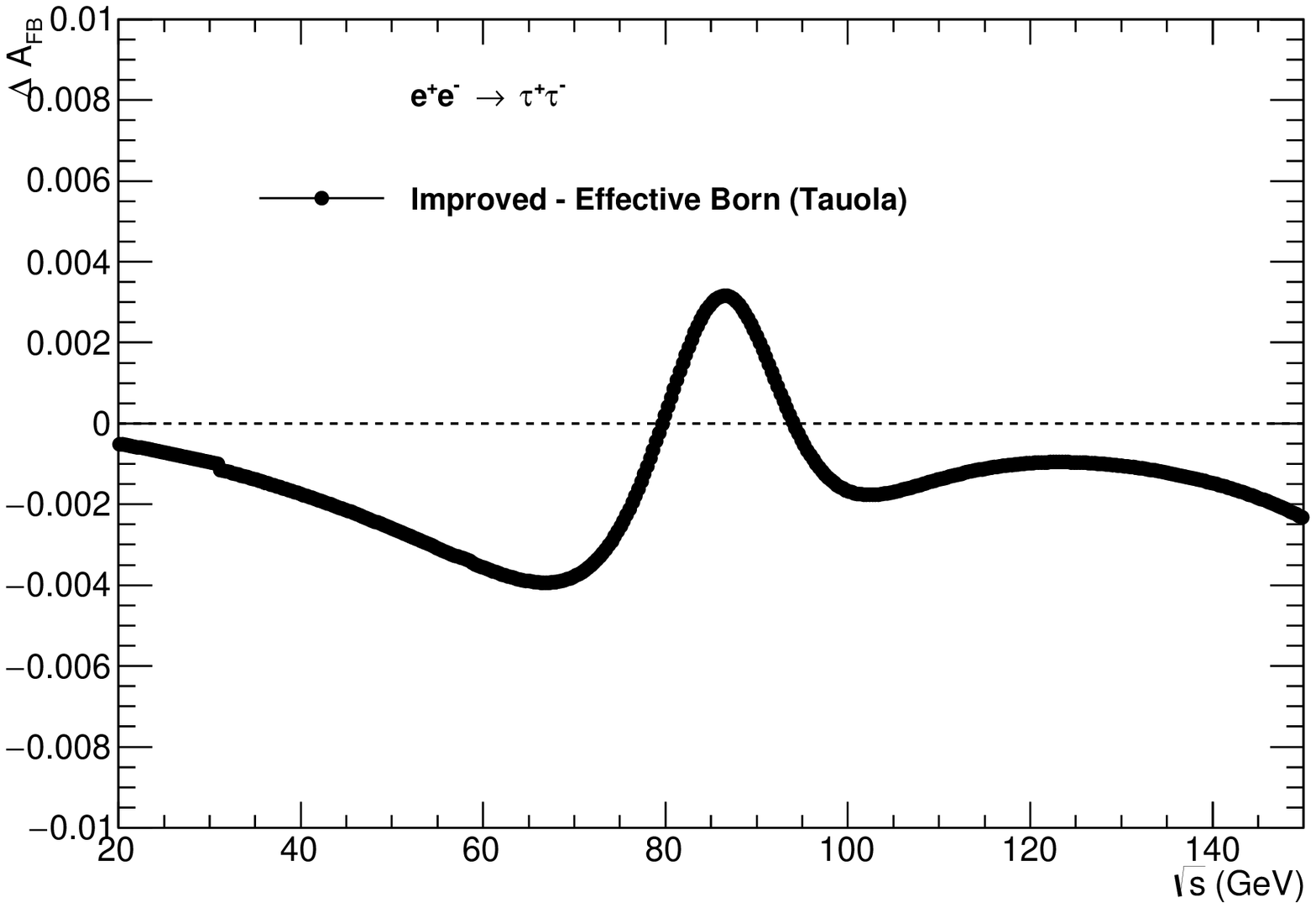}  \\
  \includegraphics[width=0.47\columnwidth]{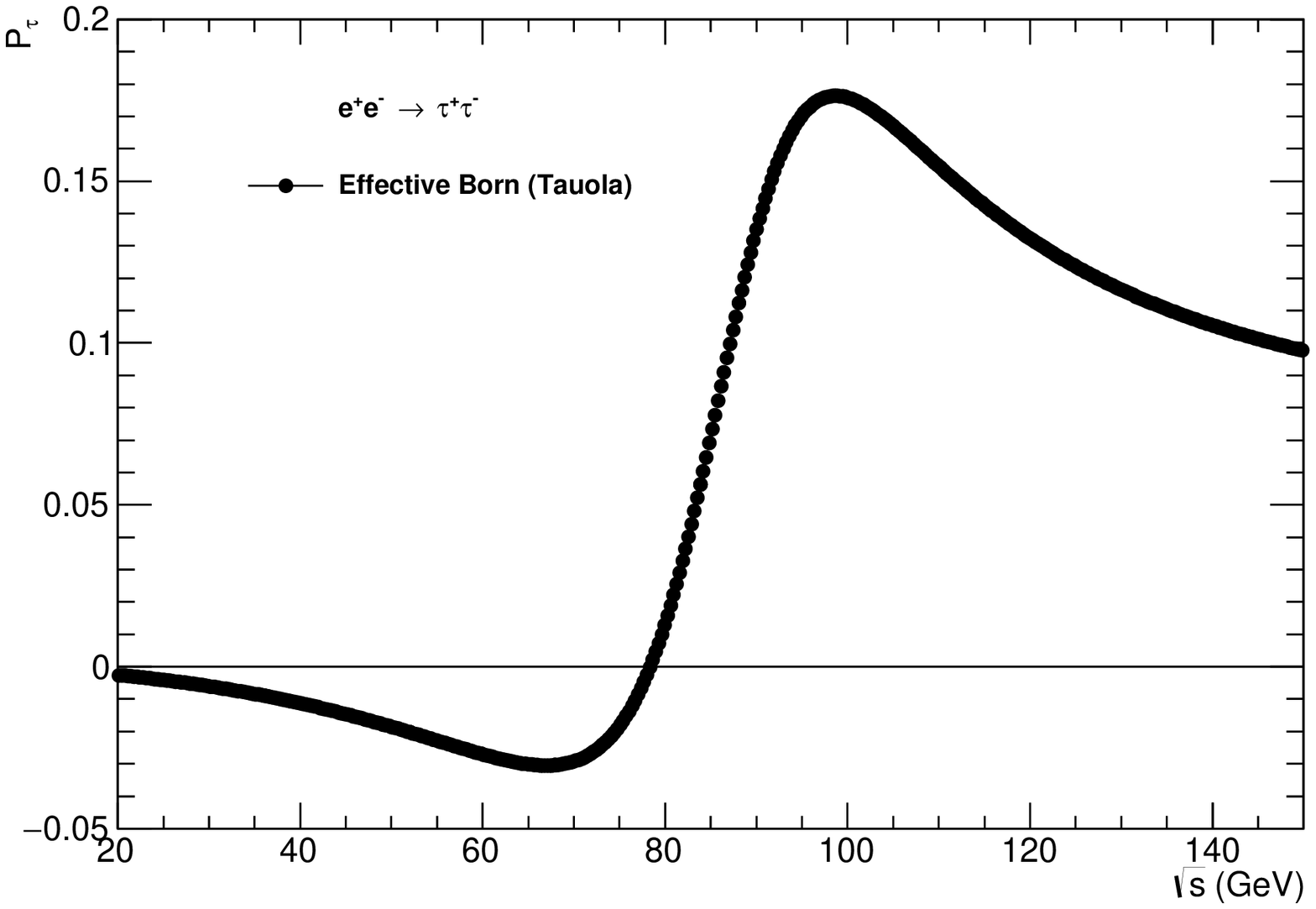} & 
  \includegraphics[width=0.47\columnwidth]{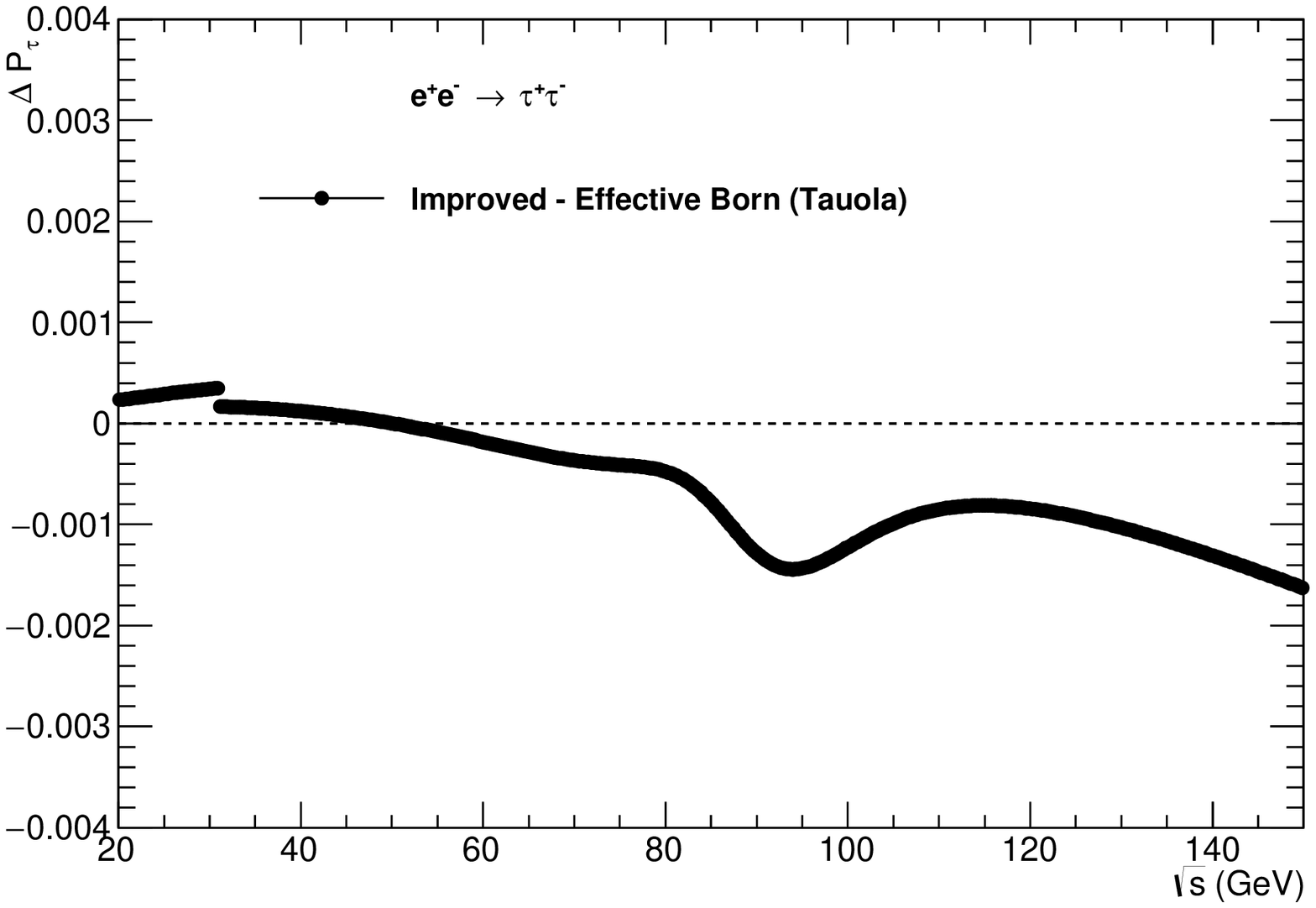}   
\end{tabular}
\caption{\small The  $\sigma^{tot}(s)$, $A_{FB}(s)$ and  $P_{\tau}(s)$
  (left side, top, center and bottom plots respectively) of
  {\tt TauSpinner} calculation with Effective Born (Tauola/LEP as installed in Tauola since December 2019).
   In the right side plots  results  of Improved Born calculations with
  EW form-factors   
  from {\tt DIZET 6.45}
  are compared with those of the left side plots.
  \newline
  Note the differences depicted in the right side plots  are
  enhanced in part because  input parameters of 
  {\tt Tauola}/LEP initialization, see Table \ref{Tab:BornEff}
  for details. The $M_z$ and $\Gamma_Z$ are
  slightly different than those used for Improved Born, causing sizable kink
  in the top right plot of $\sigma^{tot}$. On the  right hand side plots
  there are minor
  discontinuities  at 30 GeV, too. They are smaller than calculation precision and
  are due to granularity used for tabulation.
   \label{figTest1} }
\end{figure}

Fig.~\ref{figTest1} for the $e^+e^-$ case is shown for the start of numerical 
comparisons of Improved Born and  Effective Born
({\tt TAUOLA}/LEP initialization as specified later and  installed in
{\tt Tauola} distribution)  is shown.
Differences are not large,
but possibly not always satisfactory for precision physics.
In fact {\tt TAUOLA}/LEP Effective Born becomes insufficient
for high precision measurements, especially of hadron colliders
(where off the $Z$ peak contributions, contrary to the FCC, can not be minimized/excluded by the
fixed colliding quark energies).

\section{From Improved Born to Effective Born: numerical results.}
\label{sec:FromIBtoEF}

%
  Let us now attempt  to identify those Effective Born 
  simplifications which are of  numerical consequences
  and those which are  important from the theoretical perspective, but 
  hopefully not so much numerically. The 
  results  will be compared to Improved Born results, which are 
  the most precise ones.
  We keep all input parameters as of Improved Born
  but gradually simplify EW correcting terms.
  %


It is helpful to 
 test and  to  understand the impact of simplification steps
 from Improved $\to$ Effective Born,  on $\sigma^{tot}$, $A_{FB}$ and $P_\tau$
 and 
 for all the  elementary processes: $e^+e^-\to \tau^+\tau^-$ and
 $u\bar u (d \bar d) \to \tau^+\tau^-$.
We concentrate on $e^+e^-\to \tau^+\tau^-$ process and choose for Figs. \ref{fig:Efectiveness} and \ref{fig:v0to3}
 the energy range important for the measurement of the $Z$ boson couplings,
that is 
$M_Z\pm 5$ GeV. The $e^+e^-$ case is simpler to present and conclusions
would not differ much if instead quark level processes would be used.
For  $pp$ collisions, parton distribution
functions would make discussions obscured, or  unrealistic if dropped out.
For reference results  Improved Born and   {\tt DIZET 6.45} EW library was used.
\\
{\bf Approximation as of green points in Fig.~\ref{fig:Efectiveness}}  (marked ``complex of $Z$ peak'').
\\
In this first step of approximation introduced into Eq.~(\ref{Eq:BornEW}), the  
$s$- and $t$-dependent form-factors  are replaced with their values
at the $Z$ peak and
for  the scattering angle $\cos\theta=0$.
  %
  %
One can see, that if constant
complex couplings calculated at the $Z$ peak instead of $s,t$-dependent form-factors are used,
in the range of $M_Z$ 
$\pm$ 5 GeV the $P_\tau$, $A_{FB}$ and $\sigma^{tot}$ departs from the exact result at the peak, respectively
by up to
$8\cdot 10^{-5}$, $45\cdot 10^{-5}$ and $40\cdot 10^{-5}$. These largest differences are
at the edge of the range, where cross section is already about a factor
20 smaller than at the peak.
\\
If in addition $vv_{ef}$ was set
to 1,  additional changes are marginal.
That is why in the Figure the case of  $vv_{\ell f}=1$ is not presented. Once
$vv_{\ell f}=1$ is set,
mixing term is avoided and   effective
couplings are 
attributed separately  to incoming and  
outgoing flavour.
\\{\bf Approximation as of blue triangles in Fig.~\ref{fig:Efectiveness}} (marked as ``real v, vv=1'').\\
The second step  is to neglect imaginary parts of the vector couplings to $Z$.
They are about factor of 100 smaller than the real parts. Now, the differences
became larger, respectively up to $166\cdot 10^{-5}$, $50\cdot 10^{-5}$ and $32\cdot 10^{-5}$. That means non-negligible
degradation for $P_\tau$, corresponding,
in the language of  $\sin^2\theta_W^{eff}$ to  a $21\cdot 10^{-5}$ prediction ambiguity.
\\ {\bf Approximations as of red triangles and yellow stars in Fig.~\ref{fig:Efectiveness}}
(marked respectively `` real $\Pi_{\gamma\gamma}$'' and ``LEP2005 style''). \\
The role of the imaginary part of $\Pi_{\gamma\gamma}$ requires special attention,
particularly for $A_{FB}$.
For red triangles, with respect to previous case, imaginary parts of $\Pi_{\gamma\gamma}$ and
$\rho_{\ell f}$ are set to zero, whereas for yellow stars,
the imaginary part  of $\rho_{\ell f}$ is set to zero only.
The differences for ``real $\Pi_{\gamma\gamma}$'' (``LEP 2005 style'') are respectively
$152\cdot 10^{-5}$, $188\cdot 10^{-5}$,  $32\cdot 10^{-5}$,
($172\cdot 10^{-5}$,  $60\cdot 10^{-5}$, $30\cdot 10^{-5}$). The numerical effect of these imaginary parts,
which can not be easily absorbed in redefinition of the couplings, need to be kept in mind. 
\\
 With  ``LEP2005 style'' parametrization we still do not address
more subtle LEP time  choices used in  data analysis.
  In particular, of parametrisations
  used to separate contributions from s-channel exchange of $Z$ boson
  and virtual photon exchange interfering background.
   In practice, in ``LEP2005 style'' variant, we use formula (\ref{Eq:BornEW}) but 
   with
   $ { K_{e/f}(s,t)}\to \mathrm{Re} { K_{e/f}(M_Z^2,-M_Z^2/2)}$, ${ K_{ef}(s,t)} \to 1 $, 
   $\rho_{\ell f}\to  \mathrm{Re} \rho_{\ell f}(M_Z^2)$,
  that translates into use of flavour-dependent $\sin^2\theta_W^{e/f\; eff}(M_Z^2)$.
  For $\alpha(M_Z^2)$ the replacement  $\Pi_{\gamma \gamma}(s) \to \Pi_{\gamma \gamma}(M_Z^2)$ with a
  complex value is used.
   Now the purpose is to evaluate numerical consequences of $\alpha(s)$'s imaginary
  part. But it is also similar to what was   used
  at a time of final precision data analysis of all LEP collaborations
  combined \cite{ALEPH:2005ab}. Motivations of the choices are discussed
  in Ref.~\cite{Bardin:1999gt}. The numerical impact 
   is presented in Table 19 and is discussed in Section 5.4  of that reference,
   see also Section 5.4.4 of \cite{ALEPH:2005ab}.

   Figure \ref{fig:Efectiveness} is accompanied with the extensive
   Table~\ref{Tab:KEYGSWr} in Appendix~\ref{app:variantseff}.
   In total, results of eleven initializations variants are used for the Table.
   Of those, four are used in  the figure. The variants, with gradually
   introduced simplifications to Improved Born EW corrections, were chosen.
   Most of the results were obtained with semi-analytic scripts of
   {\tt TauSpinner} package, described in  Appendix \ref{app:B}. Details of the 
   initializations are depicted in Table \ref{Tab:KEYGSW}.
\\ {\bf Our main observations: }\\
(i) The  $\Pi_{\gamma\gamma}$ imaginary part, formally contributing at higher orders,
was included in calculations for final LEP time data analysis. Its impact is largest for $A_{FB}$,
whereas for $P_\tau$ imaginary parts of $v_e$, $v_f$  couplings are 
more important. 
(ii)
 The form-factors replacement with constant
 effective couplings is numerically less important than when their imaginary parts are
 dropped. Also, the closer to the $Z$ peak one goes, the smaller the
 disturbing of the Effective Born picture from photon exchanges. The same is true for the
 complex part of 
 $\rho_{\ell f}$. Overall, numerical impact on observables is not universal and distinct sets
 of effective couplings
 might be needed for each of our test observables to match best  the result
 of the Improved Born approximation.

 \subsection{  The  {\tt v0, v1, v2} variants of  Effective Born.  }
\label{sec:variantsof}

 The formulae for Improved Born Eq. (\ref{Eq:BornEW})  and for Effective Born  
 Eq. (\ref{Eq:Borneff}) differ  with subtle, but numerically important
 details. We evaluate  numerical effects again with the help of
 options in {\tt TauSpinner} explained in Appendix \ref{app:variantseff}, in particular
 in Table \ref{Tab:KEYGSW}.

 One can ask the question how close can one approach Improved Born results,
 with the effective ones, without breaking features necessary for matching
with calculations of  strong interactions. Details of listed below  
  variants for Effective Born are provided in
Table  \ref{Tab:BornEff}. Non-essential details are delegated to  Appendix
\ref{app:6XXEW}.
For completeness the reference starting points, {\tt TAUOLA}/LEP parametrization and 
the  EW LO Born
parametrization in EW $\alpha(0)$ scheme, are provided.
The consecutive three variants {\tt v0, v1, v2} of Effective Born are ordered with their ability to approximate better the {\it Improved Born} results,
  but at the same time, variants {\tt v1} and {\tt v2} may be less straightforward to implement into some programs
  designed for strong interaction in $pp$ collisions.

\begin{itemize}
\item
The {\tt v0} variant is using formula (\ref{Eq:Borneff}) for spin amplitude, with
$\alpha (s) = \alpha (M_Z^2)$,  $s^2_W = \sin^2\theta_W^{eff} (M_Z^2)$,
but with $\rho_{\ell f}$= 1.0. 
\item
  The {\tt v1} variant is using formula (\ref{Eq:Borneff}) for spin amplitude, parameters
  are set as for {\tt v0} parametrization, but   $\rho_{\ell f} \ne 1$.
\item
  The {\tt v2} variant is using formula (\ref{Eq:Borneff}) for spin amplitude, parameters
  are set as if both $s^2_W$ and $\rho_{\ell f}$ were flavour-dependent, and equal at the 
  $Z$-pole to the {\tt Dizet 6.45} predicted ones. See Table \ref{Tab:BornEff} and Table \ref{Tab:Dizet_6.XX_coupl}.
\item The {\tt TAUOLA}/LEP variant differs from {\tt v0} by
  numerical values of   $\alpha$,  $s^2_W = \sin^2\theta_W^{eff} (M_Z^2) $.
  Also $M_Z$ and $\Gamma_Z$ differ. It is worth to point
  that Eq.~(\ref{Eq:Borneff}) remains the same as for the Effective Born
  used at LEP 1 times.
\end{itemize}

Table~\ref{Tab:BornEff}  details {\tt v0,v1,v2} variants of  {\it Effective Born}.
Note that depending on the activated Born  variant numerical values
for $\sin^2\theta_W^{eff}$, $\rho_{\ell f}$ and $\alpha $ may vary. The flavour dependence for
$\sin^2\theta_W^{eff}$ and  $\rho_{\ell f}$ may appear, too.

We now present numerical results useful to evaluate robustness of
the {\it Effective Born} picture, where effective couplings are
 used for describing EW effects, and compare it with results of the {\it Improved Born} picture.
We will  use
{\it Improved Born} predictions as a reference for  $e^+e^-$ and $q\bar q$ cases\footnote{ In Appendix,
Table \ref{Tab:KEYGSWr}, two versions of  the {\it Improved Born} are used, with weak boxes 
included and not. This is important for large energy ranges.
For $e^+e^-$ we concentrate mostly on the region of the $Z$ pole where the impact
of EW boxes is marginal.
We demonstrate the quantitative impact
of the $s,t$-dependence, which can not be absorbed into effective couplings.}.
As we will see,
Effective Born  {\tt v2} variant works quite well around the $Z$ pole, for
 the line-shape and forward-backward asymmetry too. It may be not as
straightforward to implement into strong interaction Monte Carlo programs
as is the case of {\tt v0} or {\tt v1}. That is why we will keep 
attention to all variants which differ numerically but may 
pose smaller or bigger problems for consistency of strong interaction calculations.

These Effective Born variants differ from
Improved Born: constants instead of form-factors are used. This is  partly compensated with
 adjustments of input parameters.
In this way, in the following comparisons,  we evaluate limits of $\sin^2\theta_W^{eff}$  
and Born-level EW formula for the interpretation of $Z-l-l$ couplings measurements.
 
\begin{table}
 \vspace{2mm}
 \caption{The EW parameters used for:  the  EW LO Born in $\alpha(0)$ scheme,
   and for  variants of effective Born. 
   The $G_{\mu}$ = $1.1663887 \cdot 10^{-5}$ GeV$^{-2}$, $M_Z$ = 91.1876 GeV
   ($M_Z$ = 91.1887 GeV for {\tt Tauola}/LEP) and
   ${\mathscr K}_f, {\mathscr K}_e, {\mathscr K}_{\ell f}$ = 1.}.
 \label{Tab:BornEff}
  \begin{center}
    \begin{tabular}{|l|l|l|l|l|}
      \hline\hline
  Effective Born     &     EW LO                          &      Effective Born       &   Effective Born  &   Effective Born                    \\
    {\tt TAUOLA}/LEP &      $\alpha(0)$ scheme            &      {\tt v0}             &   {\tt v1}        &   {\tt v2} \\
        \hline\hline
    $\alpha$   = 1/128.6667471  &     $\alpha$  = 1/137.03599     &     $\alpha$   = 1/128.9503022  &     $\alpha$   = 1/128.9503022       &   $\alpha$       = 1/128.9503022  \\
    $s^2_W$    = 0.23152         &    $s^2_W$    = 0.21215         &      $s^2_W$    = 0.231499      &      $s^2_W$    = 0.231499            &  ${s^2_W}^{\ell}$  = 0.231499      \\
                                &                                 &                                 &                                      &   ${s^2_W}^{up}$   = 0.231392      \\
                                &                                 &                                 &                                      &   ${s^2_W}^{down}$  = 0.231265      \\
      $\rho_{\ell f}$  = 1.0      &      $\rho_{\ell f}$  = 1.0       &     $\rho_{\ell f}$  = 1.0  &     $\rho_{\ell f}$  = 1.005                &   $\rho_{\ell up}$  = 1.005403       \\
                                &                                 &                                 &                                      &   $\rho_{\ell down}$   = 1.005889     \\
      \hline 
    \end{tabular}
  \end{center}
\end{table}

\begin{figure}[htp!]
\begin{tabular}{ccc}
  \includegraphics[width=0.37\columnwidth]{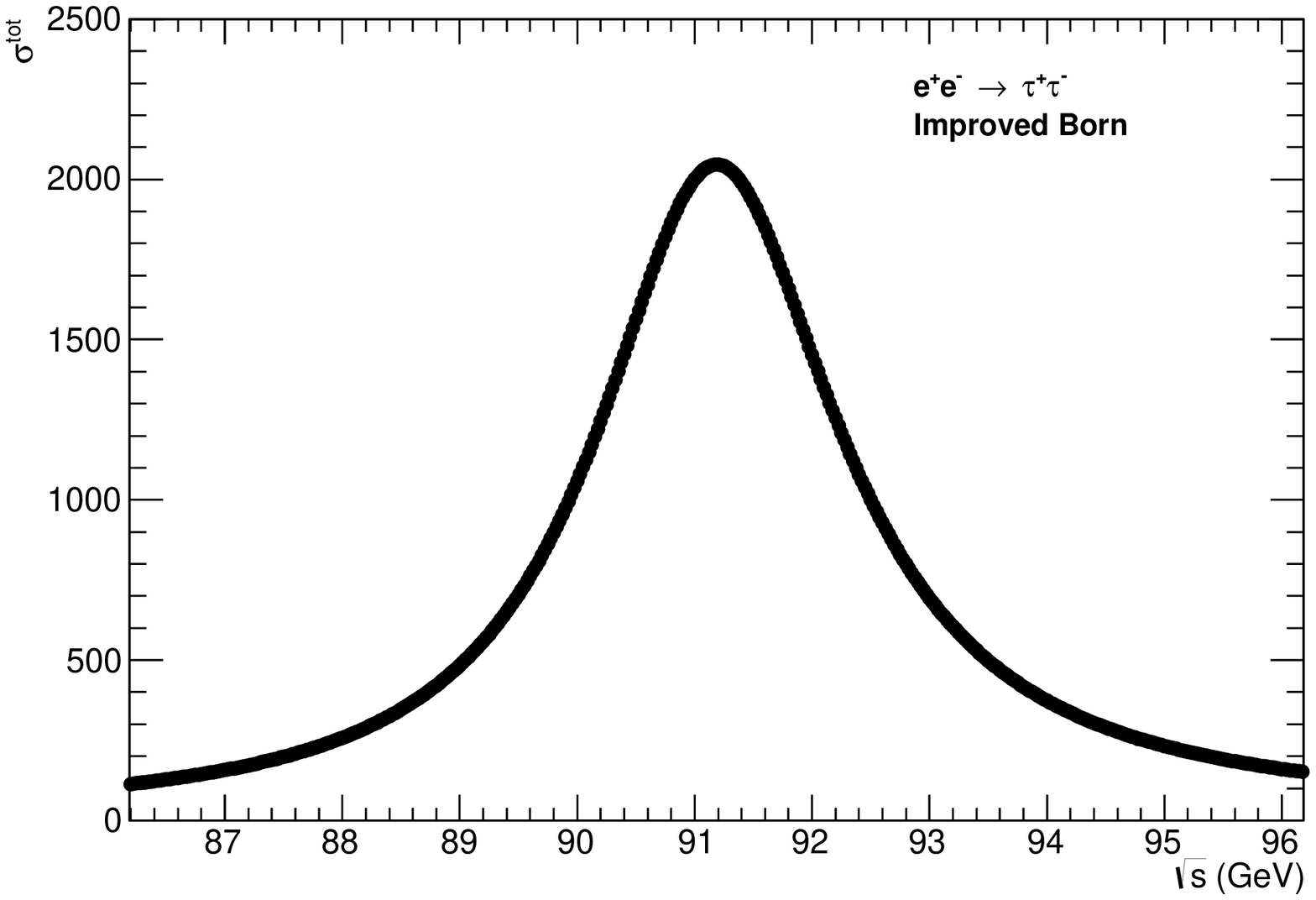} & 
  \includegraphics[width=0.37\columnwidth]{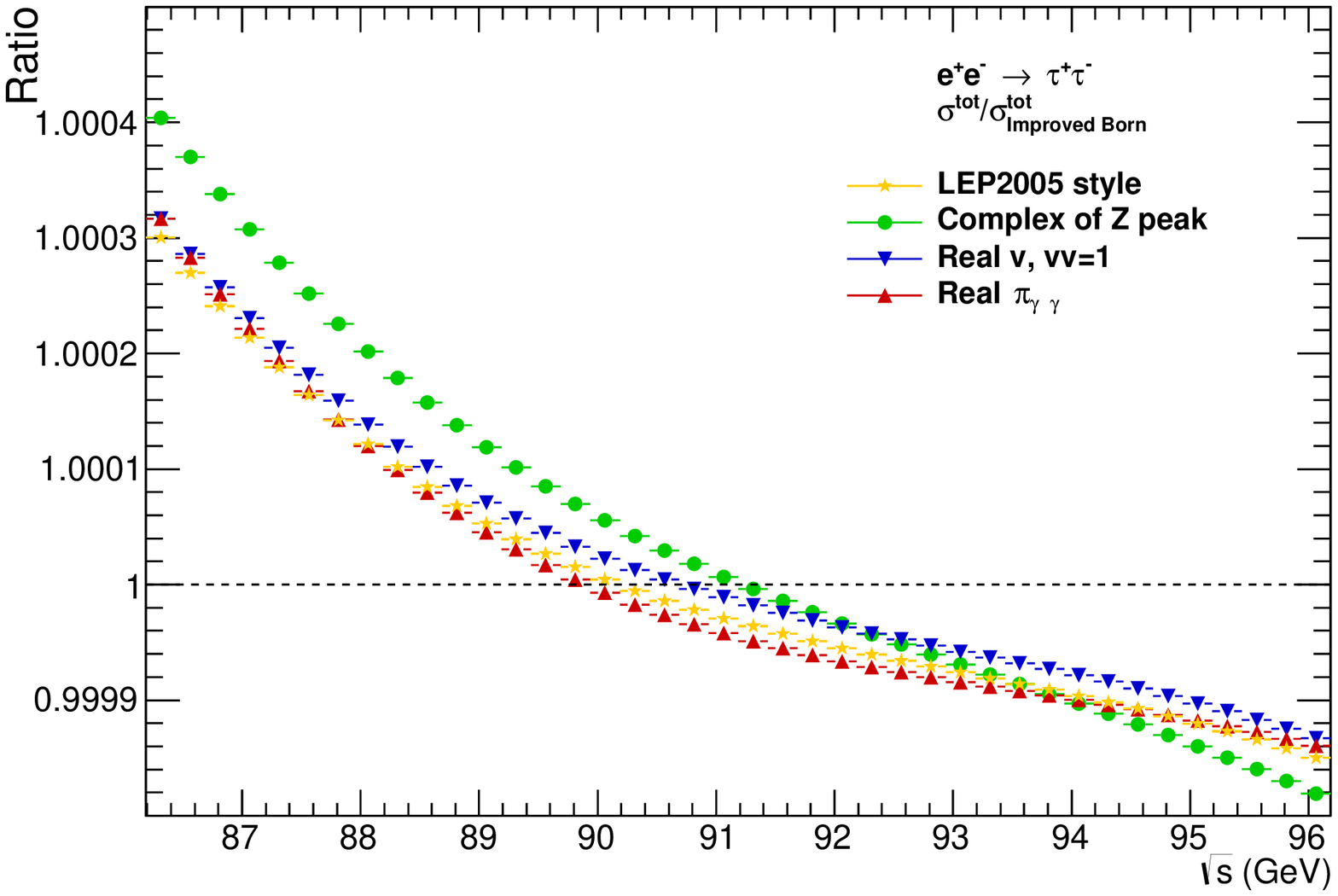} \\
  \includegraphics[width=0.37\columnwidth]{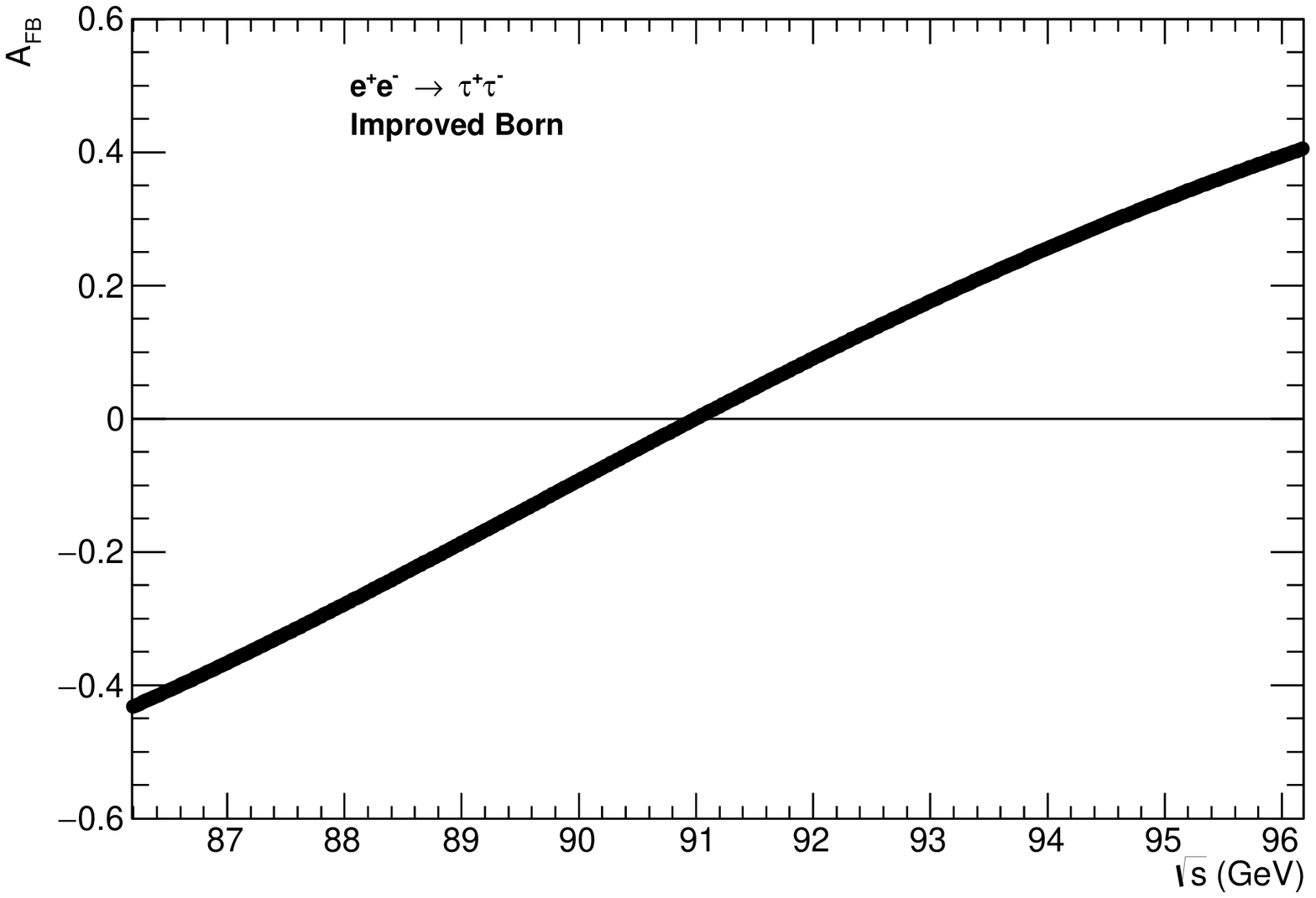} & 
  \includegraphics[width=0.37\columnwidth]{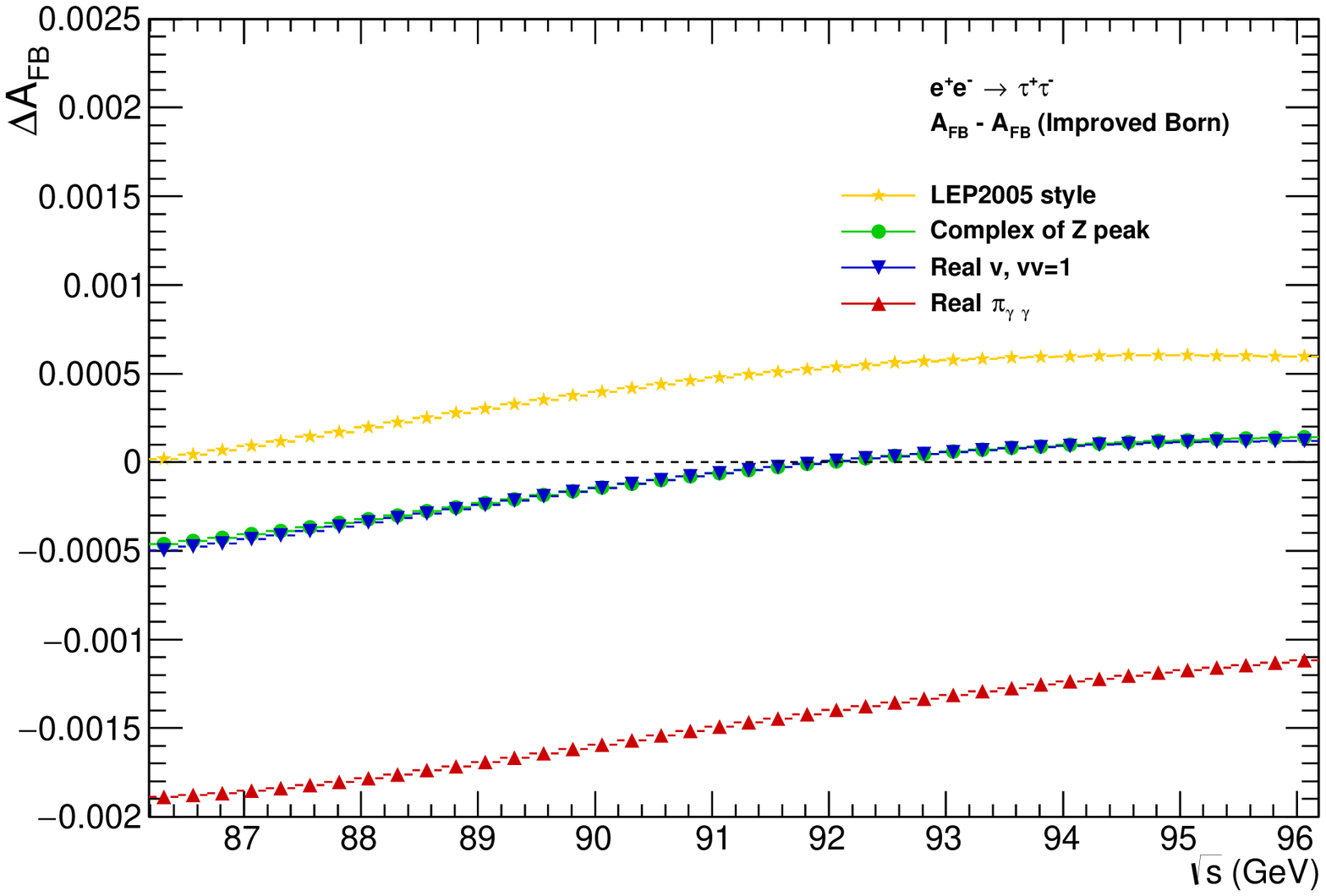} \\
  \includegraphics[width=0.37\columnwidth]{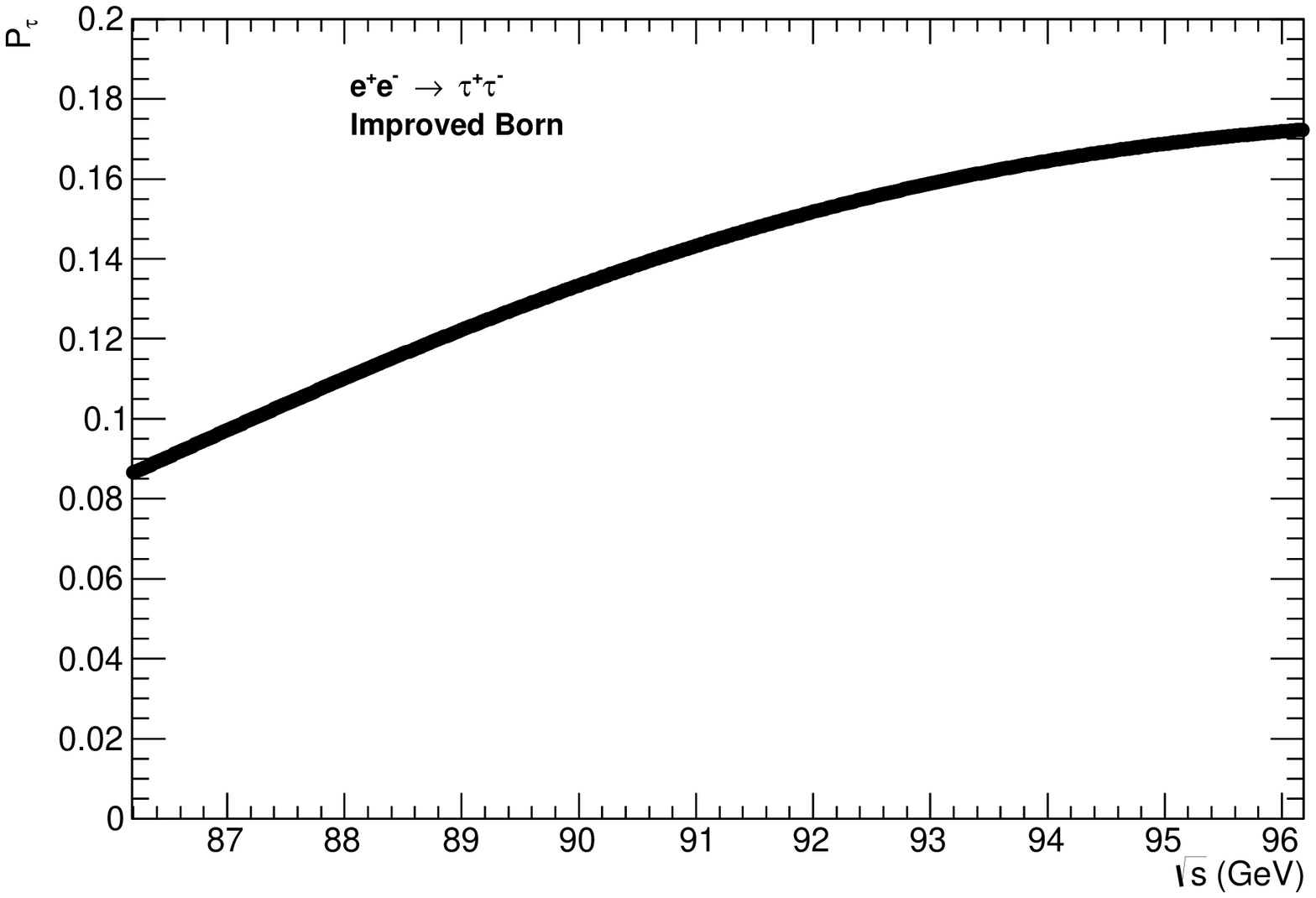} & 
  \includegraphics[width=0.37\columnwidth]{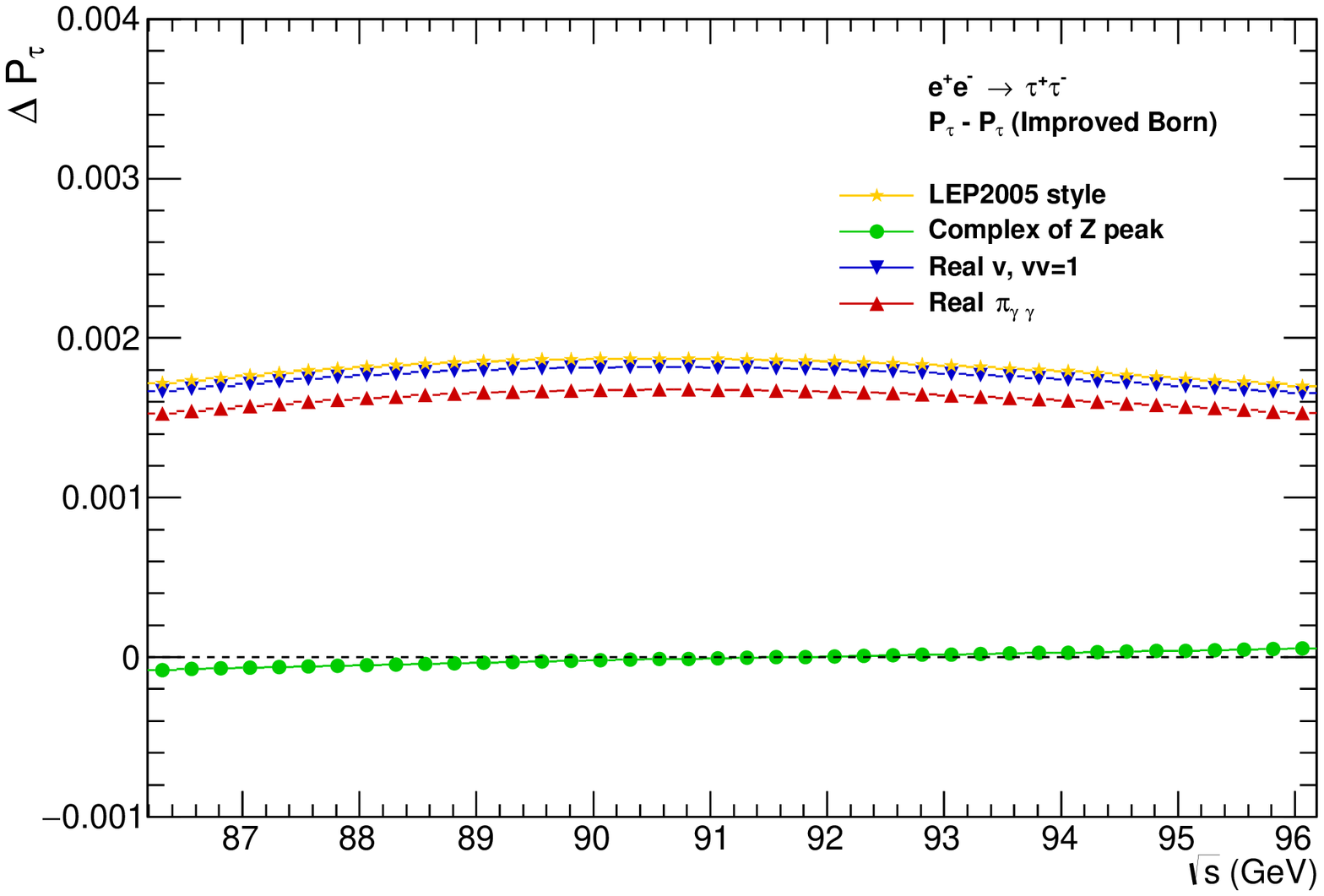}
\end{tabular}
\caption{  Left side plots, Improved Born-level and vicinity
  of the $Z$ peak: $\sigma^{tot}$ (top), $A_{FB}$ (middle)
  and $P_\tau$ (bottom).
  Right side plots enumerate, with  ratios or differences the  effects of
  simplifications
  with respect to Improved Born results.
  Green points: instead of form-factors, their constant values calculated
  at s=$M_Z^2$, t=$-M_Z^2/2$ are used. \;
  Blue triangles: as for green ones, but in addition $vv_{\ell\; f}=1$
  and only real parts of  $v_e$, $v_f$ are used. 
  Red triangles: as in blue triangles, but only real parts of $\Pi_{\gamma\gamma}$ and $\rho_{\ell f}$ are
  taken into account. 
  Yellow stars: with respect to red triangles imaginary parts of
  $\Pi_{\gamma\gamma}$ are switched back on.
    \label{fig:Efectiveness}}
\end{figure}

\begin{figure}[htp!]
\begin{tabular}{ccc}
  \includegraphics[width=0.37\columnwidth]{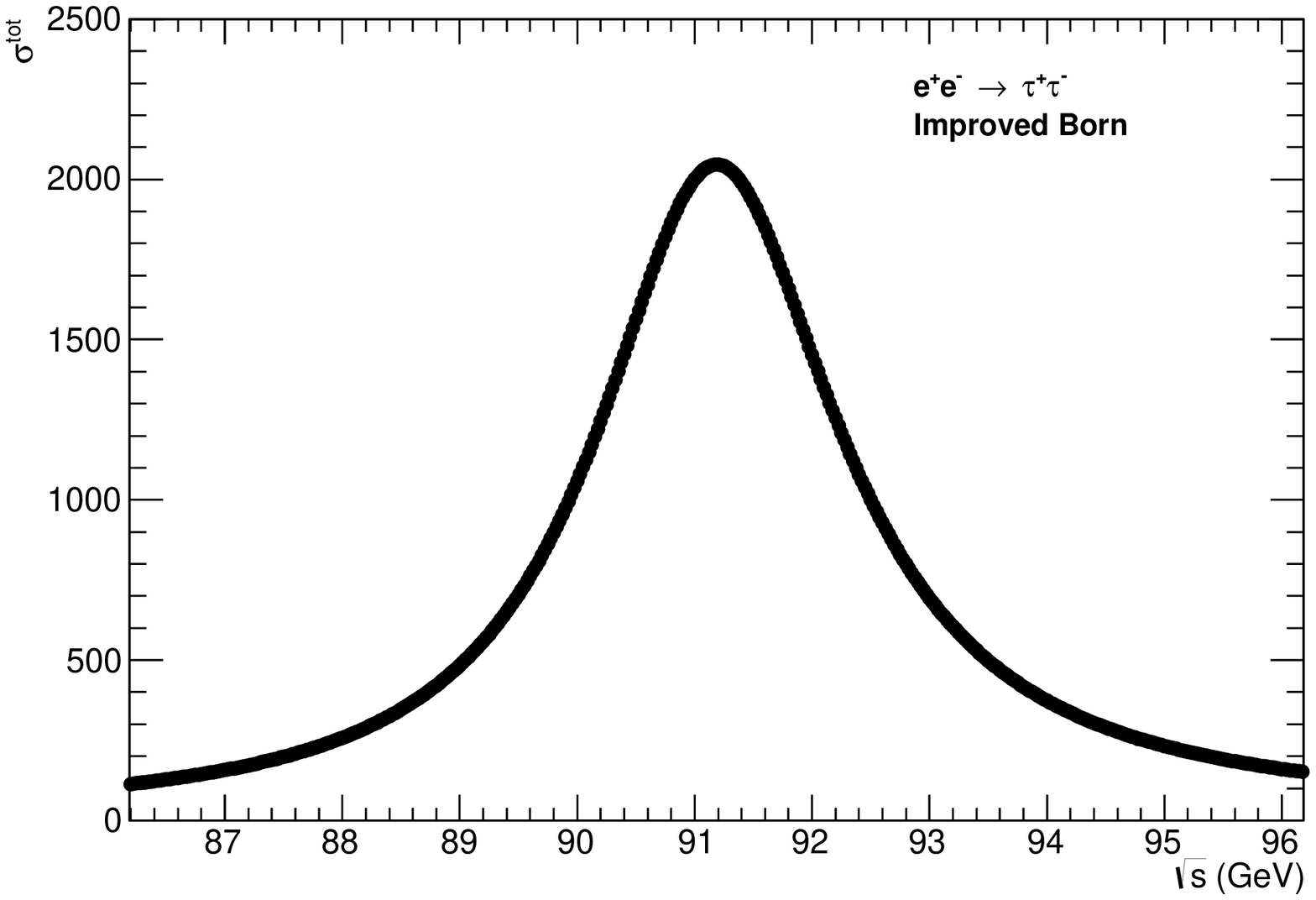} & 
  \includegraphics[width=0.37\columnwidth]{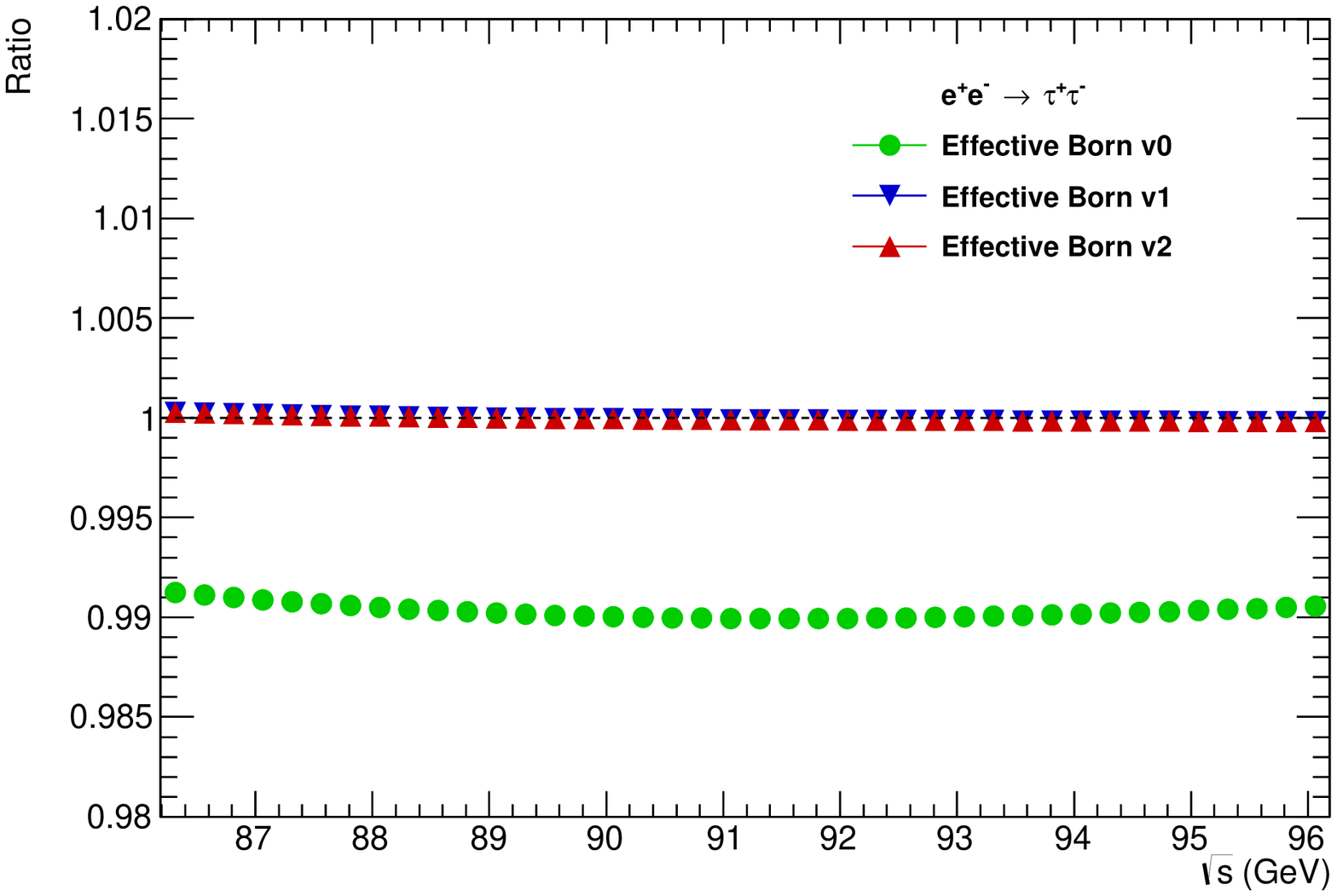} \\
  \includegraphics[width=0.37\columnwidth]{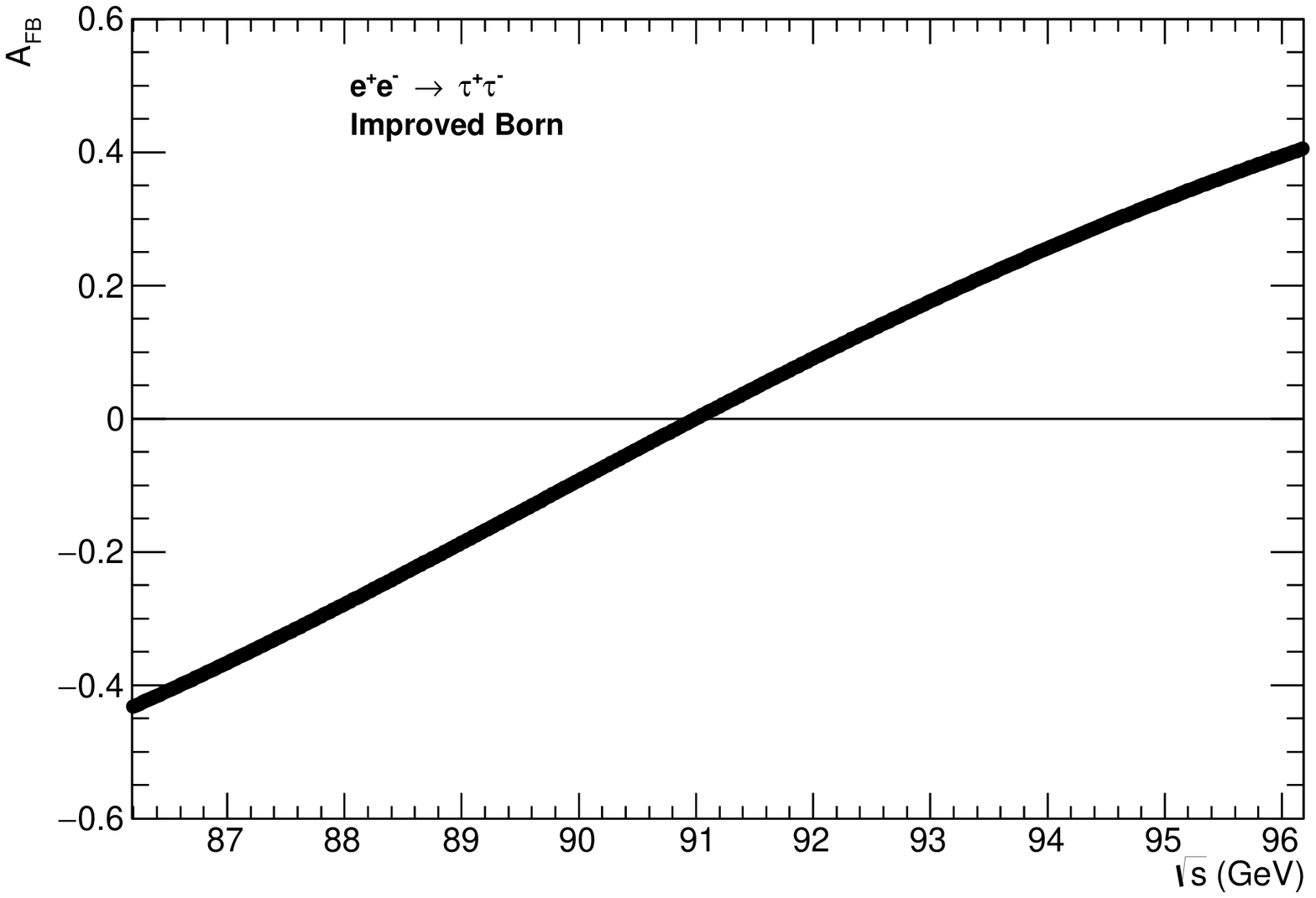} & 
  \includegraphics[width=0.37\columnwidth]{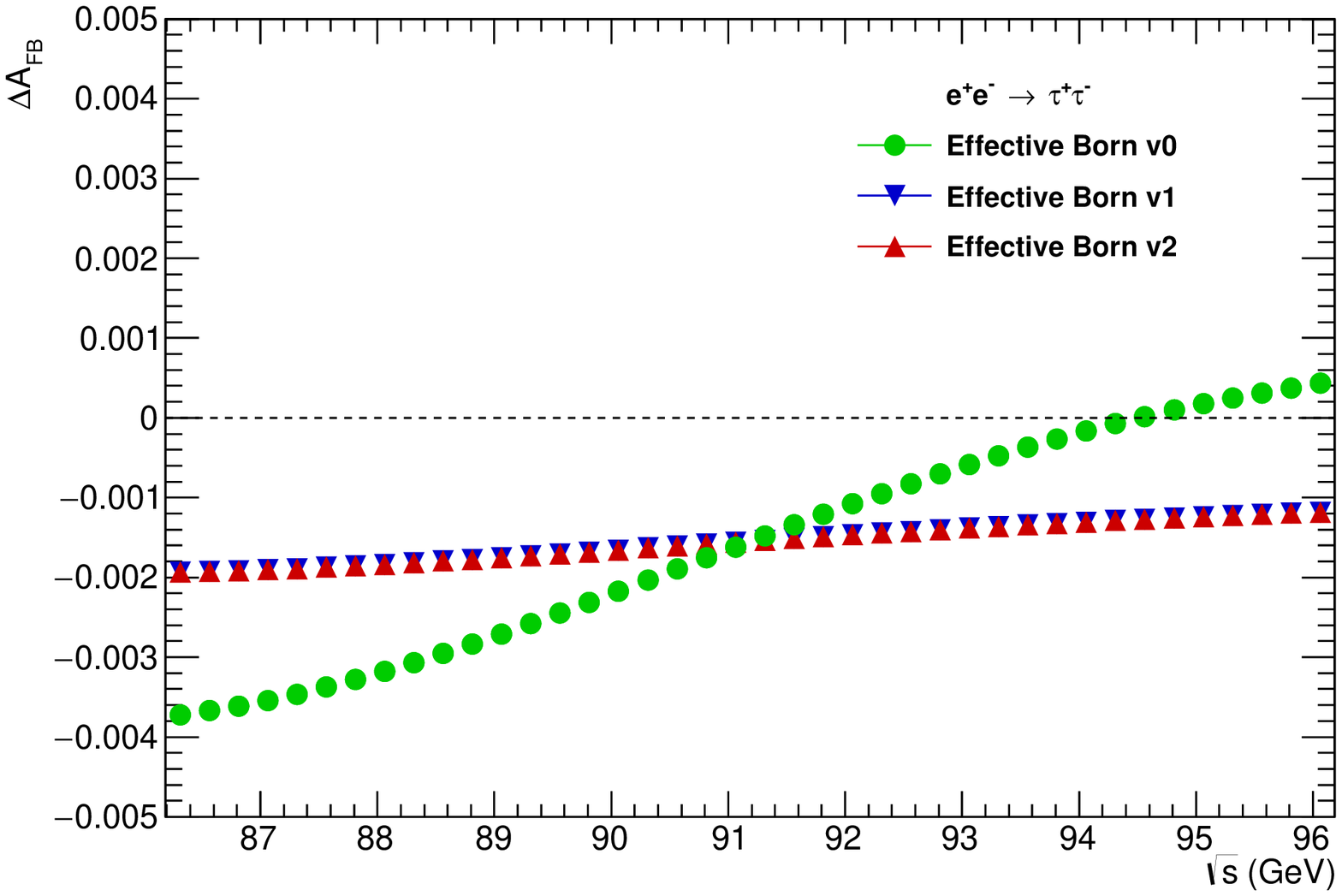} \\
  \includegraphics[width=0.37\columnwidth]{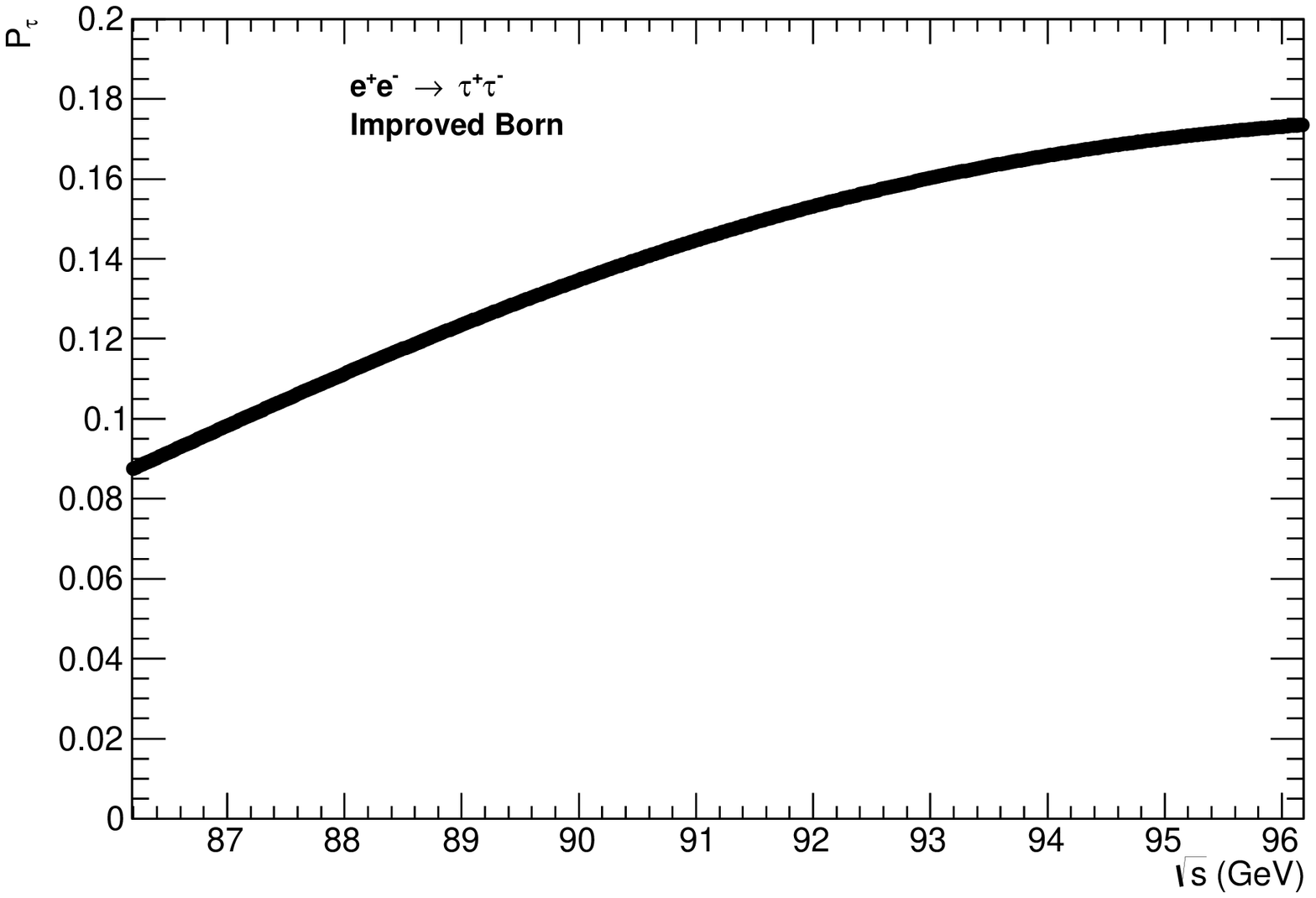} & 
  \includegraphics[width=0.37\columnwidth]{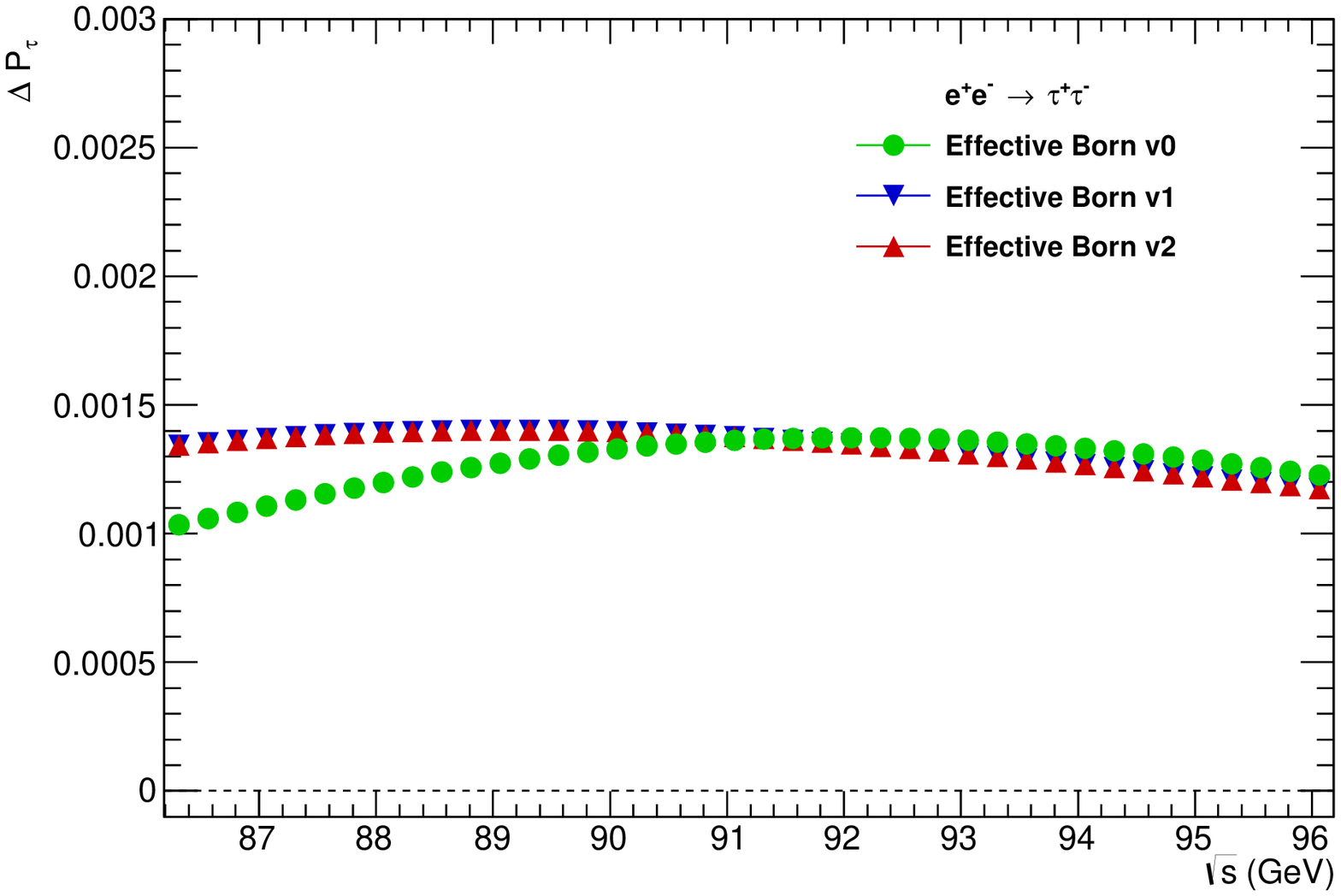}
\end{tabular}
\caption{  Left side plots, Improved Born in the vicinity
  of the $Z$ peak: $\sigma^{tot}$ (top), $A_{FB}$ (middle)
  and $P_\tau$ (bottom), as in Fig~\ref{fig:Efectiveness}.
  Right side plots enumerate, with  ratios or differences the  effects of
  Effective Born simplifications
  with respect to Improved Born.
  Green points: Effective Born {\tt v0}, \;
  Blue triangles:  Effective Born {\tt v1} \;
  Red (rotated) triangles:  Effective Born {\tt v2}. 
  \label{fig:v0to3}}
\end{figure}

Let us now turn our attention to  numerical results.
 Fig.~\ref{fig:v0to3}  is similar to
Fig.~\ref{fig:Efectiveness} but serve different purpose. It enumerates the 
performance of {\tt v0,v1,v2} Effective Born with respect to the Improved Born.  
  The shifts of {\tt v0} with respect to Improved Born respectively for $P_\tau$, $A_{FB}$
  and  $\sigma^{tot}$ read
  $140\cdot 10^{-5}$, $370\cdot 10^{-5}$ and $1000\cdot 10^{-5}$. This is already
substantially better than EW at LO\footnote{
If Effective Born {\tt v0} would be used, but with LO EW parameters
the shifts on our tests observables with respect to Improved Born would be
about a factor 100 larger.
}.
Obviously,
normalization of the $Z$ exchange needs to be corrected further. 
If $\rho_{\ell f}$ is used of variant {\tt v1},
differences with respect to  Improved Born reduce to:
$140\cdot 10^{-5}$, $200\cdot 10^{-5}$ and $50\cdot 10^{-5}$.
With the {\tt v2}  setting, reinstalling flavour dependence of couplings, we obtain results for the Effective Born which
differ from  the EW corrected ones by
$140(40)\cdot 10^{-5}$, $200(40)\cdot 10^{-5}$ and $60(20)\cdot 10^{-5}$.
Numbers in brackets were obtained with virtual $\gamma$ contribution switched off\footnote{
  This hints that Effective Born may work better for $pp$ collisions than for $e^+e^-$, because
  of smaller electric charge of quarks than of leptons.}.

From these results, particularly for $P_{\tau}$, we may  
 conclude that approaches that rely on effective couplings may not work well for the 
  $\sin^2\theta_W^{eff}$ precision tag up to about   $~20\cdot 10^{-5}$.  For further improvement 
revisiting EW effects in their complexity is required. Use of numerically adapted 
 Eq. (\ref{Eq:Borneff}) constants, which originally in Eq. (\ref{Eq:BornEW}) were multiplied by
 form-factors, does not suffice.  For high
 precision, the
 picture of effective couplings is not universal: while appropriate for 
$A_{FB}$ the choice may be not optimal  for $P_\tau$.

 \subsection{Case of $pp\to ll$ processes at LHC}
 \label{sec:atLHC}

Let us now discuss properties of these benchmark observables distributions
and how numerically significant is change from Improved Born 
to the  Effective  Born   approximation
 in the $pp$ case contrary to $e^+e^-$,  when semi analytic methods were
applied simulated event sample and
{\tt TauSpinner} reweighting are used. 

\begin{figure}
  \begin{center}                               
    {
  \includegraphics[width=7.5cm,angle=0]{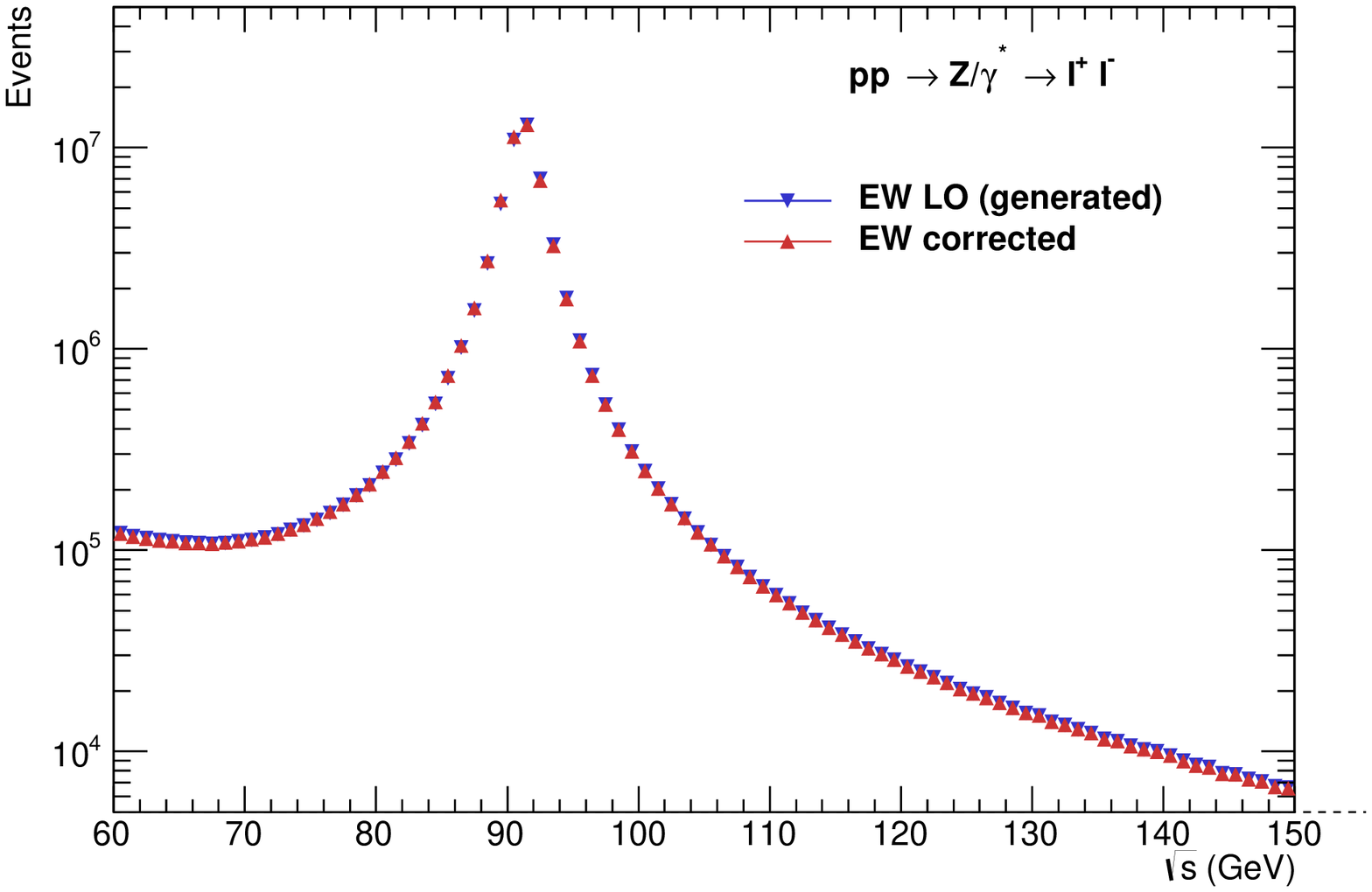}
  \includegraphics[width=7.5cm,angle=0]{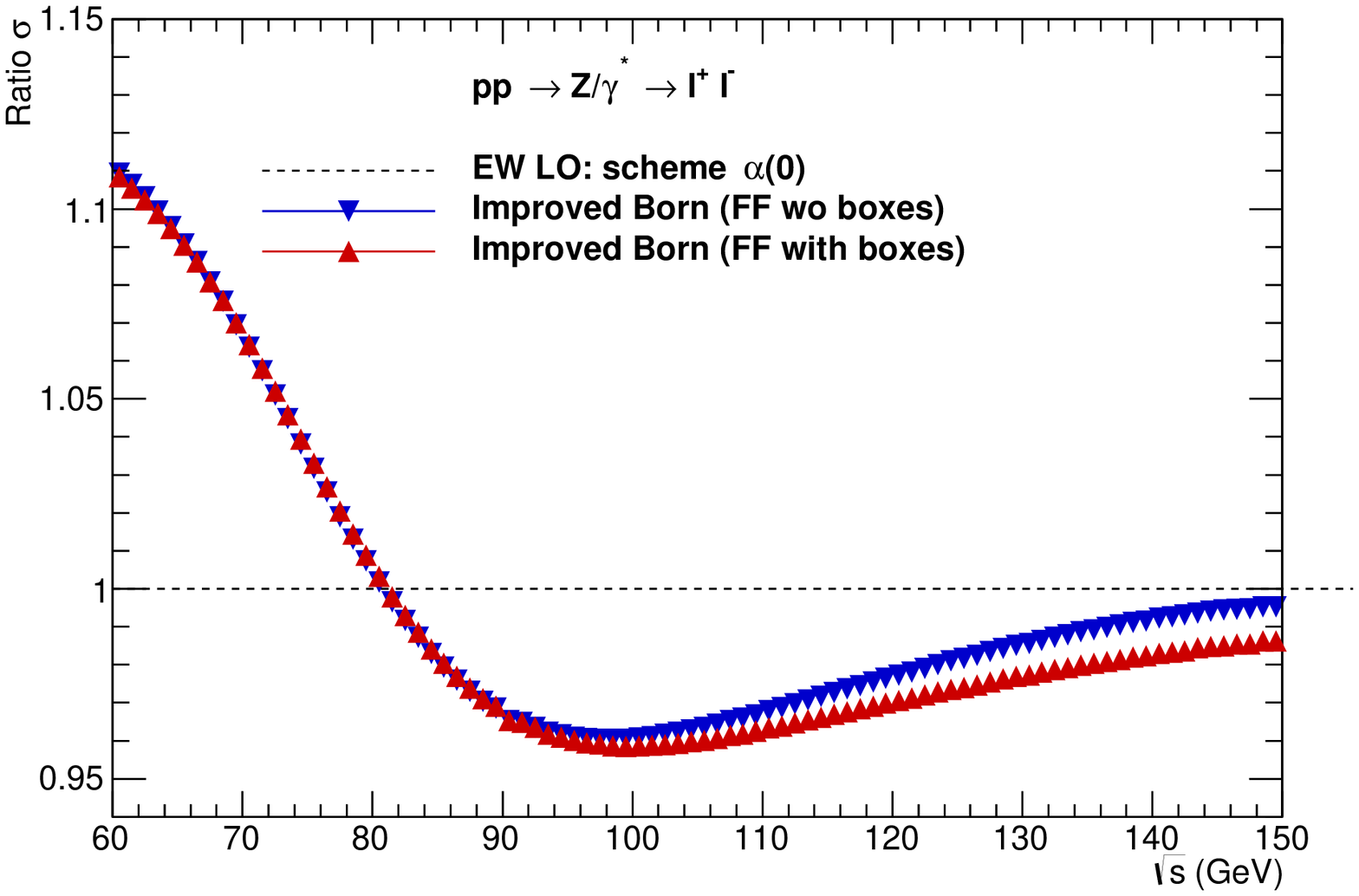}
  \includegraphics[width=7.5cm,angle=0]{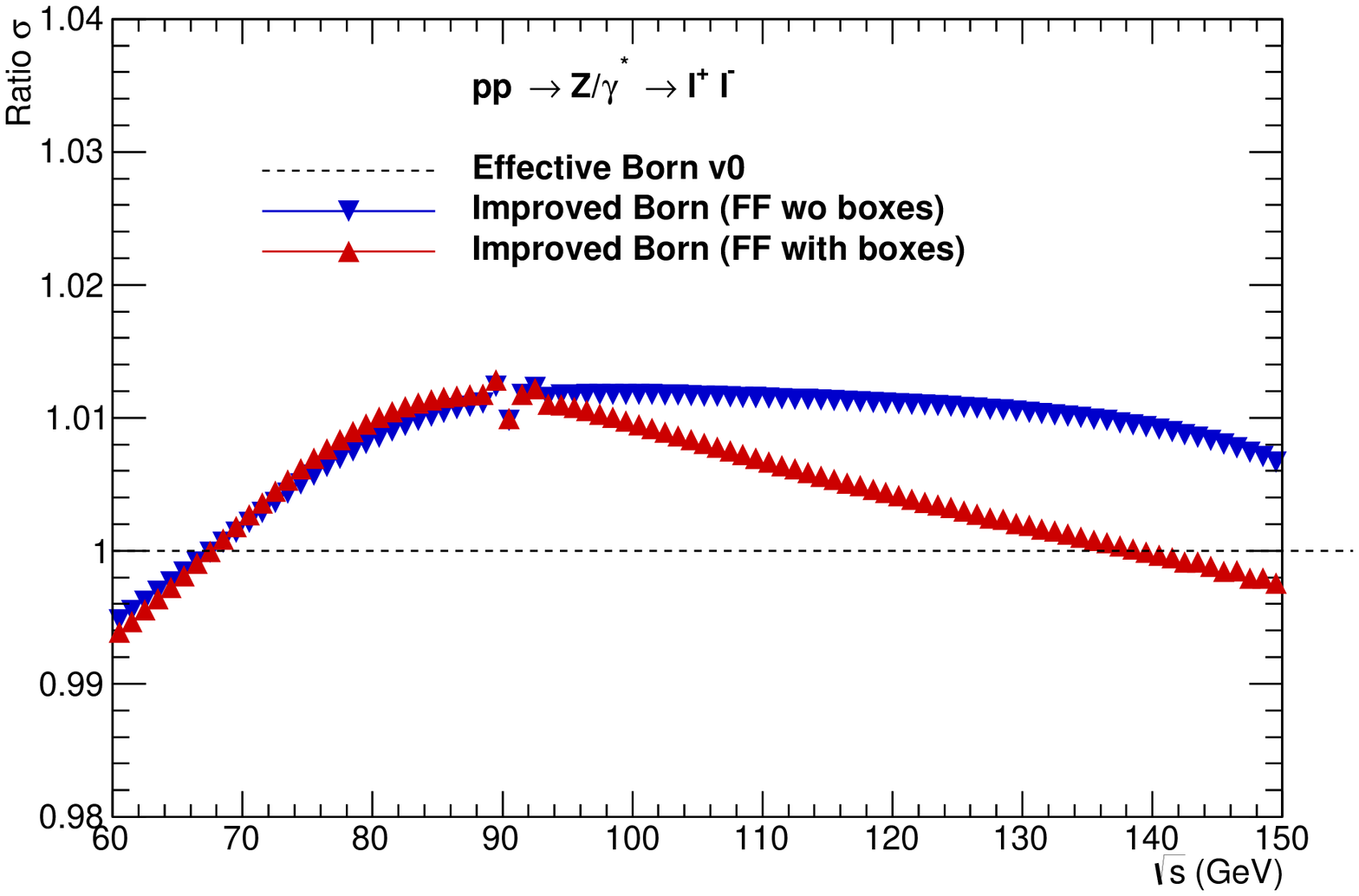}
  \includegraphics[width=7.5cm,angle=0]{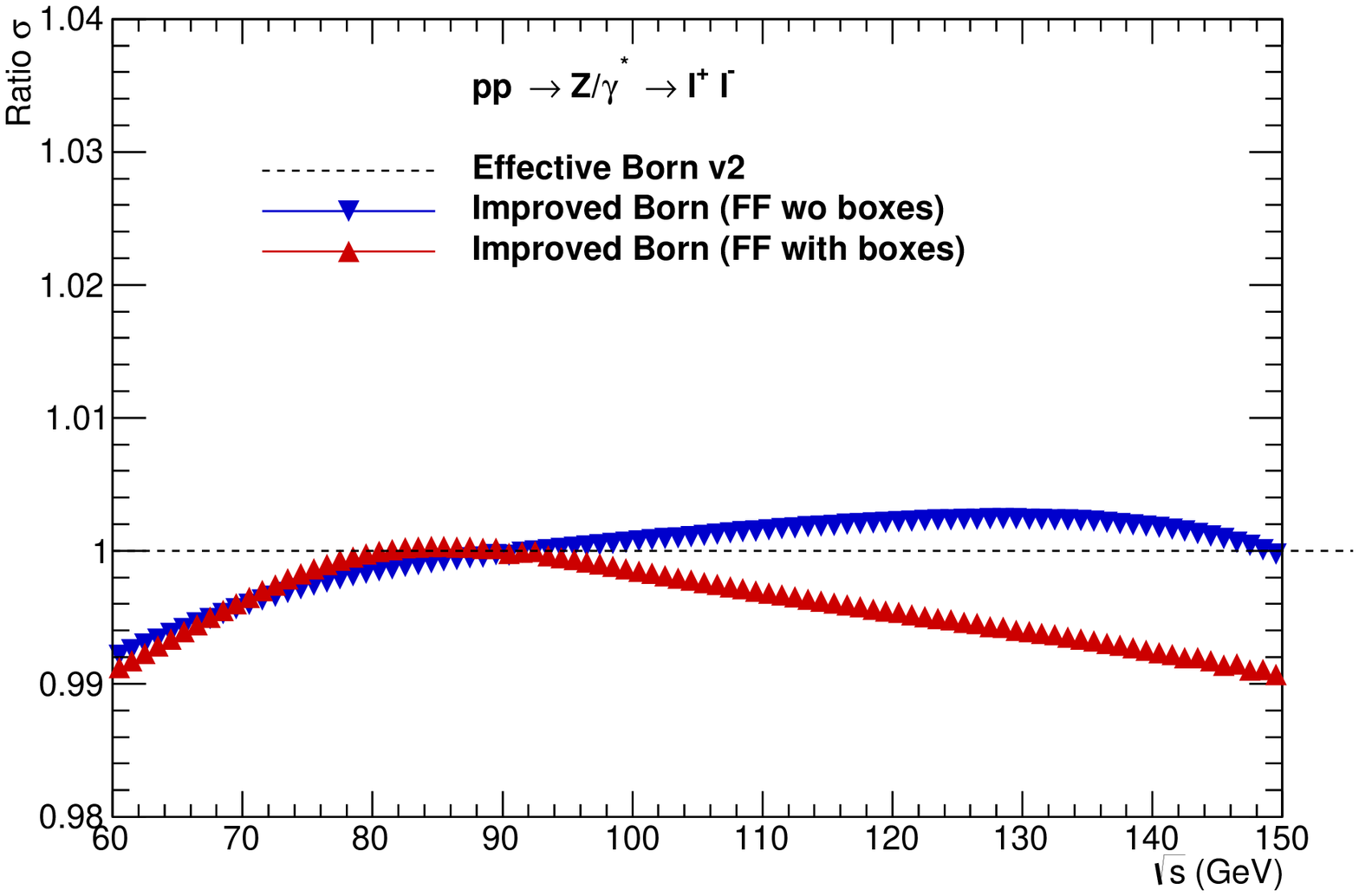}
}
\end{center}
  \caption{ Top-left: $Z$ line-shape distribution as generated with {\tt Powheg+MiNLO} (blue triangles)
    and after reweighting introducing all EW corrections discussed (red triangles). The points
    are barely distinguishable. Ratios of Improved Born results (with and without EW boxes) to Effective Born 
    in: (i) EW LO $\alpha(0)$ scheme
    are given in top-right,
    (ii) in bottom-left  to  Effective Born {\tt v0} and
    and (iii) in bottom-right plots to Effective Born {\tt v2}.
\label{Fig:Zlineshape} }
\end{figure}

In Fig.~\ref{Fig:Zlineshape} (top-left) distributions of generated and
EW corrected $Z$-line-shape (through  $\sigma^{tot}$) are shown for the $pp$ collision case. The EW weight
is calculated 
using $\cos \theta^*$ definition of the scattering angle as defined in \cite{Richter-Was:2018lld}.
On the logarithmic scale the difference is barely visible.
In the following plots of the same figure we study it in more details.
The ratios of the $Z$ line-shape distributions with gradually introduced
EW corrections are shown. We intend to evaluate the size of
complete {\it Improved Born} predictions with respect to variants of {\it Effective Born}. That is
why for reference predictions (denominator of the weights) the following:
(i) EW LO $\alpha(0)$ (top-right plot),
(ii) Effective Born {\tt v0} (bottom-left plot)  and (iii) Effective Born {\tt v2} (bottom right plot), are used.
For numerators, Improved Born of form-factors without/with box diagram contributions are  used. 
At the $Z$-pole, complete EW corrections of Improved Born give for $\sigma$
about 0.01\% different results from  the one of Effective Born {\tt v2}.
It demonstrates  that if 
for event generation an EW LO matrix element  is used
with   effective variant {\tt v2}
parametrization, the size of missing EW effects will be significantly
reduced.

Similar conclusions can be drawn from 
Table~\ref{Tab:EWnormcorr}, where the numerical impact of EW corrections on the
normalization i.e.
ratios of the $pp$ cross sections integrated
in the range $ 81 < m_{ee} <101$~GeV and $89 < m_{ee} < 93$~GeV, are given.
Total EW corrections for 
EW LO $\alpha(0)$ cross section are about 0.035, while for 
 the Effective Born {\tt v0} it is of about 0.01 and for
 Effective Born~{\tt v2} is of about 0.0001. The main improvement
 of {\tt v2} with respect to {\tt v0} is thanks to  $\rho_{\ell f} \ne 1$ introduced
 already for {\tt v1}.

\begin{table}
 \vspace{2mm}
 \caption{EW corrections to cross sections $\sigma^{tot}$ in the specified mass windows.
   {\tt DIZET 6.45} form-factors and running width was used in re-weighting
   of LHC $pp \to Zj;\;Z \to l^+l^-$ events simulated at 8 TeV.
   From the first two lines magnitude of EW corrections with respect to lowest order, $\alpha(0)$ scheme can be read off.
   Following three lines demonstrate precision of Effective Born variants
   with respect to Improved Born.
 \label{Tab:EWnormcorr}}
 \begin{center}
    \begin{tabular}{|l|c|c|}
        \hline\hline
         Corrections to cross section                       & $ 89 < m_{ee} < 93$ GeV &  $81 < m_{ee} < 101$ GeV  \\ 
         \hline \hline
         $\sigma^{tot}$(Improved Born, no boxes)/$\sigma$(EW LO $\alpha(0)$)   &  0.96505 &  0.96626     \\
         \hline 
         $\sigma^{tot}$(Improved Born, with boxes)/$\sigma$(EW LO $\alpha(0))$   &  0.96510 &  0.96631     \\
         \hline \hline
         $\sigma^{tot}$(Eff. Born {\tt v0})/$\sigma$(Improved Born, with boxes)   &  1.01142 &  1.01135    \\
         \hline
         $\sigma^{tot}$(Eff. Born {\tt v1})/$\sigma$(Improved Born, with boxes)   &  1.00130 &  1.00132     \\
         \hline
         $\sigma^{tot}$(Eff. Born {\tt v2})/$\sigma$(Improved Born, with boxes)   &  0.99989 &  0.99987     \\
    \hline
 \end{tabular}
  \end{center}
\end{table}

\begin{figure}
  \begin{center}                               
{
  \includegraphics[width=7.5cm,angle=0]{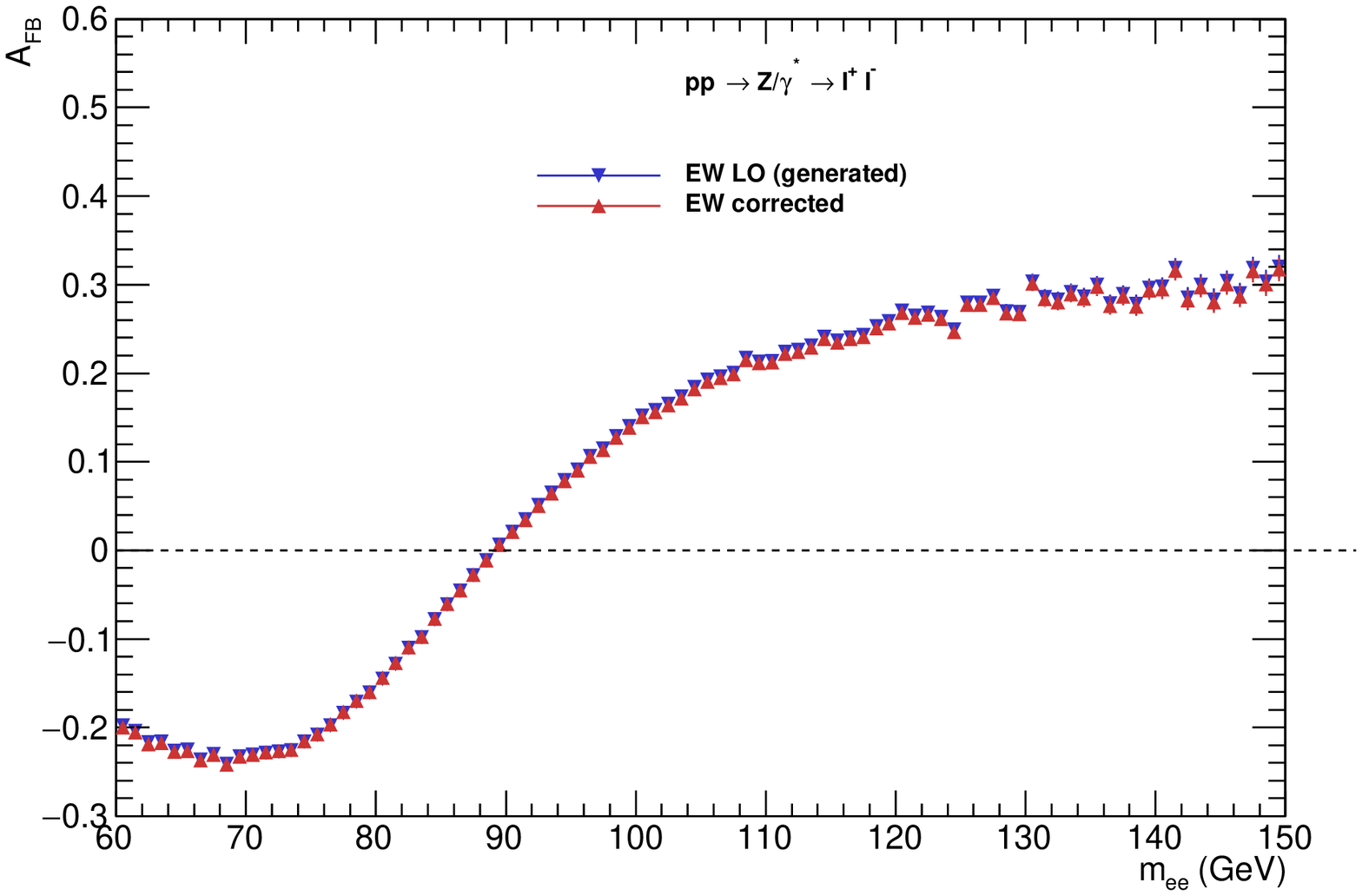}   
  \includegraphics[width=7.5cm,angle=0]{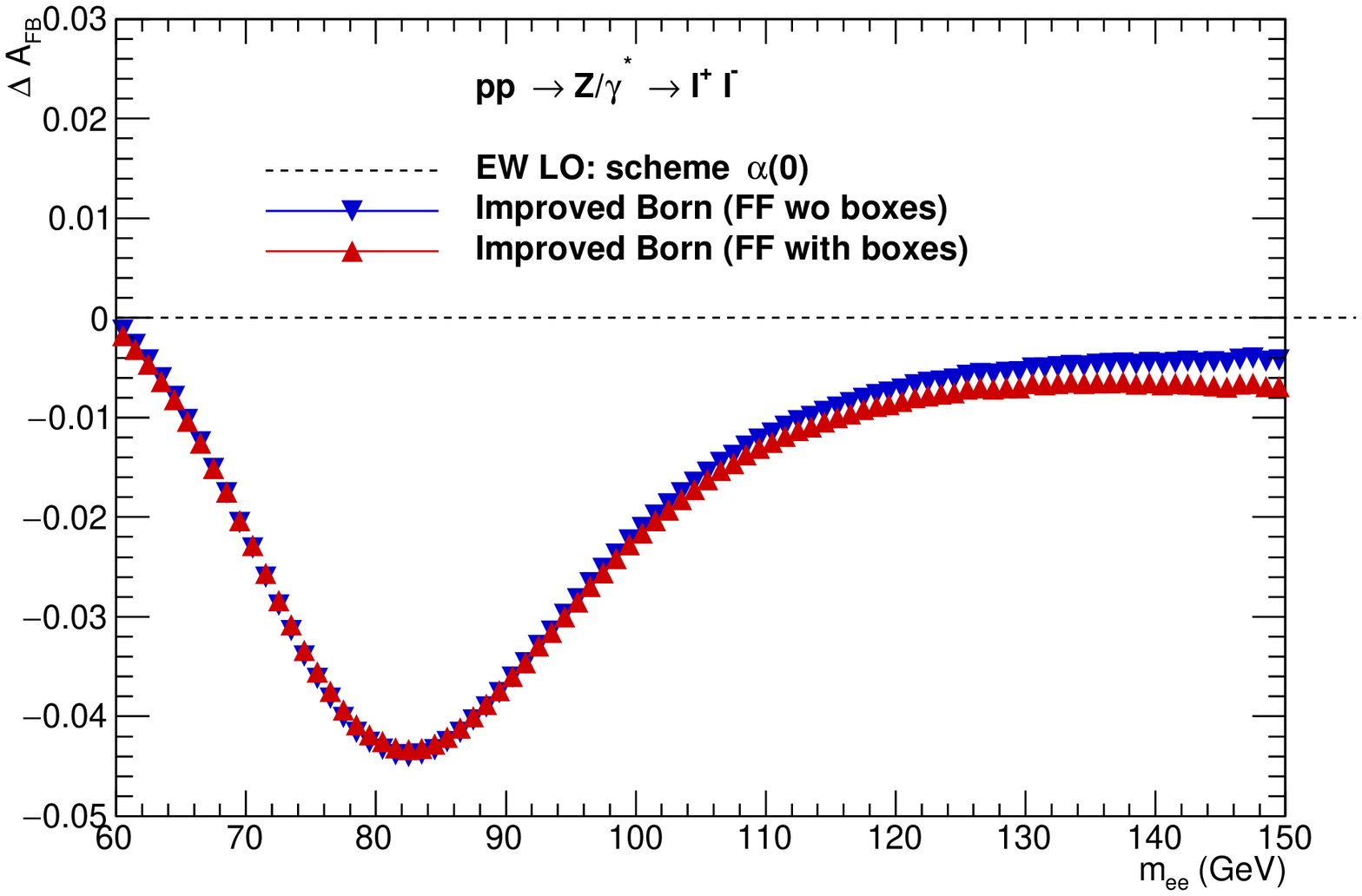}
  \includegraphics[width=7.5cm,angle=0]{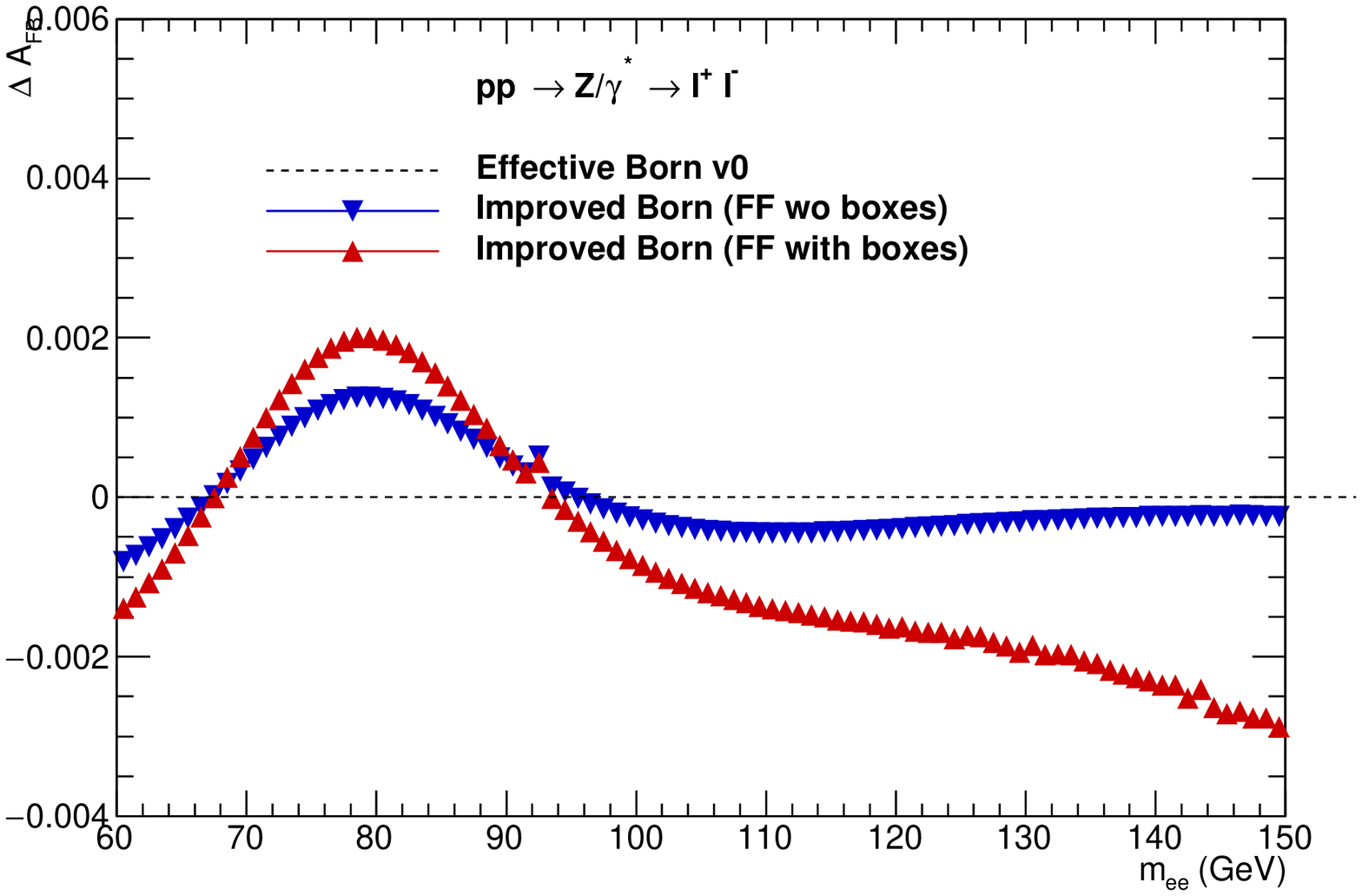}
  \includegraphics[width=7.5cm,angle=0]{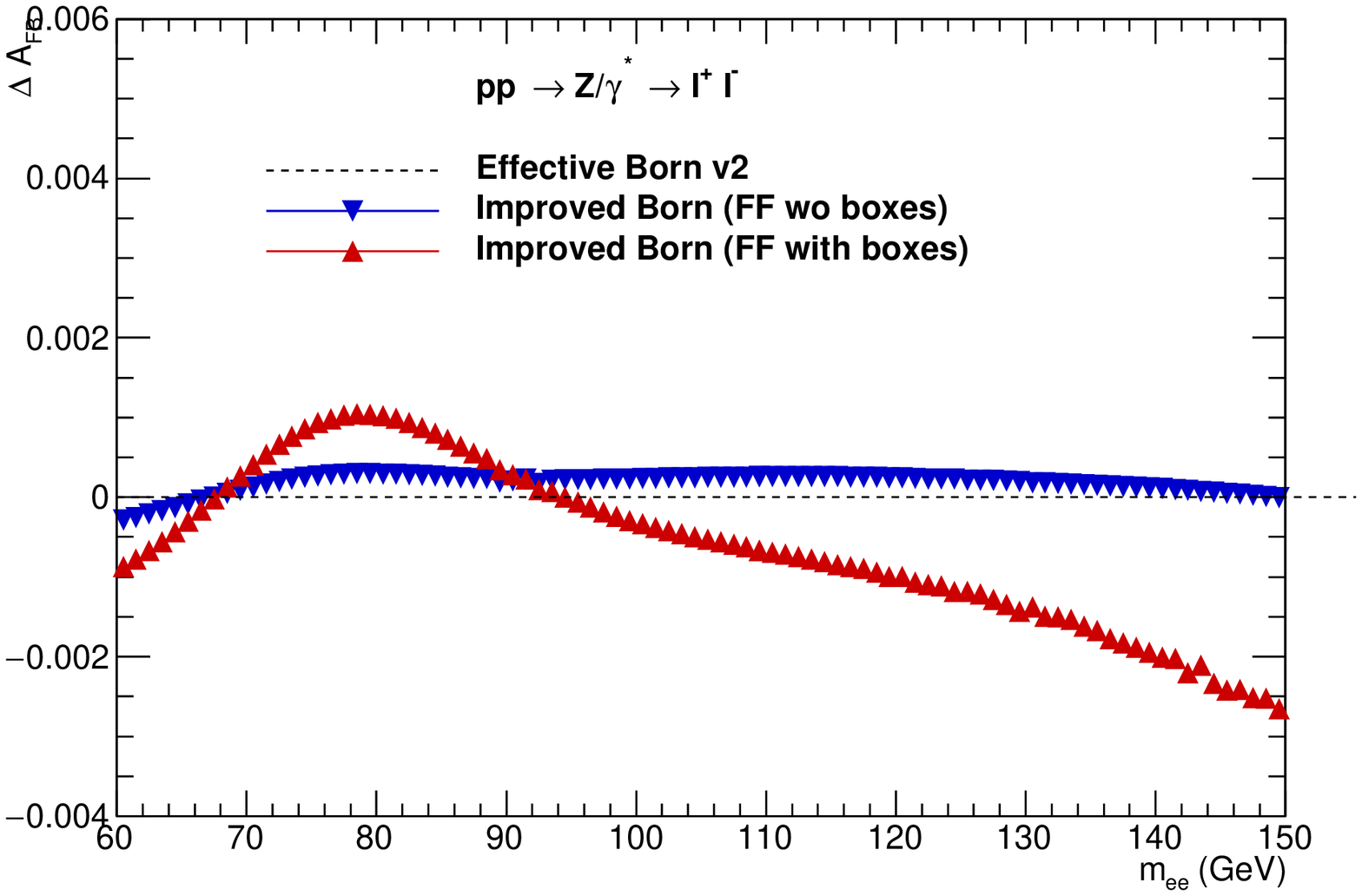}
}
\end{center}
  \caption{Top-left: the $A_{FB}$ distribution as generated in {\tt Powheg+MiNLO} sample (blue triangles)
    and after reweighting introducing all EW corrections (red triangles).
    The two choices are barely distinguishable. The differences $\Delta A_{FB} = A_{FB} - A_{FB}^{ref}$,
    of Improved Born results (with and without EW boxes) to
    Effective Born 
    in: (i) EW LO $\alpha(0)$ scheme
    are given in top-right,
    (ii) in bottom-left  to  Effective Born {\tt v0} and
    and (iii) in bottom-right plots to Effective Born {\tt v2}.
%
\label{Fig:Afb} }
\end{figure}

Let us now turn our attention to the EW corrections for the forward backward asymmetry  $A_{FB}$. Again 
for the $pp \to Z/\gamma^*\to l^+l^-$ process, energy range from  60 to 150 GeV was chosen which is of interest for EW effects.
As in the case of cross section,
shape and size of the corrections depend on whether
box exchange diagrams are included in the Improved Born result.
In top-left plot of Fig.~\ref{Fig:Afb}, the $A_{FB}$ distribution,
as generated (EW LO) and superimposed with  EW corrected result is shown.
The points for the two cases are practically indistinguishable.
Further three  plots of the figure, with the difference
$\Delta A_{FB} = A_{FB} - A_{FB}^{ref}$ provide details.
For the  reference $A_{FB}^{ref}$, the three versions  of the Effective Born
detailed in Table \ref{Tab:BornEff} are used again:
(i) EW LO $\alpha(0)$,
(ii) {\tt v0} and (iii) {\tt v2}.
The EW corrections  for   $A_{FB}$ of EW LO Born with $\alpha(0)$ scheme, integrated around the  $Z$-pole,
necessary to reproduce Improved Born result can reach 
 -0.03514, see Tab.~\ref{Tab:AFBEWcorr}.
The Effective Born {\tt v0} reproduces Improved Born up to $\Delta A_{FB}$ of about
-0.0004, while  the Effective Born {\tt v2} up to  -0.0002. The {\tt v2} variant is again better
by a factor of two than the {\tt v0} one.

All that points to the limitation of  real constants Effective Born and its parametrization with 
 $\sin^2\theta_W^{ eff}(M_Z)$ at about $20 \cdot 10^{-5}$ or so, even if
 $\alpha(M_Z)$ and $\rho_{lf}(M_Z)$ is used.
Note that even if $P_\tau$ is not particularly suitable for $pp$ collision
measurements, it weakly depends on the production process and that is why it is
suitable for numerical $\sin^2\theta_W^{ eff}(M_Z)$ ambiguities evaluation in general case.
That is why previous subsection results are of the relevance for $pp$ too.
It is worth noting that they also  point to $20 \cdot 10^{-5}$ as an ultimate precision tag,
 

\begin{table}
 \vspace{2mm}
 \caption{The difference in forward-backward asymmetry, $\Delta A_{FB}$, in the specified mass windows.
   {\tt DIZET 6.45} form-factors and running width was used in re-weighting
   of LHC $pp \to Zj;\;Z \to l^+l^-$ events simulated at 8 TeV.
   From the first two lines magnitude of EW corrections with respect to lowest order, $\alpha(0)$ scheme.
   Following three lines demonstrate precision of Effective Born variants
   with respect to Improved Born.
 }
 \label{Tab:AFBEWcorr}
 \begin{center}
    \begin{tabular}{|l|c|c|}
        \hline\hline
         Corrections to $A_{FB}$                       & $ 89 < m_{ee} < 93 $ GeV &  $81 < m_{ee} < 101 $ GeV  \\ 
         \hline \hline
         $A_{FB}$(Improved Born, no boxes) - $A_{FB}$(EW LO $\alpha(0)$)   &  -0.03491 &  -0.03515     \\
         \hline 
         $A_{FB}$(Improved Born, with boxes) - $A_{FB}$(EW LO $\alpha(0))$   &  -0.03489 &  -0.03514     \\
         \hline \hline
         $A_{FB}$(Eff. Born {\tt v0}) - $A_{FB}$(Improved Born, with boxes)   &  -0.00039 &  -0.00042    \\
         \hline
         $A_{FB}$(Eff. Born {\tt v1}) - $A_{FB}$(Improved Born, with boxes)   &  -0.00042 &  -0.00042     \\
         \hline
         $A_{FB}$(Eff. Born {\tt v2}) - $A_{FB}$(Improved Born, with boxes)   &  -0.00022 &  -0.00024     \\
    \hline
 \end{tabular}
  \end{center}
\end{table}

\section{Electroweak corrections in {\tt TauSpinner}: library versions and initializations}
\label{sec:onLIbraries}

In the present section  we  address the impact of the {\tt DIZET} library variants,
which have by now a life-time of more than
three decades.
The  versions of the {\tt DIZET} EW correction library, which are
used in our numerical discussions are presented briefly in Appendix~\ref{app:kkmc};   details are given in Ref.~\cite{Arbuzov:2020}.
Specification
of   initializations are collected in  Appendix \ref{app:iniflags}.
One may wonder if the last version would not suffice. However, availability
of the software used for the solutions of legacy measurements is of some value.
That is why in Ref.~\cite{Arbuzov:2020} several versions
of the present and past EW  {\tt DIZET} library were collected.
On the other hand, archived with \cite{koralz4:1994} less popular calculations
of the past will not receive  our attention.

Each of the four versions of {\tt DIZET} library of EW effects comes
with a wealth of options, which may be activated with their input flags.
These options can be used
to evaluate the importance of the particular improvement introduced over the years.
The graphical programs to monitor the changes are available in
{\tt TAUOLA/TauSpinner/examples} directory. The {\tt tau-reweight-test.cxx}
can be used to demonstrate how events can be corrected with the weight
representing improvement from {\tt TAUOLA} Effective Born of its constant
couplings to the one of Improved Born of formula (\ref{Eq:BornEW})
with form-factors interpolated  from the text files with tables prepared with
{\tt KKMC} interface to {\tt DIZET}.

The default for {\tt anomalous Born} function, introduced for the first time
 in Ref.~\cite{Banerjee:2012ez}, is not anymore a dummy but is  
 now the one of EW Improved Born, which 
uses the EW form-factors tables (if available). The new sub-directory {\tt Dizet-example}
collects programs and scripts for 
form-factors graphic representation.
Plots of form-factors can be drawn, as a function of energy, scattering angle
and flavour of incoming partons (it can be an electron-positron pair as well).
The integrated over angle partonic cross section $\sigma^{tot}$, $A_{FB}$ and $P_\tau$ can be
graphically presented. Comparison plots can be prepared,
either with the help of the {\tt FFdrawDwa.C} script to compare results
with EW  form-factors obtained with  variants of {\tt DIZET} initialization, or
with  {\tt FFdraw.C} to compare  Improved Born and Effective Born of the choice 
  as implemented in {\tt TAUOLA} package. For technical details see Appendix
  \ref{app:B}. An example results for comparison of
  Effective Born as encapsulated in {\tt TauSpinner/Tauola} (version of December 2019) defaults and Improved Born
  with EW form-factors of {\tt DIZET 6.45}  were shown in
  Fig.~\ref{figTest1}. 
   One should note that differences between semi-analytical results obtained from Improved Born and Effective Born,
   even in case when detailed tuning of parameters is not performed,
   are not large from the perspective of many applications.


\begin{figure}[htp!]
\begin{tabular}{ccc}
  \includegraphics[width=0.47\columnwidth]{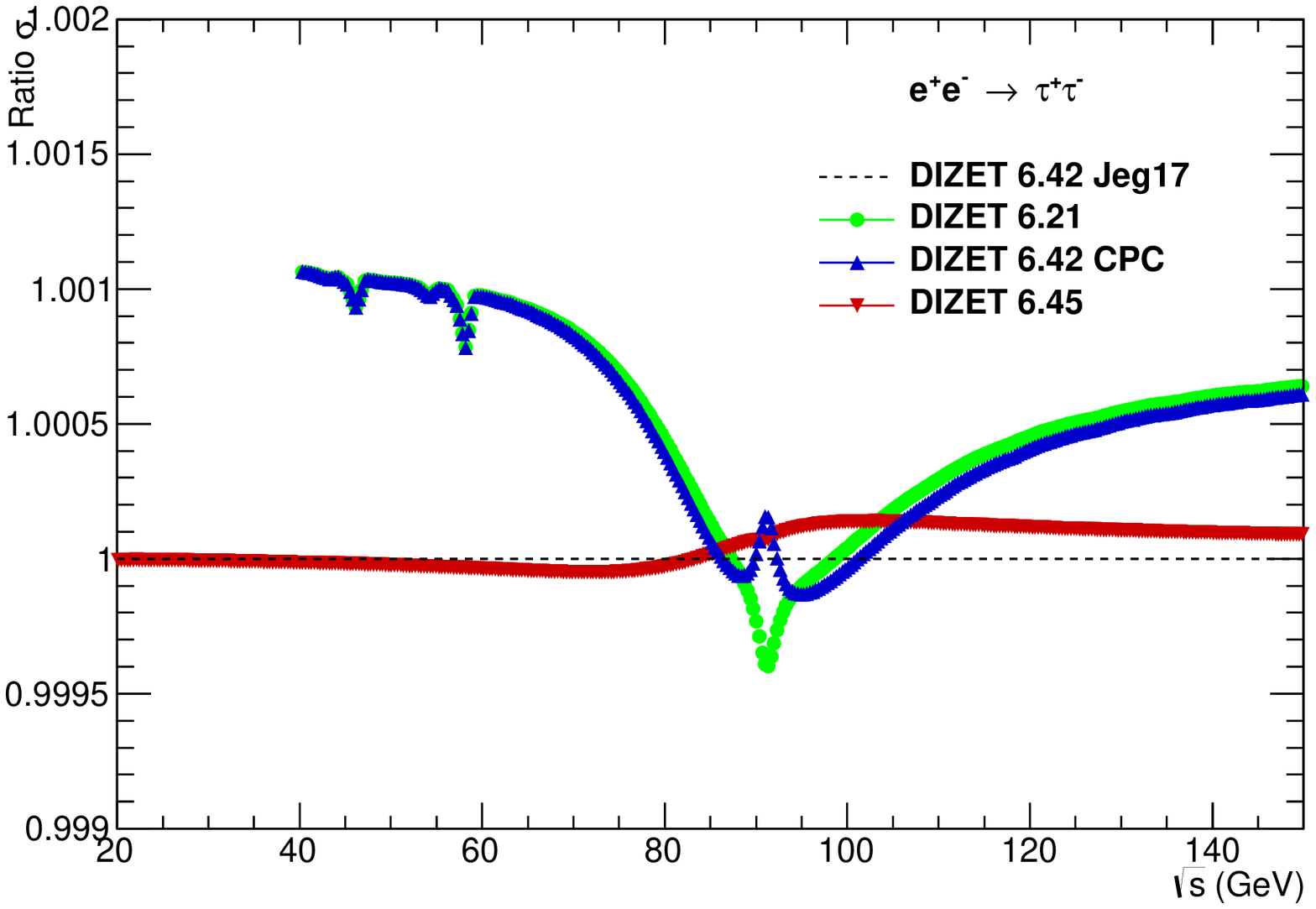} &
    \includegraphics[width=0.47\columnwidth]{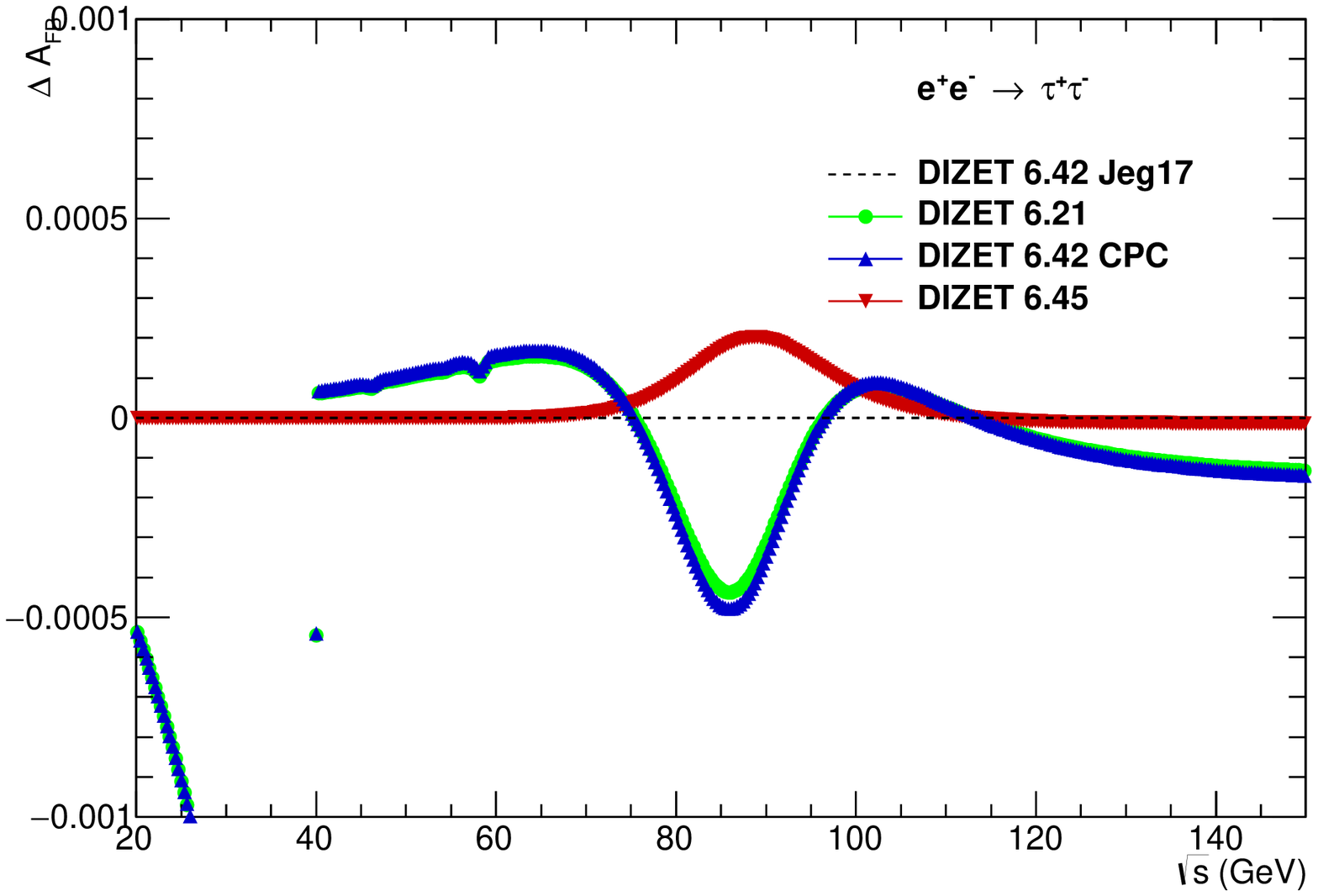} \\
  \includegraphics[width=0.47\columnwidth]{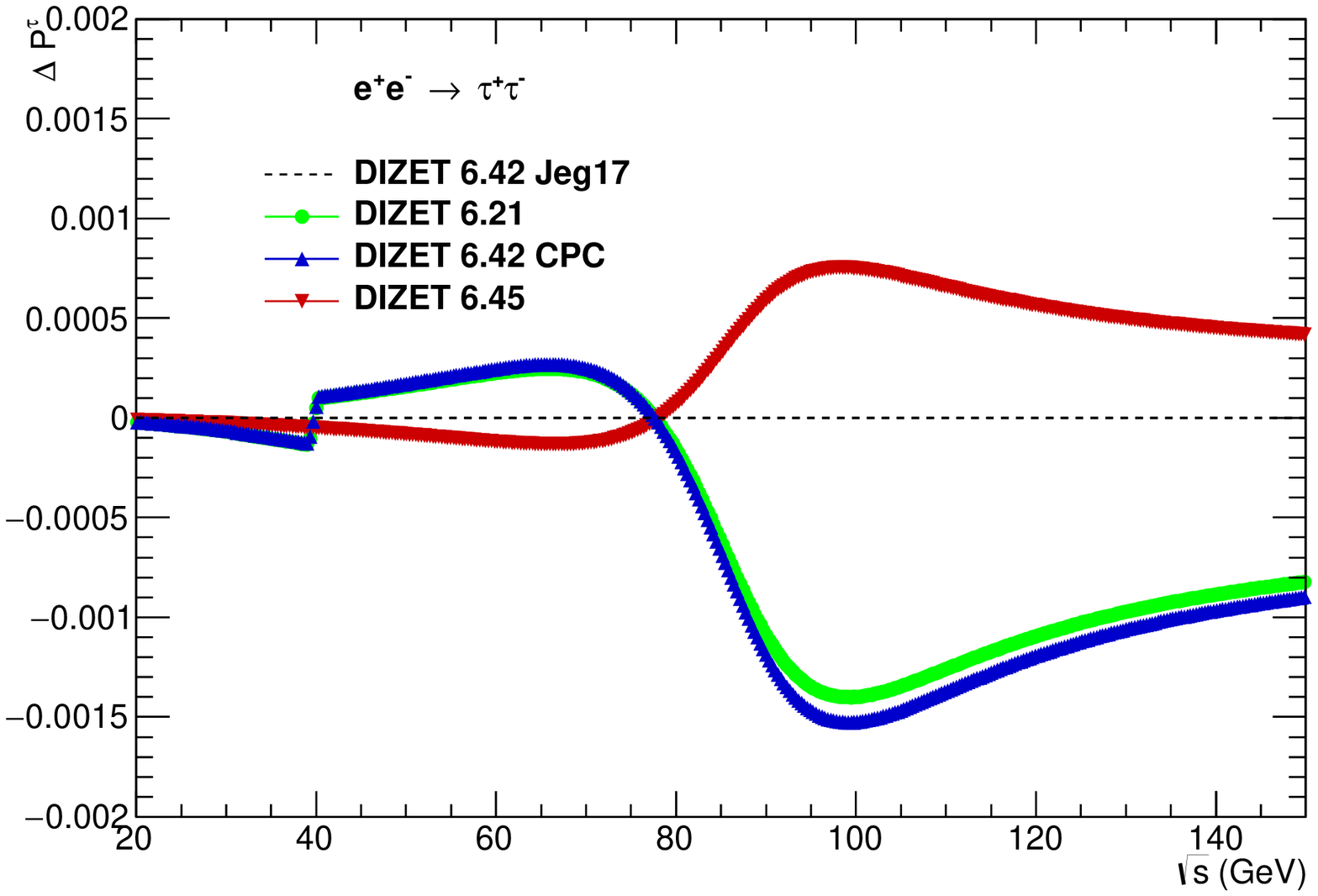} 
\end{tabular}
\caption{ Comparison of   $\sigma_{tot}(m_{\tau^+ \tau^-})$, $A_{FB}(m_{\tau^+ \tau^-})$ and
  $P_{\tau}(m_{\tau^+\tau^-})$, obtained from {\tt TauSpinner} calculations of
  Improved Born and 
  EW tables calculated  with the {\tt DIZET} libraries. As a reference,
  version 6.42 improved
  with photon vacuum polarization of Ref.~\cite{Jegerlehner:2019lxt} is used.
  \;\;
  Note, that interface of photon vacuum polarization of {\tt DIZET} 6.42 and
  6.21 prevented its calculation below 40 GeV. This, and other minor parameter variation in particular of
  $M_Z$ and $\Gamma_Z$, lead to  bumps on the plots which do  not need to be investigated now, and even for the precision
  tests of the SM at LHC, as they are smaller than  precision requirements.
  On the other hand, proper adjustment
  for the effects important at (and around) the $Z$ peak need to be performed.
  The corresponding EW-corrected $M_W$, $\sin^2\theta_W^{eff}$ and $\alpha$ at the $Z$ pole are collected in
  Table \ref{Tab:Dizet_6.XX_coupl}. 
  Similar results can be obtained from {\tt TauSpinner} for the quark level Effective Born and Improved Born predictions.
    \label{figTest2}}
\end{figure}
%
  
  We concentrate when presenting comparison results on $e^+e^- \to \tau^+\tau^-$ production process as its
  phenomenology represents LEP time reference to present day projects, in particular for  LHC measurements. Also, parton level cross sections,
  necessary for
  LHC phenomenology, are obscured by hadronic interaction effects, thus are
  more complex for interpretation and require simultaneous evaluation of
  hadronic interactions.

 Numerical results, as in  previous
section, are  monitored with  $\sin^2\theta_W^{eff}$, $P_{\tau}$, $A_{FB}$ and
parton level $\sigma^{tot}$. 
In  Fig.~\ref{figTest2} and Table~\ref{Tab:Dizet_6.XX_coupl} it  is shown
how results depend on the library version. The presentation in
Table~\ref{Tab:Dizet_6.XX_coupl} includes predictions on $\alpha(M_Z^2)$, $M_W$, $\Delta r$,
and $\sin^2\theta_W^{eff}$. Further results  are
 delegated to Appendices.

  By inspection of Table \ref{Tab:Dizet_6.XX_coupl} one can conclude that the choice of EW library variant is not of great importance,
  unless precision better than $20 \cdot 10^{-5}$ on $\sin^2\theta_W^{eff}$ is required. Even if precision requirements are not very demanding one should keep in mind that below 40 GeV
  in older versions of {\tt DIZET} the hadronic part of  $\Pi_{\gamma\gamma}(s)$ was set to zero. See also
  Fig.~\ref{figTest1} for minor discontinuity at 30 GeV due to edge of tabulation zones. Further 
  details on the impact of change of options/flags  of {\tt DIZET 6.45}
  are collected in
  Tables~\ref{Tab:Dizet_dalf5h_update}, \ref{Tab:Dizet_AMT4_printout} of
  Appendix~\ref{sec:App-A-2}.
  
\begin{table}
 \vspace{2mm}
 \caption{Predictions for different versions of {\tt DIZET 6.}
   explained in Appendix \ref{app:6XXEW}. The $\Delta r$, $\Delta r_{em}$ represent
 corrections to $M_W$ calculations, see Eq.~(\ref{Eq:Dr}), where $s_W^2=1-\frac{M_W^2}{M_Z^2}$.} 
 \label{Tab:Dizet_6.XX_coupl}
  \begin{center}
    \begin{tabular}{|l|c|c|c|c|}
        \hline\hline
        Parameter  &  {\tt DIZET 6.21} &  {\tt DIZET 6.42} &  {\tt DIZET 6.42}     &  {\tt DIZET 6.45}\\
                   &                    &     CPC           &  (Jeg. 2017) &      \\
         \hline\hline
         $\alpha (M_Z^2)$                            &  0.007759954      & 0.007759954    &  0.0077549256  & 0.0077549256    \\
         $1/\alpha (M_Z^2)$                          &  128.86674175     & 128.86674175   &  128.95030206  & 128.950302056    \\
         \hline 
         $M_W$ (GeV)                                 &  80.3560012       &  80.3535973    &   80.3621285   & 80.3589358    \\
         $\Delta r$                                  &  0.03676619       & 0.03690875     &  0.03640232    & 0.03633354   \\
         $\Delta r_{rem}$                             &  0.01168031       & 0.01168001     &  0.01168106    & 0.01168393    \\
         \hline
         $s^2_W$                                     &  0.22345780       & 0.22350426     &  0.22333937    & 0.22340108     \\
         \hline
         $\sin^{2}\theta_W^{eff\ lepton}(M_Z^2)$         &   0.23173519      & 0.23174233     &  0.23157947    & 0.23149900     \\
         $\sin^2\theta_W^{eff\ up-quark}(M_Z^2)$         &   0.23162861      & 0.23174233     &  0.23147298    & 0.23139248    \\
         $\sin^2\theta_W^{eff\ down-quark}(M_Z^2)$       &   0.23150149      & 0.23174233     &  0.23134599    & 0.23126543    \\
   \hline
 \end{tabular}
  \end{center}
\end{table}

\subsection{Parametric uncertainties}
\label{sec:numerical}
The precision  of the  EW calculations depends not only on the EW
scheme used for the calculations (see e.g. \cite{Alioli:2016fum,Dittmaier:2014qza},
but also on imposed set of input parameters and
corresponding parametric uncertainties. Parametric uncertainties are defined
as the ones due to ambiguities of EW calculation inputs,
such as $m_t$, $M_W$ or $\Delta \alpha_h^{(5)}(s)$.
That is the reason why precision of these input parameters (taken from
measurements), specially $M_W$ or $\sigma^{tot}_{e+e- \to hadrons} (s)$
(for $\alpha_{QED}(M^2_Z)$), is of importance.
For clarity of the presentation this topic is covered only in Appendices
\ref{sec:App-A-2}, \ref{sec:App-A-3},
in particular in Tables   \ref{Tab:Dizet_dalf5h_update},
\ref{Tab:Dizet_AMT4_printout},
\ref{Tab:Dizet_dalf5h_uncertainty},
\ref{Tab:Dizet_mt_uncertainty}.
We show
 how some phenomenologically sound quantities  depend
 on  initialization ambiguities for $\alpha(0)$ EW scheme used in the {\tt DIZET} library.
 In particular how do they depend on:
(i) distinct $\Delta \alpha_h^{(5)}(s)$ parametrization,
(ii) uncertainty from changing $\Delta \alpha_h^{(5)}(M_Z^2)$ by $\pm$ 0.0001,
 (iii) uncertainty due to top mass change by $\pm 0.5$ GeV.
 The estimated total parametric uncertainty for EW  $\alpha(0)$ scheme used for  $\sin^2\theta_W^{eff}(M_Z^2)$ is about 0.00005.

To summarize, these example results can be useful by themselves but they 
represent a precondition for the choice of the EW schemes and their inputs.
Large parametric uncertainties
could indicate that some schemes are not optimal. That is why
they contribute to evaluation of reliability for the 
Effective Born (effective couplings) concept, too.


\section{Summary}
\label{sec:summary}

One loop EW corrections play important role in the precision tests of
the Standard Model. At the same time, other effects, related to special classes
of higher order corrections had to be taken into account. That is the reason, why
special libraries of EW corrections were developed, maintained and gradually
improved. Over the last 30 years the {\tt DIZET} library was established as a
prominent one.
Special role was played by the so-called EW
$\alpha(0)$ scheme. This scheme and {\tt DIZET} library
found large spectrum of applications, not only in phenomenology of LEP $e^+e^-$ collisions
but $p\bar p$ and $p p$  Tevatron and LHC experiments as well.
An attempt to archive distinct versions of EW libraries is provided in
\cite{Arbuzov:2020}. These results and methods need to be reproducible at the time of future
{\tt FCC} or similar experiments. Negligible differences of the past  may play important role
for future higher precision projects.
From our investigations, see Subsection \ref{sec:variantsof}, \ref{sec:atLHC},
we can conclude
that effective  couplings approach can be useful for  $\sin^2\theta_W^{eff}$ precision
up to about $20 \cdot 10^{-5}$.
Beyond that, Improved Born without simplifications is needed.

In principle, {\tt DIZET}
relies on one-loop calculations, but it is supplemented with dominant higher
order terms. Presented implementation of {\tt TauSpinner} weights enable discussion of
particular classes of higher order effects.
In the present paper we have explained how the  numerical impact of some
effects of EW results can be imprinted into broad spectrum
of simulation samples where EW loop effects are missing, or impact of their 
initialization is to be studied. We have installed useful for that purpose algorithms into  {\tt TauSpinner} library. 
Example numerical results are focused on center-of-mass system energy
dependence of total cross sections, forward-backward
asymmetry of leptons, and $\tau$ lepton polarization. 
Those were studied for  $e^+e^-$ and $pp$ collisions.

New {\tt TauSpinner} algorithms have potential  to improve e.g. EW
effects in simulation samples obtained from programs predominantly of strong interactions.
We have shown results of re-weighting with different levels of sophistication for implementation of
EW corrections.
Let us point out that from the perspective
of forthcoming efforts on higher-order high precision EW calculations,
{\tt TauSpinner/DIZET} algorithms  may be
used as a set of methods for evaluation which contributions (and to which order) 
need to be taken into account to attain requested precision level.

\centerline{\bf\large Acknowledgments}

E.R-W. would like to thank Daniel Froidevaux
and colleagues
from ATLAS Collaboration Standard Model Working Group for numerous inspiring
discussions on the applications of presented here implementation of EW
corrections to the $\sin^2\theta_W^{eff}$ measurement at LHC.
\providecommand{\href}[2]{#2}

\appendix

\section{DIZET  EW corrections}
\label{app:6XXEW}

The {\tt DIZET} package relies on {\it on-mass-shell} (OMS) normalization
scheme~\cite{Bardin:1980fe,Bardin:1981sv}, thus the
($G_\mu,\ \alpha(0)$, $M_Z$) are the principal
input parameters, dependence on $m_h$, top quark and lepton masses are
numerically less important.
The OMS normalization scheme input
includes masses
of all fundamental particles, both fermions and bosons,
electromagnetic coupling constant $\alpha(0)$ and strong coupling $\alpha_s(M_Z)$. The OMS is used  with modifications.
The dependence on the ill-defined masses of the light quarks {\it u, d,c, s} and {\it b} is solved
by dispersion relation, for details see \cite{Bardin:1999yd}. Another exception is $W$-boson mass $M_W$, which
still can be predicted with better theoretical uncertainties than
experimentally measured values, exploiting the very precise
knowledge of the Fermi constant in $\mu$-decay $G_{\mu}$.
%
The discussed above EW scheme is in the literature often called EW $\alpha(0)$ scheme
\cite{ALEPH:2005ab}.
%
%
%
%
The $M_W$ is calculated iteratively from the  equation

\begin{equation}
  M_W = \frac{M_Z}{\sqrt{2}} \sqrt{ 1 + \sqrt{ 1 - \frac{ 4 A^2_0}{M^2_Z ( 1 - \Delta r)}}},
  \label{Eq:MWiterative}
\end {equation}
where
\begin{equation}
  A_0 = \sqrt{ \frac{\pi \alpha(0)}{\sqrt{2}G_{\mu}}}.
\end {equation}

The Sirlin's parameter $\Delta r$ \cite{Sirlin:1980nh}
\begin{equation}
  \label{Eq:Dr}
  \Delta r = \Delta \alpha(M_Z) + -\Delta r_L + \Delta r_{em}
\end {equation}
is also calculated iteratively, and the definition of $ \Delta r_{L}, \Delta r_{em} $ involve resummation and higher order corrections.
Since this term implicitly depends on $M_W$ and $M_Z$ iterative procedure is needed.
The resummation term in formula (\ref{Eq:Dr}) is not formally justified by renormalisation group arguments,
correct generalization is to compute higher order corrections, see more discussion in ~\cite{Bardin:1999yd}.
The electromagnetic coupling evolves from Thomson limit  and for $Z$-boson energy scale receives corrections
\begin{equation}
  \Delta \alpha(M_Z) = \Delta \alpha_h^{(5)}(M_Z) +\Delta \alpha_l(M_Z)
  +\Delta \alpha_t(M_Z) + \Delta\alpha^{\alpha\alpha_s}(M_Z) 
\end{equation}

The hadronic vacuum polarization correction is contained in the quantity denoted as $\Delta \alpha_h^{(5)}(M_Z)$,
which is treated as one of the input parameters. It can be either computed from quark masses or, preferably, fitted to
experimental low energy $e^+ e^- \to hadrons$ data \cite{Jegerlehner:2017zsb}.
The leptonic loop correction $\Delta \alpha_l(M_Z)$ is calculated analytically.
 Both $\Delta \alpha_h^{(5)}(M_Z)$ and $\Delta \alpha_l(M_Z)$
 are significant, respectively about 0.0275762 and 0.0314976, the remaining terms are rather marginal,
 respectively about $-5 \cdot 10^{-5}$ and  $-1 \cdot 10^{-5}$.
 

In the OMS renormalisation scheme the weak mixing angle is defined uniquely through the gauge-boson masses:
\begin{equation}
  \label{Eq:sw2onshell}
  \sin^2\theta_W = s^2_W = 1 - \frac{M^2_W}{M^2_Z}.
\end{equation}
With this scheme, measuring $\sin^2\theta_W$ would be equivalent to indirect measurement of $M^2_W$ through the
relation (\ref{Eq:sw2onshell}).


\subsection{Initialization flags and input parameters}
\label{app:iniflags}

The recommended sets of flags are quite stable since 1995, new options consider updated parametrisations
of the vacuum polarization hadronic corrections $\Delta \alpha_{had}^{(5)}$ (flag {\tt IHVP}), updated calculations for two
loop fermionic corrections (flag {\tt IAMT4}) and updated three-loop corrections (flag {\tt IAFMT}).

In Table~\ref{Tab:Dizet_6.XX_flags} we collected information on the initialization flags recommended for different versions
of {\tt DIZET 6.XX}. For detailed information about meaning of the individual flags see {\tt DIZET 6.XX}
documentations~\cite{Bardin:1989tq, Bardin:1999yd, Arbuzov:2005ma}.

Let us here just explain those, for which recommended values have changed since version {\tt DIZET 6.21}:
\begin{itemize}
\item
Switch for Hadronic vacuum polarization corrections $\Delta \alpha_{had}^{(5)}$:\\
{\tt IHVP}  = 1   parametrization of \cite{Eidelman:1995ny}\\
{\tt IHVP}  = 5   parametrization of \cite{Jegerlehner:2017zsb}
\item
Switch for resummation of the leading $O(G_f m^2_t)$ EW corrections:\\
{\tt IAMT4} = 4  with two-loop sub-leading corrections and resummation~\cite{Degrassi:1994tf,Degrassi:1995mc,Degrassi:1996mg,Degrassi:1999jd}\\
{\tt IAMT4} = 5  with fermionic two-loop corrections to $M_W$ ~\cite{Freitas:2000gg,Freitas:2000nv,Freitas:2002ja} \\
{\tt IAMT4} = 6  with complete two-loop corrections to $M_W$~\cite{Awramik:2003rn} and fermionic two-loop corrections
                 to $sin^2\theta_W^{eff\ lep}$~\cite{Awramik:2004ge}\\
{\tt IAMT4} = 7  with complete two-loop corrections to  $sin^2\theta_W^{eff\ lep}$ and  $sin^2\theta_W^{eff\ lb}$~\cite{Awramik:2006uz,Dubovyk:2016aqv}\\
{\tt IAMT4} = 8  with complete two-loop corrections to  $sin^2\theta_W^{eff}$~\cite{Dubovyk:2019szj}
\item
Switch for three-loop corrections $O(\alpha \alpha^2_s)$ to the EW $\rho$ parameter:\\
{\tt IAFMT} = 1  corrections $O(G_{\mu} m^2_t \alpha^2_s)$ included~\cite{Avdeev:1994db}\\
{\tt IAFMT} = 2  corrections $O(G_{\mu} m^2_t \alpha^2_s)$, $O(G_{\mu} M^2_Z \alpha^2_s + log (m^2_t))$ included
{\tt IAFMT} = 3  corrections $O(G_{\mu} m^2_t \alpha^2_s)$, $O(G_{\mu} M^2_Z \alpha^2_s + log (m^2_t))$
                 and  $O(G_{\mu} M^2_Z/m^2_t \alpha^2_s)$ included
\end{itemize}

Since LEP time physics measurements evolved,
and as a consequence initialization parameters as well.
For  the recent status summary
see last edition by Particle Data Group~\cite{Tanabashi:2018oca}.
The Higgs boson has been discovered at LHC and its mass measured with precision of 25 MeV~\cite{Aaboud:2018wps}.
The W boson mass in known at LHC  with precision better than 18 MeV ~\cite{Aaboud:2017svj} and the top mass is known with precision
much better than 1 GeV ~\cite{Aaboud:2018zbu}.  
  
In Tables~\ref{Tab:EWtreelevel_oms} and~\ref{Tab:masses_init} we collected initialization parameters:
masses and couplings, used of the paper numerical evaluation.
The exact values of some of them, which serve as benchmark values for different comparisons, has been chosen as such to be
fully compatible with the ongoing studies of the LHC EW Working Group~\cite{LHCEWWG}.

\begin{table}
 \vspace{-1mm}
 \caption{ {\tt DIZET} initialization flags: different versions defaults.} 
 \label{Tab:Dizet_6.XX_flags}
  \begin{center}
    \begin{tabular}{|l|l|c|c|c|l|}
        \hline\hline
        Input NPAR()  &  Internal flag  & {\tt DIZET} 6.21 & {\tt DIZET} 6.42 &  {\tt DIZET} 6.45 & Comments \\
                      &                 & Defaults in~\cite{Bardin:1989tq} &  Defaults in~\cite{Arbuzov:2005ma}&  & \\
  \hline\hline 
  NPAR(1)  & IHVP  &  1  &  1  &  5  & $\Delta \alpha_{had}^{(5)}$ param. from~\cite{Jegerlehner:2017zsb} in v6.45 \\
  NPAR(2)  & IAMT4 &  4  &  4  &  8  & New development in v6.42, v6.45 \\
  NPAR(3)  & IQCD  &  3  &  3  &  3  &  \\
  NPAR(4)  & IMOMS &  1  &  1  &  1  & $M_W$ calculated with formula~(\ref{Eq:MWiterative}) \\
  NPAR(5)  & IMASS &  0  &  0  &  0  &  \\
  NPAR(6)  & ISCRE &  0  &  0  &  0  &  \\
  NPAR(7)  & IALEM &  3  &  3  &  3  &  \\
  NPAR(8)  & IMASK &  0  &  0  &  0  & Not used since v6.21 \\
  NPAR(9)  & ISCAL &  0  &  0  &  0  &  \\
  NPAR(10) & IBARB &  2  &  2  &  2  &  \\
  NPAR(11) & IFTJR &  1  &  1  &  1  &  \\
  NPAR(12) & IFACR &  0  &  0  &  0  &  \\
  NPAR(13) & IFACT &  0  &  0  &  0  &  \\
  NPAR(14) & IHIGS &  0  &  0  &  0  &  \\
  NPAR(15) & IAMFT &  1  &  3  &  3  &  \\
  NPAR(16) & IEWLC &  1  &  1  &  1  &  \\
  NPAR(17) & ICZAK &  1  &  1  &  1  &  \\
  NPAR(18) & IHIG2 &  1  &  1  &  1  &  \\ 
  NPAR(19) & IALE2 &  3  &  3  &  3  &  \\
  NPAR(20) & IGREF &  2  &  2  &  2  &  \\
  NPAR(21) & IDDZZ &  1  &  1  &  1  &  \\
  NPAR(22) & IAMW2 &  0  &  0  &  0  &  \\
  NPAR(23) & ISFSR &  1  &  1  &  1  &  \\
  NPAR(24) & IDMWW &  0  &  0  &  0  &  \\
  NPAR(25) & IDSWW &  0  &  0  &  0  &  \\
  \hline
  \end{tabular}
  \end{center}
\end{table}

\begin{table}
 \vspace{2mm}
 \caption{The EW parameters used at tree-level EW, with on-mass-shell definition (LEP convention).} 
 \label{Tab:EWtreelevel_oms}
  \begin{center}
    \begin{tabular}{|l||c|}
      \hline\hline
      Parameter & ($\alpha(0), G_{\mu}, M_Z$) \\
      \hline\hline
      $M_Z$  (GeV)        & 91.1876   \\
      $\Gamma_Z$  (GeV)   & 2.4952    \\
      $\Gamma_W$  (GeV)   &  2.085    \\
      $1/\alpha$          & 137.035999139  \\
      $\alpha$            & 0.007297353    \\
      $G_{\mu}$   (GeV$^{-2}$) & 1.1663787 $\cdot 10^{-5}$ \\
      \hline 
      $M_W$   (GeV)       & 80.93886         \\
      $s^2_W$             & 0.2121517      \\
      \hline
      $ \alpha_s(M_Z)$    & 0.12017890  \\
      \hline
    \end{tabular}
 \end{center} 
 \vspace{2mm}
 \caption{Values of fermions and Higgs boson masses used for calculating EW corrections. 
 \label{Tab:masses_init}}
  \begin{center}
    \begin{tabular}{|l|c|l|}
        \hline\hline
        Parameter      & Mass (GeV)          & Description \\
         \hline\hline
        \hline
      $m_e$             &  5.1099907e-4           &  mass of electron \\
      $m_{\mu}$          &  0.1056583              &  mass of muon \\
      $m_{\tau}$         &  1.7770500              &  mass of tau \\
      $m_{u}$           &  0.0620000              &  mass of up-quark \\
      $m_{d}$           &  0.0830000              &  mass of down-quark \\
      $m_{c}$           &  1.5000000              &  mass of charm-quark \\
      $m_{s}$           &  0.2150000              &  mass of strange-quark \\
      $m_{b}$           &  4.7000000              &  mass of bottom-quark \\
      $m_t$             &  173.0                  &  mass of top quark \\
      $m_H$             &  125.0                  &  mass of Higgs boson \\
   \hline
  \end{tabular}
  \end{center}
\end{table}

\subsection{Numerical results}
\label{sec:App-A-2}

The {\tt DIZET} library, when invoked, provide tabulated $s,t$-dependent
form-factors. It calculates also $M_W$, Stirling parameter $\Delta r, \Delta r_{em}$ and 
flavour-dependent $\sin^2\theta_W^{eff}$ at $Z$ peak.   
In Table~\ref{Tab:Dizet_6.XX_coupl} we have collected numerical results on predicted masses and couplings,
including EW corrections. Those values come directly as control printout from {\tt DIZET 6.XX} code.
In total, evolution of the implemented EW corrections, lead to shift in the predicted $M_W$ by +3 MeV, on-shell $s^2_W$ by -0.00005
and $\sin^2\theta_W^{eff\ lepton}$ by -0.00020. Let us comment on this evolution:
\begin{itemize}
\item
The change in  $\alpha (M_Z^2)$ is due to improvements in the theoretical predictions and experimental low-energy measurements
over last 25 years, and following update in the used parametrization from~\cite{Eidelman:1995ny} to~\cite{Jegerlehner:2017zsb}.
\item  
The $\Delta r$ and  $\Delta r_{rem}$, which are displayed separately, represent gauge invariant corrections to $M_W$ calculation
as shown in formulas~(\ref{Eq:MWiterative}) and~(\ref{Eq:Dr}). The $\Delta r$ is affected by options
used for calculating $\Delta r_L$,
which depends on the flag  {\tt AMT4} used. It also depends on the parametrization of $\Delta \alpha(M_Z)$.
The sensitivity of the  $\Delta r_{rem}$ to all changed introduced between {\tt v6.21} and {\tt v6.45} is almost negligible.  
\item
As a consequence of different predicted $M_W$, the on-shell $s^2_W$ has evolved as well.
\item
Evolution of $\sin^2\theta_W^{eff\ f}$, illustrated with  Fig.~\ref{Fig:sw2effpp}, comes from changing  $s^2_W$ and  ${\mathscr K}^f(s,t)$ form-factors. It
impact  $P^\tau$, $A_{FB}$, $\sigma^{tot}$ too.
\end{itemize}

In Table~\ref{Tab:Dizet_dalf5h_update}, we document impact of  changing only parametrization of $\Delta \alpha^{(5)}_{had}(s)$, with
other parameters and flags unchanged. Dominant effect comes from EW corrections to $M_W$, which shifts its value by + 8.4 MeV, reflected
in change of $s^2_W$  by -0.00016. The impact on the form-factors is less significant and final shift in the
${\sin^{2}\theta_W^{eff}}^{lepton}(M_Z^2)$ is of -0.00023.

In Table~\ref{Tab:Dizet_AMT4_printout}, we document impact of  changing only two-loop corrections to $M_W$, with
other parameters and flags unchanged. The resulting shift on  $M_W$ is smaller, -2.9 MeV only, resulting in + 0.00006 shift
on $s^2_W$ and, while multiplied with form-factors which also have changed,
correspondingly in -0.00008 shift on $\sin^2\theta_W^{eff\ lepton}(M_Z^2)$.

\begin{table}
 \vspace{2mm}
 \caption{The {\tt DIZET 6.45} predictions for two different parametrisations of $ \Delta \alpha_h^{(5)}(M_Z^2)$.
   Other flags as in Table~\ref{Tab:Dizet_6.XX_flags}.} 
 \label{Tab:Dizet_dalf5h_update}
  \begin{center}
    \begin{tabular}{|l|c|c|c|}
        \hline\hline
        Parameter  &  $ \Delta \alpha_h^{(5)}(M_Z^2)$ = 0.0280398 &  $\Delta \alpha_h^{(5)} (M_Z^2)$ = 0.0275762 & $\Delta$\\
                   &   (param. Jegerlehner 1995)                &  (param. Jegerlehner 2017) & \\
         \hline\hline
         $\alpha (M_Z^2)$                              &  0.0077587482      & 0.0077549256  &   \\
         $1/\alpha (M_Z^2)$                            &  128.88676996    & 128.95030224  &  \\
         \hline 
         $M_W$ (GeV)                                   &  80.350538        &  80.358936    &  +8.4 MeV  \\
         $\Delta r$                                    &  0.03690873       &  0.03640338   &     \\
         $\Delta r_{rem}$                               &  0.01168001       &  0.01167960   &     \\
         \hline
         $s^2_W$                                       &   0.22356339       & 0.22340108    &  - 0.00016 \\
         \hline
         $sin^2\theta_W^{eff\ lepton}(M_Z^2)$              &   0.23166087       & 0.23149900    & - 0.00023 \\
         $sin^2\theta_W^{eff\ up-quark}(M_Z^2)$             &   0.23155425       & 0.23139248    & - 0.00016 \\
         $sin^2\theta_W^{eff\ down-quark}(M_Z^2)$           &   0.23142705       & 0.23126543    & - 0.00016 \\
   \hline
 \end{tabular}
  \end{center}
 \vspace{2mm}
 \caption{The {\tt DIZET 6.45} predictions with improved treatment of two-loop corrections. Other flags as in Table~\ref{Tab:Dizet_6.XX_flags}.  } 
 \label{Tab:Dizet_AMT4_printout}
  \begin{center}
    \begin{tabular}{|l|c|c|c|}
        \hline\hline 
        Parameter                                  & AMT4= 4        &  AMT4 = 8        & $\Delta$ \\
         \hline\hline
         $\alpha (M_Z^2)$                          &  0.0077549256113      &    0.0077549256002       &   \\
         $1/\alpha (M_Z^2)$                        &  128.95030206       &    128.95030224       &   \\
         \hline 
         $M_W$ (GeV)                               &   80.361846       &    80.358936      &  - 2.9 MeV  \\
         $\Delta r$                                &   0.03640338      &    0.03640338     &      \\
         $\Delta r_{rem}$                           &    0.01167960     &    0.01167960     &     \\
         \hline
         $s^2_W$                                   &   0.22333971      &    0.22340108      & + 0.00006  \\
         \hline
         $sin^{2}\theta_W^{eff\ lepton}(M_Z^2)$        &   0.23157938      &    0.23149900      & -0.00008 \\
         $sin^2\theta_W^{eff\ up-quark}(M_Z^2)$        &   0.23147290      &    0.23139248      & -0.00008  \\
         $sin^2\theta_W^{eff\ down-quark}(M_Z^2)$      &    0.23134590     &    0.23126543      & -0.00008 \\
   \hline
 \end{tabular}
  \end{center}
\end{table}

\subsection{Parametric uncertainties on $sin^{2}\theta_W^{eff}(M_Z^2)$ predictions}
\label{sec:App-A-3}
We have studied dominant parametric uncertainties from $\Delta \alpha_h^{(5)} (M_Z^2)$ and $m_t$
for  $\sin^{2}\theta_W^{eff}(M_Z^2)$ prediction.
Recent  detailed discussion on the parametric uncertainties of SM parameters
can be found in \cite{Blondel:2019vdq}. 
Both components  of $\sin^{2}\theta_W^{eff}(M_Z^2)=$\\ Re${\mathscr K}^f(M_Z^2,\frac{-M_Z^2}{2}) \cdot s^2_W$  definition
are sensitive to parametric uncertainties.

\begin{itemize}
\item
In Table~\ref{Tab:Dizet_dalf5h_uncertainty} we show impact of changing $\Delta \alpha_h^{(5)} (M_Z^2)$ $\pm$ 0.0001,
which is the uncertainty of the parametrization of~\cite{Jegerlehner:2017zsb}.
The resulting uncertainty on $sin^{2}\theta_W^{eff}(M_Z^2)$ is of $\pm$ 0.000035. 
\item
In Table~\ref{Tab:Dizet_mt_uncertainty} we show impact of changing $m_t \pm$ 0.5 GeV, which is
roughly the anticipated uncertainty of the measurements at LHC~\cite{Aaboud:2018zbu}.
The resulting uncertainty on $sin^{2}\theta_W^{eff}(M_Z^2)$ is of $\pm$ 0.000016.
\end{itemize}

The total parametric uncertainty, added in quadrature, on  $sin^{2}\theta_W^{eff}(M_Z^2)$ is about $\pm 0.00005$.

\begin{table}
 \vspace{2mm}
 \caption{The {\tt DIZET 6.45} predictions: uncertainty from changing $ \Delta \alpha_h^{(5)}(M_Z^2)= 0.0275762$ (param.~\cite{Jegerlehner:2017zsb} ),
       by $\pm$ 0.0001.   } 
 \label{Tab:Dizet_dalf5h_uncertainty}
  \begin{center}
    \begin{tabular}{|l|c|c|c||c|}
        \hline\hline
        Parameter  &  $ \Delta \alpha_h^{(5)}(M_Z^2)$ - 0.0001 &  $\Delta \alpha_h^{(5)} (M_Z^2)$ = 0.0275762 &  $ \Delta \alpha_h^{(5)}(M_Z^2)$ + 0.0001 & $\Delta/2$ \\
         \hline\hline
         $\alpha (M_Z^2)$                              &    0.0077541016       &   0.0077549256   & 0.0077557498     & \\
         $1/\alpha (M_Z^2)$                            &  128.96400565       &   128.95030224   & 128.93659846   & \\
         \hline 
         $M_W$  (GeV)                                  &     80.360747        &   80.358936      & 80.357124     & 1.8 MeV \\
         $\Delta r$                                    &     0.03629414        &  0.03640338      & 0.03651261    &   \\
         $\Delta r_{rem}$                               &     0.01167983        &   0.01167960     & 0.01167938    &    \\
         \hline
         $s^2_W$                                       &   0.22336607          &    0.22340108    & 0.22343610    & 0.000035  \\
         \hline
         $sin^{2}\theta_W^{eff\ lepton}(M_Z^2)$             &  0.23146409           &  0.23149900      &  0.23153392   & 0.000035\\
         $sin^2\theta_W^{eff\ up-quark}(M_Z^2)$             &  0.23135758           &  0.23139248      &  0.23142737   & 0.000035 \\
         $sin^2\theta_W^{eff\ down-quark}(M_Z^2)$           &  0.23123057           &  0.23126543      &  0.23130029   & 0.000035 \\
   \hline
 \end{tabular}
  \end{center}
 \vspace{2mm}
 \caption{The {\tt DIZET 6.45} predictions: uncertainty from changing top-quark mass $m_t$ = 173.0 GeV by $\pm 0.5$ GeV.} 
 \label{Tab:Dizet_mt_uncertainty}
  \begin{center}
    \begin{tabular}{|l|c|c|c||c|}
        \hline\hline
        Parameter  &  $ m_t$ - 0.5 GeV &  $m_t$ = 173.0 GeV &  $m_t$ + 0.5 GeV & $\Delta/2$ \\
         \hline\hline
         $\alpha (M_Z^2)$                          &  0.0077549221     &   0.0077549256    &  0.0077549291  &   \\
         $1/\alpha (M_Z^2)$                        &  128.95036003   &    128.95030224   &  128.95024461   &    \\
         \hline 
         $M_W$  (GeV)                              &  80.355935       &   80.358936       &  80.361941  & 3 MeV \\
         $\Delta r$                                &  0.03658500      &   0.03640338     &   0.03622132  &   \\
         $\Delta r_{rem}$                           &  0.01167011     &   0.01167960     &    0.01168907  &    \\
         \hline
         $s^2_W$                                   &   0.22345908      &    0.22340108     &   0.22334300   & 0.000058  \\
         \hline
         $sin^{2}\theta_W^{eff\ lepton}(M_Z^2)$         &   0.23151389      &    0.23149900     &   0.23148410  & 0.000016\\
         $sin^2\theta_W^{eff\ up-quark}(M_Z^2)$         &   0.23140736      &   0.23139248      &   0.23137758  & 0.000016 \\
         $sin^2\theta_W^{eff\ down-quark}(M_Z^2)$       &   0.23128031      &   0.23126543      &    0.23125053  & 0.000016 \\
   \hline
 \end{tabular}
  \end{center} 
\end{table}

\section{Technical documentation of upgrades for {\tt TAUSPINNER} electroweak
re-weighting code} \label{app:B}
In {\tt TauSpinner} the Improved  Born of (\ref{Eq:BornEW}) is coded as default for its {\tt nonSMBorn} function.
If form-factors are available, look-up tables present, then they will be used for re-weighting algorithm.
At  default, it will be then assumed\footnote{If it is not the case, the weight will not be
  appropriate if {\tt TauSpinner} initialization is not adjusted.}
that sample was generated with Effective Born
{\tt Tauola}/LEP  variant, see Table~\ref{Tab:BornEff}.  A wealth of options is available for EW form-factors
calculation, see Section \ref{app:iniflags}, Table~\ref{Tab:Dizet_6.XX_flags} and Section \ref{app:kkmc}.
Choice of options simplifying Improved Born to the cases closer, or to effective Born itself,  are listed in
Table \ref{Tab:KEYGSW}.

To monitor in a quick manner look-up tables with form-factors, root scripts
{\tt FFdraw.C} and  {\tt FFdrawDwa.C} are
provided, see Appendix \ref{sec:onLIbraries}.
These scripts provide semi-analytical results for 
Born-level cross section,  $A_{FB}$ and $A_{pol}$ as a function
of centre-of-mass energy, for incoming $e^+e^- (\mathrm{or}\; u\bar u, d \bar d)$ pairs. Two versions can be compared;
e.g.  EW improved Born with default Effective {\tt Tauola/LEP} Born of {\tt TauSpinner},
or of two variants of EW-initialization.

In the distribution tar-ball~\cite{tauolaC++} of {\tt Tauola/TauSpinner} there are two example
main programs, which demonstrate how re-weighting of EW effects is implemented
(on the {\tt LHE} and {\tt HepMC} event formats) and  how analytically these effects can be monitored. These programs,
respectively {\tt tau-reweight-test.cxx }
and {\tt Dizet-example/table-parsing-test.cxx} need explanation 
because they evolved with time and more
initialization options were introduced.  In
particular, to evaluate numerical
differences between Effective Born and EW Improved Born as well as of the
intermediate variants; the steering flag for the variants is {\tt keyGSW}.
Note, that to re-weight, program needs to
flip for each event, initialization between two variants and save some
intermediate results to avoid massive recalculations. The semi-analytical
{\tt table-parsing-test.cxx } is obviously much faster. It is useful for study 
of small effects and not only to checks of the correctness for EW form-factors tabulation.

Let us collect technical details  for these  programs and dependencies between routines:

\begin{itemize}
\item  {\tt Dizet-example/table-parsing-test.cxx}.
This semi-analytic program is fast, does not require continuous 
interchanging between variants of initialization and is suitable to study small variation for predictions
due to initialization fine tuning. It can be set for incoming
$e^+e^- (\mathrm{or}\; u\bar u, d \bar d)$ pairs.
It is obviously unable to tackle experimental selection. 
\item
  For {\tt tau-reweight-test.cxx } event re-weighting is demonstrated
  ({\tt LHE} or {\tt HepMC} format can be used) and
for each event, two variants of Born are calculated. That is why flag {\tt keyGSW}
can not be set  once; method {\tt calculateWeightFromParticlesH()} is executed
twice\footnote{ The  {\tt getWtNonSM()} returns weight for
  correction with respect to calculation of default SM
  calculation as initialized in {\tt tauola universal interface}.
  It may be also suitable to activate call on  {\tt EWreInit(1/2)}, for both variants
  of {\tt calculateWeightFromParticlesH()} calculation. The {\tt EWreInit(1/2)}
is expected to be adjusted by the user to actual needs.}
and ratio of the result is used.\\
At the beginning {\tt EWanomInit} (defined locally in demo) is called and
defines initialization variant  with the help of 
   {\tt ExtraEWparamsSet()} method. \\
Next, for each event in the loop,
{\tt   EWreInit} (it is expected to be adjusted by the user) re-initializes {\tt keyGSW} and other parameters,
with the, otherwise dummy,  call on
   {\tt sigbornsdelt}.
 \item One should note that   {\tt sigbornsdelt}  call  in {\tt tau-reweight-test.cxx } returns dummy variable. The call is to pass  {\tt keyGSW}
   only. Returned value is used   in {\tt Dizet-example/table-parsing-test.cxx}
   though.
\item
In all the code of  {\tt src/ew\_born.cxx} and  {\tt src/tau\_reweight\_lib.cxx}
the {\tt keyGSW} is not used. Corresponding routines are EW variants independent.     
\end{itemize}

The {\tt src/initwsw.f}  is the file where routines
for calculating Effective- and/or EW Improved-Born are placed.

In the  {\tt src/nonSM.cxx}  file, the method
     {\tt  default\_nonSM\_bornZ()} resides. 
     The  class data member variable {\tt m\_keyGSW} is used for {\tt keyGSW},
      it is accessed  with the help of the
     {\tt keyGSWGet(keyGSW)} method;   options are collected in
     Table~\ref{Tab:KEYGSW}. In Table \ref{Tab:gswnames} names of variables corresponding to
     the form-factors, introduced in
     Eq.~(\ref{Eq:BornEW}) of Improved Born, which are also  used in {\tt KKMC} and in {\tt Dizet-example}
     scripts of  {\tt TauSpinner},
     are explained. These form-factors are used when 
      {\tt t\_bornew\_} calculation is invoked.

\begin{table}
 \vspace{2mm}
 \caption{EW form-factors of formula~(\ref{Eq:BornEW}) calculated by {\tt DIZET} and provided by {\tt KKMC} tabulation code.
   Names as used in: formula~(\ref{Eq:BornEW}),
   {\tt KKMC} tabulation code and in {\tt Dizet-example} are collected. In Effective Born  Eq.~(\ref{Eq:Borneff})
   numerical constants  replace form-factors of Improved Born ~(\ref{Eq:BornEW}).
   Depending on Effective Born version (see Table~\ref{Tab:BornEff})  $\sin^2\theta_W $ of Improved Born is replaced by
   $\sin^2\theta_W^{eff} $ or by flavour-dependent variants $\sin^2\theta_W^{eff\; l/up/dow} $, the
   ${ \rho_{\ell f}(s,t)}$ is then replaced
   with constant $\rho_{\ell f}$. Similarly $\alpha(0)$ is replaced with $\alpha(M_Z)$  } 
 \label{Tab:gswnames}
  \begin{center}
    \begin{tabular}{|l|c|c|l|}
        \hline\hline
        Form-factor  &  in {\tt KKMC}  &  in {\tt Dizet example}     &  in Effective Born\\
         \hline\hline
         ${ \rho_{\ell f}(s,t)}$    &  {\tt GSW(1)}      & {\tt FF1}   & 1     \\
         {\tt v1 }                        &                    &             & 1.005  \\
         {\tt v2}                         &                    &             & 1.005403\; (up)    \\
                                      &                    &             & 1.005889\; (down)  \\
         ${ K_e(s,t)}$            &  {\tt GSW(2)}      & {\tt FF2}   & 1      \\
         ${ K_f(s,t)}$            &  {\tt GSW(3)}      & {\tt FF3}   & 1      \\
         ${ K_{ef}(s,t)}$          &  {\tt GSW(4)}      & {\tt FF4}   & 1      \\
         $--$                         &  {\tt ---}         & {\tt ---}   &  -     \\
         ${ \Pi_{\gamma \gamma}(s)}$ &  {\tt GSW(6)}      & {\tt FF6}   & 1      \\
   \hline
 \end{tabular}
  \end{center}
\end{table}

    From the point of view of EW calculations most of the
    variants are set and stored in {\bf  {\tt EWtables.cxx} }.
    In particular
    class/file  variable {\tt m\_keyGSW} is stored among other variables of
    initialization. To access or modify
    {\tt ExtraEWparamsGet} and 
    {\tt ExtraEWparamsSet}  are prepared. These parameters
    used in  {\tt initEWff} passed with {\tt ExtraEWparamsGet} in particular
     {\tt keyGSW}  is passed to {\tt initwkswdelt\_}
    
    In the {\bf  {\tt EWtables.cxx}, } the code 
    for {\tt sigbornswdelt} and {\tt AsNbornswdelt} are stored. Through these
    methods 
    {\tt keyGSW} is passed to {\tt t\_bornew\_}. The {\tt AsNbornswdelt};
    a near clone of {\tt sigbornswdelt}, is used in drawing scripts for asymmetries.

   It is important to note, that the functions {\tt t\_born} and {\tt t\_bornnew} which are used in {\tt tauola universal interface} \cite{Davidson:2010rw} and
   in {\tt TauSpinner}, are normalized to lowest order Born cross section, of photon exchange only and   mass effects excluded.
   That means 
 $\frac{d \sigma^{q \bar q}_{Born}}{d \cos\theta} (s, \cos\theta, p)$ is multiplied by $\frac{2s}{\pi \alpha^2}$ ($\alpha=\alpha_{QED}(0)$.
   Thus, Born  expression used in the weight calculations should approach $q_f^2q_l^2(1 + \cos^2\theta)$ at low energies.
   This condition, defines  normalization for the
   EW-improved  (or other e.g. non SM variant) of user provided Born\footnote{
      The method for non SM Born can be replaced with the
      pointer to the one of  user choice.}.
   This is important 
 if weights are used to relate cross sections obtained  from {\tt t\_born} and {\tt t\_bornnew}. It is of no importance if
 only spin  weight or weight based
 solely on user provided Born is used.
   This reference {\tt t\_born} of {\it Effective Born} (inherited from {\tt Tauola}) can be easily modified/replaced by the user,
   as well as is the case with independently initialized  {\tt t\_bornnew} of {\it Improved Born}.

In {\tt TauSpinner} option to
improve precision of generated MC events and reweight from ``fixed'' to ``running'' width propagator, see Appendix
\ref{app:Zwidth} for explanation,
is available. For that {\tt TauSpinner} initialization with {\tt KEYGSW=11}, {\tt KEYGSW=10}
and {\tt KEYGSW=2} is prepared. For  {\tt KEYGSW=11}, {\tt KEYGSW=12}
and {\tt KEYGSW=13} results corresponding to Effective Born, variants {\tt v0,\;v1,\;v2} respectively can be
obtained, see Table \ref{Tab:KEYGSW}.  Let us explain now  meaning of all other 
{\tt sigbornswdelt(mode,ID, s, cc, SWeff, DeltSQ, DeltV, Gmu, alfinv, AMZ00, GAM00, KEYGSW)} input parameters.
{\tt mode} can be set to 0 or 1. In the second case {\tt SWeff,  AMZ00, GAM00} for  $\sin^2\theta_W^{eff}, M_Z,\Gamma_Z$
will be overwritten with values stored in EW tables calculated  with the {\tt DIZET} library.
{\it ID=0,1,2} denotes that calculation is performed respectively for outgoing lepton  or down/up quark pair. 
The {\tt s, cc} denote Mandelstam variable and scattering angle. Anomalous coupling $\delta_{S2W}$ and $\delta_V$ of
\cite{Richter-Was:2018lld} Appendix B, are initialized with   {\tt DeltSQ,\; DeltV} respectively, finally also  $G_F$, $1/\alpha$ with  {\tt Gmu,\; alfinv}.
\noindent
The {\tt AsNbornswdelt() } feature the same set of input parameters, but returns difference for cross section of forward and
backward hemispheres instead of the sum.

For important technical details, {\tt README} files and comments in the code of
the distribution tar-ball \cite{tauolaC++} can be helpful.

\section{ Initialization of variants for EW Improved Born }
\label{app:variantseff}
In the previous appendix we have completed presentation of some easy to activate in {\tt TauSpinner}
options. In many cases, it is sufficient to change some well defined keys and/or input parameters such as
$Z$ boson mass or some couplings. This of course shifts the results. The $\sigma^{tot}$, $A_{FB}$
and $P_\tau$ predictions at the $Z$-pole are collected in Table \ref{Tab:KEYGSWr}, for incoming $e^+e^-$, up or down quarks. 

One can see that some options e.g. {\tt KEYGSW=0,2,4,10} are prepared for technical tests, rather than for
evaluation of physics ambiguities, while  other options are more useful.
All these options are useful for test of particular parts of EW predictions obtained with a given
version of EW form-factors. This supplements discussed earlier options of EW form-factors
initialization, see Appendix \ref{app:iniflags} and Table \ref{Tab:Dizet_6.XX_flags}.
Further, of more historical nature
tests, with EW form-factors
calculated with older versions of EW {\tt DIZET} library  presented in Appendix \ref{app:kkmc},
are collected in Fig.~\ref{figTest2} and Table.~\ref{Tab:Dizet_6.XX_coupl}. Results of the
present Appendix are to supplement discussion of reliability and limitation of the Effective Born variants as compared
with Improved Born, in general and in the context of particular applications.

The Effective Born, variants {\tt v0, v1, v2}, require change of input parameters. Then, {\tt mode=0} and 
{\tt KEYGSW} at 11, 12 or 13 should be respectively set. One should notice some shifts of Table \ref{Tab:KEYGSWr}
results with respect to the ones presented e.g. in Tables \ref{Tab:EWnormcorr}, \ref{Tab:AFBEWcorr}.
Note, that  we do not average over energy ranges and 
incoming quark flavours now.

\begin{table}
 \vspace{2mm}
  \begin{center}
    \caption{ Initialization variants  for
      non-standard  Born of quark level Drell Yan $2 \to 2$ processes.
      It can be used to impose with the event weight
      EW loop effects on event samples.
      Variants are steered by the {\tt keyGSW} parameter. Corresponding code
      is stored in:
      (A) -  {\tt INITWKSDELT},
      (B) -  {\tt T\_BORNEW} and   (C) -{\tt EWtables.cxx}.
      Fixed, running and fixed rescaled $\Gamma_Z$ correspond respectively
      to Eqs. (\ref{Eq:propz}), (\ref{Eq:propzfix}) and (\ref{Eq:propzst}).
      Further combination of options  can be set by the simple re-coding.
      The initialization of effective  {\tt Tauola} Born is not
      affected by these options. It is performed elsewhere.
      For {\tt KEYGSW=11,12,13}, when {\tt mode=0} results of Effective Born variants {\tt v0, v1, v2}
      can be obtained.
               \label{Tab:KEYGSW}
    }
  \vspace{2mm}
    \begin{tabular}{|c|l|c|c|}
    \hline
    KEYGSW   &  A: {\tt VVCor} & B: propagator & C: form-factors            \\
    \hline
    0  &  1 & photon propagator off, fixed $\Gamma_Z$  & all {\tt FFi}= 1        \\
    1  & on    & running $\Gamma_Z$,                      &  all {\tt FFi} from EW tables      \\
    2  &  1 & fixed $\Gamma_Z$                        &  all {\tt FFi}= 1      \\
    \hline
    3  &  1 & running $\Gamma_Z$                      &  all {\tt FFi}= 1 but {\tt FF6}$=\Pi_{\gamma\gamma}(M_Z^2)$          \\
    4  &  1 & running   $\Gamma_Z$                    &  all {\tt FFi}= 1 but {\tt FF6,FF1}$= \rho_{\ell f}(M_Z^2,-M_Z^2/2), \Pi_{\gamma\gamma}(M_Z^2)$  \\
    5  &  1    & running $\Gamma_Z$                   & all  {\tt FFi} from EW tables calculated at  $(M_Z^2,-M_Z^2/2)$   \\
    \hline
    10 &  1 & fixed $\Gamma_Z$, rescaled             & all {\tt FFi}= 1       \\
    11 &  1 & running $\Gamma_Z$                      & all {\tt FFi}= 1,   can be used for Effective Born $v0$      \\
    \hline
    12 &  1 & running $\Gamma_Z$                      &  {\tt FFi} set as for Effective Born $v1$         \\
    13 &  1 & running $\Gamma_Z$                      &  {\tt FFi} set as for Effective Born $v2$       \\
    \hline
    \end{tabular}
  \end{center}
\end{table}
     
\begin{table}
 \vspace{2mm}
  \begin{center}
    \caption{ Numerical results for initialization variants  as
      explained in Table
      \ref{Tab:KEYGSW}. Numerical results for {\tt v0, v1, v2} are also
      provided, then in addition to {\tt KEYGSW}=11,12 or 13,
      input parameters for {\tt sigbornswdelt() }, {\tt AsNbornswdelt()} need to be adjusted and {\tt mod=0}.
        The $\alpha(M_Z^2)/\alpha(0)$ factors entering  cross sections normalization are dropped out from the $\sigma^{tot}$ ratios. \newline
                \label{Tab:KEYGSWr}
    }
\vspace{2mm}
    \begin{tabular}{|c|l|c|c|}
    \hline
    KEYGSW   &  $\frac{\sigma_{tot}^{e^+e^-}(M_Z)}{\sigma_{tot\; improved}^{e^+e^-}(M_Z)}$ &   $P_\tau^{e^+e^-}(M_Z)$ & $A_{FB}^{e^+e^-}(M_Z) $         \\
    \hline
    0   & 0.989939   & 0.1463264 &  0.0161546   \\
    1   & 1.000000   & 0.1449616 &  0.0177039   \\
    2   & 1.000736   & 0.2093134 &  0.0330312   \\
    3   & 1.001438   & 0.2094436 &  0.0349617   \\
    4   & 1.011539   & 0.2093809 &  0.0344227   \\
    5   & 1.000001  & 0.1449558 &  0.0176505   \\   
    10  & 1.002227   & 0.2093154 &  0.0330336   \\
    11  & 1.000736   & 0.2093134 &  0.0330312   \\
    \hline
11 {\tt v0} & 0.9899340    & 0.1463264 &   0.0161546  \\    
12 {\tt v1} & 0.9999098    & 0.1463351 &   0.0161556  \\
13 {\tt v2} & 0.9999098    & 0.1463351 &   0.0161556  \\    
    \hline
    \end{tabular}
    \begin{tabular}{|c|l|c|c|c|c|c|}
    \hline
    KEYGSW   &  $\frac{\sigma^{u\bar u}_{tot}(M_Z)}{\sigma^{u\bar u}_{tot\; impr.}(M_Z)}$ &   $P^{u\bar u}_\tau(M_Z)$ & $A^{u\bar u}_{FB}(M_Z) $    &  $\frac{\sigma^{d\bar d}_{tot}(M_Z)}{\sigma^{d\bar d}_{tot\; impr.}(M_Z)}$ & $P^{d\bar d}_\tau(M_Z)$ & $A^{d\bar d}_{FB}(M_Z) $        \\
    \hline
    0    &  0.989244  &  0.146859 & 0.073530 & 0.988081 & 0.1471353 & 0.1032379  \\
    1    &  1.000000  &  0.146990 & 0.074664 & 1.000000 & 0.1476778 & 0.1038546  \\
    2    &  1.009430  &  0.209984 & 0.109447 & 1.003882 & 0.2103266 & 0.1483706  \\
    3    &  1.009759  &  0.210529 & 0.110548 & 1.003976 & 0.2107218 & 0.1488000  \\
    4    &  1.020676  &  0.210387 & 0.110274 & 1.015833 & 0.2106413 & 0.1487130  \\
    5    &  0.999999  &  0.146974 & 0.074633 & 0.999999 & 0.1476667 & 0.1038425  \\   
    10   &  1.010939  &  0.209986 & 0.109449 & 1.005384 & 0.2103272 & 0.1483712  \\
    11   &  1.009430  &  0.209985 & 0.109447 & 1.003882 & 0.2103266 & 0.1483706  \\
    \hline
 11 {\tt v0} &  0.989244  &  0.146859 & 0.073530 & 0.988081 & 0.1471353 & 0.1032379 \\
 12 {\tt v1} &  0.999244  &  0.146863 & 0.073532 & 0.998087 & 0.1471360 & 0.1032383 \\
 13 {\tt v2} &  1.000127  &  0.146863 & 0.073573 & 1.000039 & 0.1471361 & 0.1032550 \\
    \hline
    \end{tabular}
  \end{center}
\end{table}

\section{The $s$-dependent Z-boson width}
\label{app:Zwidth}

In formula (\ref{Eq:BornEW}) for the definition of $Z$ propagator running width is used:
\begin{equation}
  \chi_Z(s) =  \frac{1}{s - M_Z^2 + i \cdot \Gamma_Z \cdot s / M_Z}.\label{Eq:propz}
\end{equation} 
The form-factors of eq.~(\ref{Eq:BornEW}) are calculated for the on mass-shell (nominal) value of $ M_Z$. The introduction of so-called $s$-dependent
width is equivalent to partial resummation to higher orders of dominant loop correction: the boson $s$-dependent self-energy.
In fact such  resummation, running $Z$ width,   was used in many analyses of LEP I era.

However, in  Monte Carlos and strong interaction calculations of LHC era, the  $Z$ propagator of constant width is often used:
\begin{equation}
  \chi^{'}_Z(s) =  \frac{1}{s - M_Z^2 + i \cdot \Gamma_Z \cdot M_Z}. \label{Eq:propzfix}
\end{equation}

One can ask the  question, how  analytic forms of (\ref{Eq:propz}) and (\ref{Eq:propzfix}) translate to each other.
In fact, this well known translation is known at least since Ref.~\cite{Bardin:1988xt} published more than 30 years ago,
but let us readdress it 
for the reference again. 
From  Eq.~(\ref{Eq:propz}), we obtain Eq.~(\ref{Eq:propzfix}) if  the following redefinitions are used
 \begin{eqnarray}
  \chi_Z(s)&=&  \frac{1}{s(1+ i \cdot \Gamma_Z  / M_Z) - M_Z^2 }\nonumber \\
&=& \frac{(1- i \cdot \Gamma_Z  / M_Z)}{s(1+  \Gamma_Z^2  / M_Z^2) - M_Z^2(1- i \cdot \Gamma_Z  / M_Z) } \nonumber \\
&=& \frac{(1- i \cdot \Gamma_Z  / M_Z)}{(1+  \Gamma_Z^2  / M_Z^2)}\frac{1}{s - \frac{M_Z^2}{1+  \Gamma_Z^2  / M_Z^2}+  i \cdot \frac{ \Gamma_Z   M_Z}{1+  \Gamma_Z^2  / M_Z^2} }\nonumber \\
&=& N_Z^{'}\frac{1}{s - {{M^{'}_Z}}^2+  i  \Gamma^{'}_Z   M^{'}_Z }\nonumber \\
M^{'}_Z&=&\frac{M_Z}{\sqrt{1+  \Gamma_Z^2  / M_Z^2}} \nonumber \\
\Gamma^{'}_Z&=& \frac{\Gamma_Z}{\sqrt{1+  \Gamma_Z^2  / M_Z^2}}\nonumber \\
N_Z^{'}&=& \frac{(1- i \cdot \Gamma_Z  / M_Z)}{(1+  \Gamma_Z^2  / M_Z^2)}= \frac{(1- i \cdot \Gamma^{'}_Z  / M^{'}_Z)}{(1+  {\Gamma^{'}_Z}^2  / {M^{'}_Z}^2)}
\label{Eq:propzst}
\end{eqnarray}

The $s$-dependent width in $Z$ propagator translates into  mass and width shift and introduction of the 
complex factor in front of the constant width $Z$ propagator.
This last point is possibly least trivial as it effectively means redefinition of $Z$ coupling.
That is why it can not be understood as parameter rescaling. It points to present in higher order relations between vacuum polarization and vertex.
Most of the changes are due to the term $\Gamma_Z^2  / M_Z^2$ except of the overall phase which result
from $1- i \cdot \Gamma_Z  / M_Z$ factor and which change the $\gamma-Z$ interference.
The shift in $M_Z$ is by about 34 MeV downwards, and the shift in $\Gamma_Z$ by 1 MeV, due to the reparametrisation of the
Z-boson propagator.

\begin{figure}
  \begin{center}                               
{
  \includegraphics[width=7.5cm,angle=0]{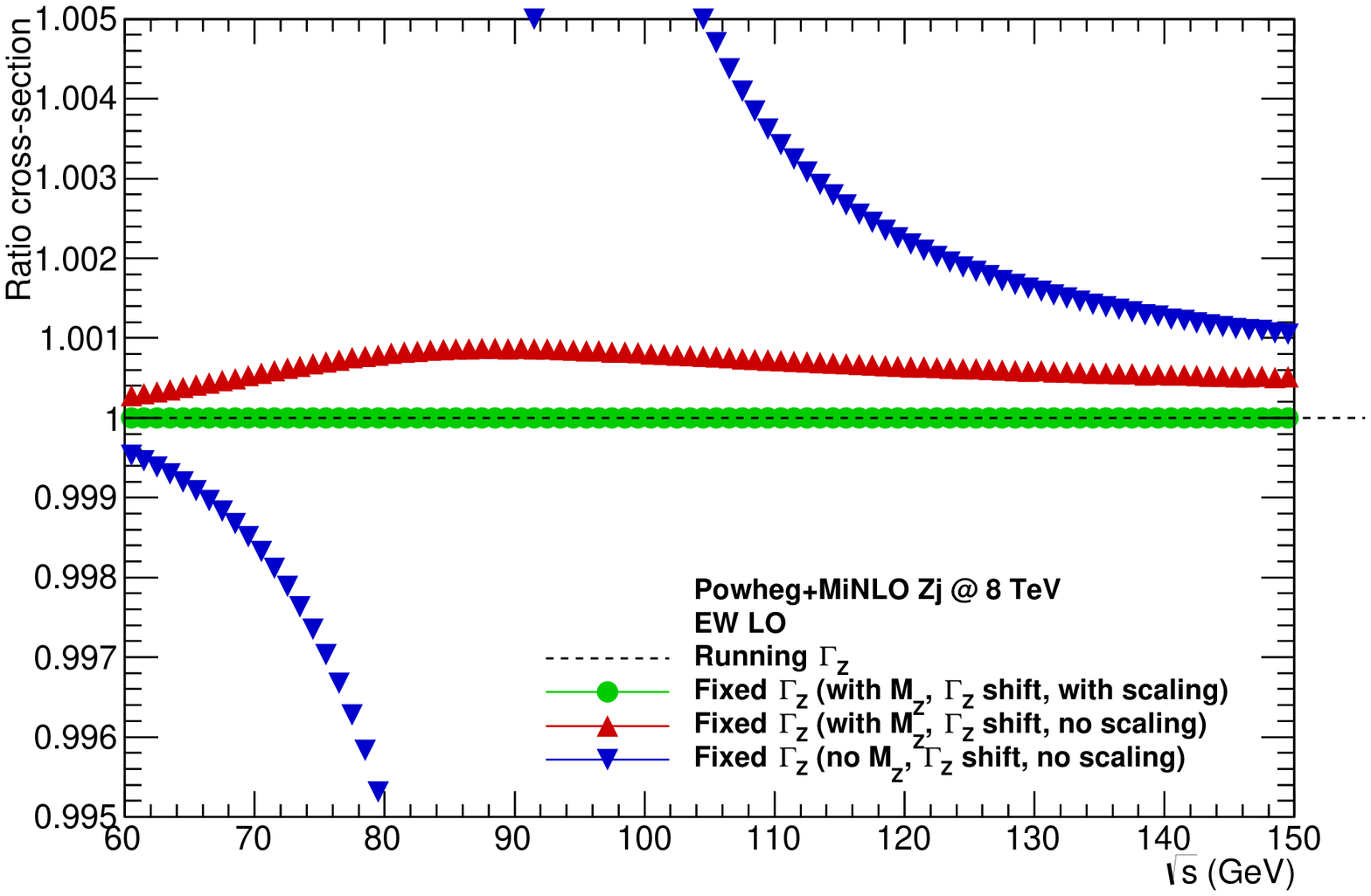}
  \includegraphics[width=7.5cm,angle=0]{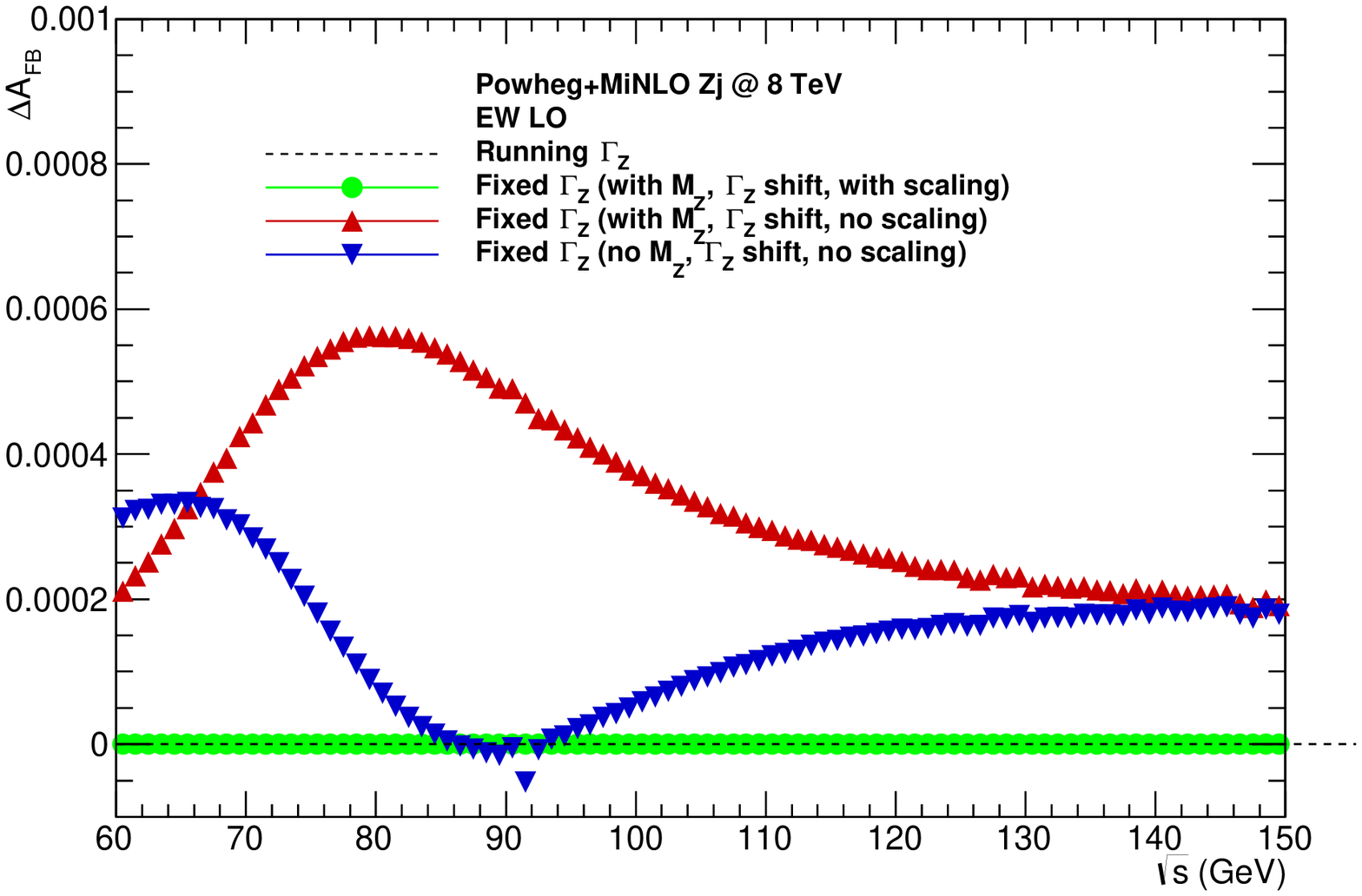}
}
\end{center}
  \caption{ Ratio of the cross sections (left) and $\Delta A_{fb}$ (right) for for EW LO but different form of Z-boson propagator,
    see text.
    The MC $pp \to Zj;\; Z \to l^+l^-$ events were used for estimations.
\label{Fig:ZGamma} }
\end{figure}

In Figure~\ref{Fig:ZGamma} shown is comparison of the cross sections and $A_{fb}$, between different implementations of $\chi_Z(s)$.
Dashed line of reference corresponds to using formula~(\ref{Eq:propz}). Green line
to complete formula~(\ref{Eq:propzst}).
Red line corresponds to  formula~(\ref{Eq:propzst}) but without $N_Z^{'}$ scaling and blue line to formula ~(\ref{Eq:propzfix}),
with nominal $M_Z$ and $\Gamma_Z$.

It is common for LHC MC generators to use formula (\ref{Eq:propzfix}) for $Z$ propagator, with $M_Z$ and $\Gamma_Z$ of nominal,
on-mass-shell values. Numerically this is better approximation than with shifted $M_Z$ and $\Gamma_Z$ but
$N_Z^{'}$ missing. This observation
is true both for EW LO and EW corrected calculations; for cross section and $A_{FB}$. Quantitative estimates
are collected in Tables \ref{Tab:GamZprop} and \ref{Tab:GamZpropAsym}.

Note that when  options of ``running $\Gamma_Z$'' and `` fixed $\Gamma_Z$'' are compared, the same EW corrections
in both cases tuned to ``running $\Gamma_Z$'' convention are used. It is beyond the scope of the paper to investigate,
how NLO+HO corrections,  calculated with the fixed width/ pole mass convention, gradually mitigate (as they should) discrepancy
observed at EW LO level between Eq. \ref{Eq:propz} and Eq. (\ref{Eq:propzst}) definition of $Z$ propagator without $N_Z^{'}$ included.

  \begin{table}
 \vspace{2mm}
 \caption{Ratio of the cross sections ($\sigma$), calculated  with different form of $Z$-boson propagator and 
   integrated over outgoing lepton pair
   mass windows.
   It is shown for  EW LO and EW NLO+HO predictions of $O(\alpha(0)$  EW scheme for $pp$ collisions at 8 TeV center of mass energy,
   while EW NLO+HO corrections are tuned to running $\Gamma_Z$ convention.
 \label{Tab:GamZprop}}
 \begin{center}
    \begin{tabular}{|l|c|c|c|c|c|}
        \hline\hline
        ${\sigma(Fixed)}/{\sigma(Running)}$ &                  &             &              &                &                \\
         $m_{ee}$ ranges (in GeV):             & $ 90.5 - 91.5$   & $ 89 - 93$  & $ 60 - 81$   & $ 81 - 101$    & $ 101 - 150$   \\ 
        \hline \hline
         at EW LO:  & & & & &  \\
         with $M_Z, \Gamma_Z$ shift, no $N_Z^{'}$  & 1.00087 & 1.00087  & 1.00062   & 1.00086    & 1.00071    \\
         no $M_Z, \Gamma_Z$ shift, no  $N_Z^{'}$    & 0.99620 & 1.00074  & 0.99716   & 0.99977    & 1.00392    \\
         \hline 
         at EW NLO+HO: & & & & & \\
         with $M_Z, \Gamma_Z$ shift, no  $N_Z^{'}$   & 1.00113 & 1.00085  & 1.00043   & 1.00083    & 1.00075    \\
         no $M_Z, \Gamma_Z$ shift, no    $N_Z^{'}$  & 0.99746 & 1.00122  & 0.99719   & 1.00013    & 1.00392    \\
    \hline
 \end{tabular}
  \end{center}
 \vspace{2mm}
 \caption{Difference of $A_{fb}$ calculated for different form of $Z$-boson propagator and 
   for integrated  outgoing lepton pair
   mass windows.
   It is shown for EW LO and EW NLO+HO predictions of $O(\alpha(0)$  EW scheme for $pp$ collisions at 8 TeV center of mass energy,
   while EW NLO+HO corrections are tuned to running $\Gamma_Z$ convention.
 \label{Tab:GamZpropAsym}}
 \begin{center}
    \begin{tabular}{|l|c|c|c|c|c|}
        \hline\hline
        $ A_{fb}$ (Running) -  $ A_{fb}$ (Fixed) &  &   &   &    &  \\
         $m_{ee}$ ranges (in GeV):             & $ 90.5 - 91.5$   & $ 89 - 93$  & $ 60 - 81$   & $ 81 - 101$    & $ 101 - 150$   \\ 
         \hline \hline
         at EW LO & & & & & \\
         with $M_Z, \Gamma_Z$ shift, no  $N_Z^{'}$   & -0.00048  & -0.00047  & -0.00047   & -0.00047    & -0.00030    \\
         no $M_Z, \Gamma_Z$ shift, no   $N_Z^{'}$    & -0.00006  & -0.00026  & -0.00012   & -0.00040    & -0.00005    \\
         \hline 
         at EW NLO+HO & & & & & \\ 
         with $M_Z, \Gamma_Z$ shift, no  $N_Z^{'}$   & -0.00053 & -0.00053  & -0.00052   & -0.00053    & -0.00024    \\
         no $M_Z, \Gamma_Z$ shift, no  $N_Z^{'}$     & -0.00007  & -0.00030  & -0.00026   & -0.00048    & -0.00004    \\
    \hline
 \end{tabular}
  \end{center}
\end{table}

\section{Versions of {\tt DIZET} library}\label{app:kkmc}
In the {\tt KKMC} distribution tar-ball, explained in Ref.~\cite{Arbuzov:2020coe},  the code for calculation of
EW corrections is stored in directory {\tt KK-all/dizet}.
The program stored in that directory calculates EW form-factors
and writes them into {\tt ASCII} format text files. To change the version
of the form-factors, requires simply use of these tables calculated with different
version of library or with different set of initialization parameters.
In parallel to program stored in {\tt KK-all/dizet} which calculates EW
form-factors as stored with Ref.~\cite{Jadach:2013aha} the following new ones,
{\tt dizet-6.42-cpc},
{\tt dizet-6.42} and 
{\tt dizet-6.45} were prepared and can be used following old instruction of
{\tt KKMC} documentation. There was only one change introduced,
convenient for table reading by the {\tt TauSpinner} package \cite{Richter-Was:2018lld},
the table write format for EW form-factors was modified
to {\tt KK-all/dizet/BornV.h} and if tables are used with {\tt KKMC}
the original one of {\tt KK-all/bornv/BornV.h} file should be re-installed.
This is true for all mentioned above {\tt DIZET 6.xx} variants.

\begin{itemize}
\item {\tt DIZET 6.21} ({\tt dizet}) is distributed with   {\tt KKMC} through CPC
  EW corrections and in particular vacuum polarization was not
  updated for backup compatibility. This version of library is documented in \cite{Bardin:1989tq}.
  Note that  upgrades of EW corrections within LEP experiments were not always well documented.
  This is in particular true for the photon vacuum polarization $\Pi_{\gamma\gamma}(s)$.
\item  {\tt DIZET 6.42}  ({\tt dizet-6.42-cpc}) as in published ZFITTER \cite{Arbuzov:2005ma}.  This is
  the last published/archived version of {\tt DIZET} code.  Note that as a  default $\Pi_{\gamma\gamma}$ from
  ref \cite{Eidelman:1995ny} is still mentioned but obviously it was upgraded for the final versions of LEP
  data analysis of Ref.~\cite{ALEPH:2005ab}.
\item {\tt DIZET 6.42} ({\tt dizet-6.42}) with  $\Pi_{\gamma\gamma}$ updated to {\tt hadr5n17\_compact.f} of
  Ref.~\cite{Jegerlehner:2019lxt}. Parametrization taken from author web page, dated
  Oct  8 02:19:56 2017.
\item {\tt DIZET 6.45} ({\tt dizet-6.45}) VERSION 6.45 (30 Aug. 2019) with the  vacuum polarization code  and  fermionic two loops corrections,
  {\tt AMT4} flag upgraded by {\tt DIZET } authors themselves.
  \end{itemize}

Comparison of results for these versions are given in Fig.~\ref{figTest2}
and in Table \ref{Tab:Dizet_6.XX_coupl}.

\end{document}